\documentclass[journal]{IEEEtran}
\ifCLASSINFOpdf
\else
\fi
\usepackage[ruled]{algorithm2e} 

\usepackage{multicol, blindtext} 
\usepackage{multirow}  
\usepackage{tabularx}
\usepackage{longtable}
\usepackage{longtable}
\usepackage{graphicx,caption,subcaption}
\newcommand{\tabitem}{~~\llap{\textbullet}~~}
\DeclareCaptionLabelFormat{AppendixTables}{Appendix \Alph{table}}
\usepackage{soul}
\usepackage{tikz}
\usepackage{balance} 
\def\checkmark{\tikz\fill[scale=0.2](0,.25) -- (.15,0) -- (0.7,.5) -- (.20,.15) -- cycle;} 
\usepackage{pdflscape}
\begin{document}

\title{Machine Learning for Detecting Data Exfiltration: A Review}
\author{\IEEEauthorblockN{{\bf Bushra Sabir}\IEEEauthorrefmark{1}\IEEEauthorrefmark{2}}
\and\IEEEauthorblockN{{\bf Faheem Ullah}\IEEEauthorrefmark{1}}
\and\IEEEauthorblockN{{\bf M. Ali Babar}\IEEEauthorrefmark{1}\IEEEauthorrefmark{3}}
\and\IEEEauthorblockN{{\bf Raj Gaire}\IEEEauthorrefmark{2}}\\
\and\IEEEauthorblockA{\IEEEauthorrefmark{1}CREST - The Centre for Research on Engineering Software Technologies, University of Adelaide, Australia}\IEEEauthorblockA{\IEEEauthorrefmark{2} CSIRO's Data61, Australia}\IEEEauthorblockA{\IEEEauthorrefmark{3} CSCRC - Cyber Security Cooperative Research Centre, Australia}}
\maketitle
\begin{abstract}
  \textbf{Context}: Research at the intersection of cybersecurity, Machine Learning (ML), and Software Engineering (SE) has recently taken significant steps in proposing countermeasures for detecting sophisticated data exfiltration attacks. It is important to systematically review and synthesize the ML-based data exfiltration countermeasures for building a body of knowledge on this important topic. \textbf{Objective}: This paper aims at systematically reviewing ML-based data exfiltration countermeasures to identify and classify ML approaches, feature engineering techniques, evaluation datasets, and performance metrics used for these countermeasures. This review also aims at identifying gaps in research on ML-based data exfiltration countermeasures. \textbf{Method}: We used Systematic Literature Review (SLR) method to select and review {92} papers. \textbf{Results}: The review has enabled us to (a) classify the ML approaches used in the countermeasures into data-driven, and behaviour-driven approaches, (b) categorize features into six types: behavioral, content-based, statistical, syntactical, spatial and temporal, (c) classify the evaluation datasets into simulated, synthesized, and real datasets and (d) identify 11 performance measures used by these studies. \textbf{Conclusion}: We conclude that: (i) the integration of data-driven and behaviour-driven approaches should be explored; (ii) There is a need of developing high quality and large size evaluation datasets; (iii) Incremental ML model training should be incorporated in countermeasures; (iv) resilience to adversarial learning should be considered and explored during the development of countermeasures to avoid poisoning attacks; and (v) the use of automated feature engineering should be encouraged for efficiently detecting data exfiltration attacks. 
\end{abstract}


\section{Introduction}

{Data exfiltration is the process of retrieving, copying, and transferring an individual's or company's sensitive data from a computer, a server, or another device without authorization \cite{RN1}. 
It is also called data theft, data extrusion, and data leakage. Data exfiltration is either achieved through physical means or network. In physical means, an attacker has physical access to a computer from where an attacker exfiltrates data through various means such as copying to a USB device or emailing to a specific address. In network-based exfiltration, an attacker uses cyber techniques such as DNS tunneling or phishing email to exfiltrate data.
 For instance, a recent survey shows that hackers often exploit the DNS port 53, if left open, to exfiltrate data \cite{InfoBlox}. The typical targets of data exfiltration include usernames, passwords, strategic decisions, security numbers, and cryptographic keys \cite{RN2, RN3} leading to data breaches. These breaches have been on the rise with 779 in 2015 to 1632 in 2017 \cite{RN6}.
In 2018, $53,000$ incidents of data exfiltration were reported out of which $2,216$ were successful \cite{RN5}. The victims included tech giants like Google, Facebook, and Tesla~\cite{RN6}. 
Hackers successfully exploited a coding vulnerability to expose the accounts of 50 million Facebook users. Similarly, a security bug was exploited to breach data of around 53 million users of Google+. }

{Data exfiltration can be carried out by both external and internal actors of an organization. According to Verizon report \cite{Verizon2019} published in 2019, external and internal actors were responsible for 69\% and 34\% of data exfiltration incidents respectively.
 For the last few years, the actors carrying out data exfiltration have transformed from individuals to organized groups at times even supported by a nation-state with high budget and resources. 
For example, the data exfiltration incidents affiliated with nation-state actors has increased from 12\% in 2018 to 23\% in 2019 \cite{RN5, Verizon2019}. }

{Data exfiltration affects individuals, private firms, and government organizations, costing around \$3.86 million per incident on average \cite{RN10}.
It can also profoundly impact the reputation of an enterprise. Moreover, losing critical data (e.g., sales opportunities and new products' data) to a rival company can become a threat to the survival of a company.
Since incidents related to nation-state actors supported with high budget and resources have been increasing (e.g., from 12\% in 2018 to 23\% in 2019 \cite{RN5, Verizon2019}), it can also reveal national security secrets.}
{or example, the Shadow network attack conducted over eight months targeted confidential, secret, and restricted data in the Indian Defense Ministry \cite{RN9}. Hence, detecting these attacks is essential.}

Traditionally, signature-based solutions (such as firewalls and intrusion detection systems) are employed for detecting data exfiltration.
A \emph{signature-based} system uses predefined attack signatures (rules) to detect data exfiltration attacks. 
 {For example, rules are defined that stop any outgoing network traffic that contains predefined security numbers or credit card numbers.} 
Although these systems are accurate for detecting known data exfiltration attacks, they fail to detect novel (zero-day) attacks. {For instance, if a rule is defined to restrict the movement of data containing a sensitive project's name, then it will only restrict the exfiltration of data containing a specific project's name. However, it will fail to detect the data leakage if an attacker leaks the information of the project without/concealing the name of the project. Hence, detecting data exfiltration can become a daunting task.}

{Detecting data exfiltration can be extremely \emph{challenging} during sophisticated attacks. 
It is because 
\begin{enumerate}
\item  The network traffic carrying data exfiltration information resembles normal network traffic.
For example, hackers exploit existing tools (e.g., Remote Access Tools (RAT)) available in the environment. Hence, the security analytics not only need to monitor the access attempts but also the intent behind the access that is very much similar to usual business activity.
\item The network traffic that is an excellent source for security analytics is mainly encrypted, making it hard to detect data exfiltration. 
For example, 32\% of data exfiltration uses data encryption techniques \cite{RN11}.
\item Internal trusted employees of an organization can also act as actors for Data exfiltration. SunTrust Bank Data Breach \cite{RN12} and Tesla Insider Saboteur \cite{RN13} are the two recent examples of insider attacks where trusted employees exfiltered personal data of millions of users. 
 Whilst detecting data exfiltration from external hackers is challenging, monitoring and detecting the activities of an internal employee is even more challenging.
\end{enumerate}
}

 Given the frequency and severity of data exfiltration incidents, researchers and industry have been investing significant amount of resources in developing and deploying countermeasures systems that collect and analyze outgoing network traffic for detecting, preventing, and investigating data exfiltration \cite{RN1}. 
 Unlike traditional cybersecurity systems, these systems can analyze both the content (e.g., names and passwords) and the context (e.g. metadata such as source and destination addresses) of the data.
 'at rest' (e.g., data being leaked from a database), 'in motion' (e.g., data being leaked while it is moving through a network) and 'in use' (e.g., data being leaked while it is under processing in a computational server).

 Furthermore, they detect the exfiltration of data available in different states – in rest (e.g., data being leaked from a database), in motion (e.g., data being leaked while it is moving through a network), and in use (e.g., data being leaked while it is under processing in a computational server).
 The deployment of data exfiltration countermeasures also varies with respect to the three states of data. 
 For 'at rest' data, the countermeasures are deployed on data hosting server and aims to protect known data by restricting access to the data; for data 'in motion', they are deployed on the exit point of an organizational network; similarly, to protect 'in use' data, they are installed on the data processing node to restrict access to the application that uses the confidential data, block copy and pasting of the data to CDs and USBs, and even block the printing of the data. 

 Based on the attack detection criteria, there are two main types of data exfiltration countermeasure systems: signature-based and Machine Learning (ML) based data exfiltration countermeasure \cite{RN2}.
 Given these challenges, several Machine Learning (ML) based data exfiltration countermeasures have been proposed.
\emph{ML-based} systems are trained to learn the difference between benign and malicious outgoing network traffic (classification) and detect data exfiltration using the learnt model (prediction).
They can also be trained to learn the normal patterns of outgoing network traffic and detect an attack based on deviation from the pattern (anomaly-based detection).
To illustrate this mechanism, let's suppose an organization's normal behavior is to attach files of 1 MB -10 MB with outgoing emails. An ML-based system can learn this pattern during training. Now if an employee of an organization tries to send an email with a 20 MB file attachment, the system can detect an anomaly and flag an alert to the security operator of the organization. Depending on the organizational policy, the security operator can either stop or approve the transmission of the file. 
As such, they can detect previously unseen attacks and can be customized for different resources (e.g., hardware and software). Attacks detected through ML-based data exfiltration countermeasures can also be used to update signatures for signature-based data exfiltration countermeasures.

{Over the last years, research communities have shown significant interest in ML-based data exfiltration countermeasures. 
However, a systematic review of the growing body of literature in this area is still lacking.
This study aims to fulfil this gap and synthesize the current-state-of-the-art of these countermeasures to identify their limitations and the areas of further research opportunities.}

We have conducted a Systematic Literature Review (SLR) of ML-based data exfiltration countermeasures. Based on an SLR of {92} papers, this article provides an in-depth understanding of ML 
based data exfiltration countermeasures. 
 Furthermore, this review also identifies the datasets and the evaluation metrics used for evaluating ML-based data exfiltration countermeasures. The findings of this SLR purports to serve as guidelines for practitioners to become aware of commonly used and better performing ML models and feature engineering techniques. 
 For researchers, this SLR identifies several areas for future research. The identified datasets and metrics can serve as a guide to evaluate existing and new ML-based data exfiltration countermeasures. 
This review makes the following \textbf{contributions}:
\begin{itemize}
\item Development of a taxonomy of ML-based data exfiltration countermeasures.
\item Classification of feature engineering techniques used in these countermeasures.
\item Categorization of datasets used for evaluating the countermeasures.
\item Analytical review of evaluation metrics used for evaluating the countermeasures.
\item Mapping of ML models, feature engineering techniques, evaluation datasets, and performance metrics. 
\item List of open issues to stimulate future research in this field.
\end{itemize}
\paragraph{Significance}
{The findings of our review are beneficial for both researchers and practitioners. For researchers, our review highlights several research gaps for future research.
The taxonomies of ML approaches and feature engineering techniques can be utilized for positioning the existing and future research in this area.
The identification of the most widely used evaluation datasets and performance metrics can serve as a guide for researchers to evaluate the ML-based data exfiltration countermeasures. 
Practitioners can use the findings to select suitable ML approaches and feature engineering techniques according to their requirements. 
The quantitative analysis of ML approaches, feature engineering techniques, evaluation datasets, and performance metrics guide practitioners on the most common industry-relevant practices. } 

The rest of this paper is organized as follows. Section~\ref{section2} reports the background and the related work to position the novelty of this paper. Section~\ref{section-research-methodology} presents the research method used for conducting this SLR. Section~\ref{results} presents our findings followed by Section \ref{section-discussion} that reflects on the findings to recommend best practices and highlight the areas for future research. Section \ref{section-threat} outlines the threats to the validity of our findings. Finally, Section \ref{section-conclusion} concludes the paper. 

\section{Foundation}
\label{section2}
This section presents the background of data exfiltration and Machine Learning (ML). It also provides a detailed comparison of our SLR with the related literature reviews. 
\subsection{Background}
\label{section2.1}
This section briefly explains the Data Exfiltration Life Cycle (DELC) and ML workflow in order to contextualize the main motivators and findings of this SLR.
\subsubsection{Data Exfiltration Life Cycle (DELC) }
\label{DELC}

\begin{figure*}[htb!]
\centerline{\includegraphics[width=\textwidth]{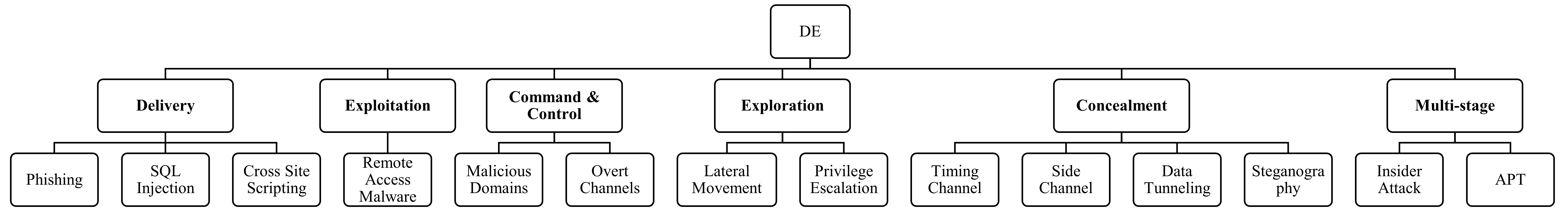}}
\caption{Mapping between DELC stages and Attack Vectors}
\label{figure2}
\end{figure*}

According to several studies \cite{RN1,RN19,RN20,RN21}, 
Data Exfiltration Life Cycle (DELC),
also known as cyber kill-chain \cite{RN18},
has seven common stages. 
{The first stage is \emph{Reconnassance}; an attacker collects information about a target individual or organization system to identify the vulnerabilities in their system.
After that, an attacker pairs the remote access malware or construct a phishing email in the \emph{Weaponization} stage. 
Rest of the stages use various targeted attack vectors \cite{RN1, RN19, RN22} to exfiltrate data from a victim. 
Fig~\ref{figure2} shows these stages and their mapping with attack vectors. 
The weapon created in the \emph{ Weaponization} stage such as an email or USB, is then delivered to a victim in the \emph{Delivery} stage. 
In this stage, an attacker uses different attack vectors to steal information from a target.}
{
These include: \emph{Phishing} \cite{RN23}, \emph{Cross Site Scripting (XSS)} and \emph{SQL Injection} attacks. 
In \emph{Phishing}, an attacker uses social engineering techniques to convince a target that the weapon (email or social media message) comes from an official trusted source and fool naive users to give away their sensitive information or download malware to their systems.
In \emph{XSS}, an attacker injects malicious code in the trusted source which is executed on a user browser to steal information. \emph{SQL Injection} is an attack in which an attacker injects SQL statements in an input string to steal information from the backend filesystem such as database.
\emph{Delivery} stage can result in either stealing the user information directly or downloading malicious software on a victims system. 
In \emph{Exploitation} stage, the downloaded malware Root Access Toolkit (RAT) \cite{RN26} is executed. The malware gains the root access and establishes a \emph{Command and Control Channel (C\&C)} with an attacker server.
Through C\&C, an attacker can control and exfiltrate data from a victim system remotely.  
C\&C channels use \emph{Malicious domains} \cite{RN29} to connect to an attacker server. Different \emph{Overt channels} \cite{RN28, RN19} such as HTTP post or Peer-to-Peer communication are used to steal data from a victim in C\&C stage.
 \emph{Exploration} stage often follows C\&C stage in which an attacker uses remote access to perform \emph{lateral movement} \cite{RN32} and \emph{privilege escalation} \cite{RN33} attacks that aid an attacker to move deeper into the target organization cyberinfrastructure in search of sensitive and confidential information.
The last stage is \emph{Concealment}; in this stage, an attacker uses hidden channels to exfiltrate data and avoid detection. 
The attack vectors \emph{Data Tunneling}, \emph{Side Channel}, \emph{Timing Channel} and \emph{Stenography} represent this stage.
\emph{Data Tunneling} attacks use unconventional (that are not developed for sending information) channels such as Domain Name Server (DNS) to steal information from a victim, while in \emph{Side Channel} \cite{RN31}, and \emph{Timing channel} \cite{RN30} an attacker gains information from physical parameters of a system such as cache, memory or CPU and time for executing a program to steal information from a victim respectively.
Lastly, in \emph{Stenography} attack, an attacker hides a victim's data in images, audio or video to avoid detection. 
Two attack vectors: \emph{Insider} attack and \emph{Advanced Persistent Attack (APT)} uses \emph{multiple stages} of DELC to exfiltrate data.
In \emph{Insider} attack, an insider (company employee or partner) leaks the sensitive information for personal gains. \emph{APT} attacks are sophisticated attacks in which an attacker persists for a long time within a victim system and uses all the stages of DELC to exfiltrate data.}

\begin{figure*}[htb!]
\centerline{\includegraphics[width=\textwidth,]{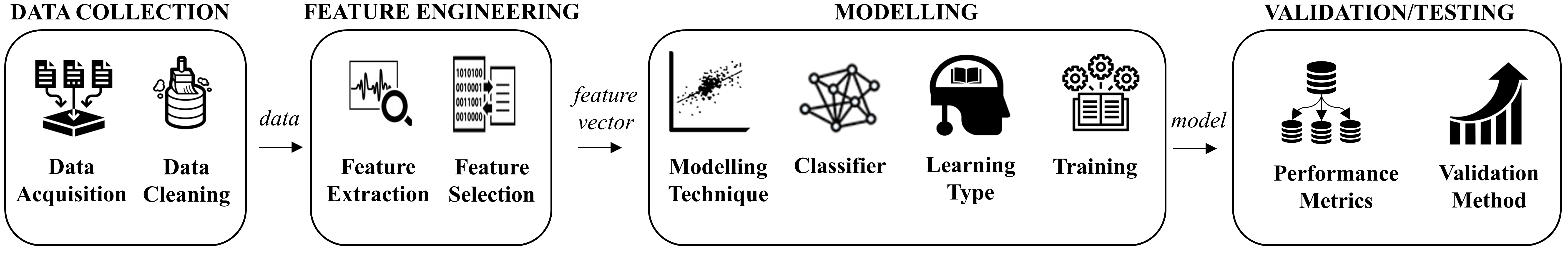}}
\caption{{Machine learning Life Cycle}}
\label{figure3}
\end{figure*}
\subsubsection{Machine Learning Life Cycle}
\label{MLLC}
Fig~\ref{figure3} shows the Machine Learning Life Cycle. Each stage is briefly discussed below (refer to \cite{RN36} for details).
\emph{Data collection}: In this phase, a dataset composed of historical data is acquired from different sources such as simulation tools, data provider companies, and the Internet to solve a problem (such as classifying malicious versus legitimate software). \emph{Data cleaning} \cite{RN29} is performed after data collection to ensure the quality of data and to remove noise and outliers. Data cleaning is also called pre-processing.
\emph{Feature Engineering}: This phase purports to extract useful patterns from data in order to segregate different classes (such as malicious versus legitimate) by using manual or automated approaches \cite{RN38}.
 In \emph{Feature selection}, various techniques (e.g., information gain, Chisquare test \cite{hira2015review}) are applied to choose the best discriminant features from an initial set of features.
\emph{Modelling}: This phase has four main elements. 
\emph{Modelling technique}: There are two main types of modelling techniques: anomaly detection and classification \cite{RN40}; Each modelling technique uses various \emph{classifiers} (learning algorithms) that learns a threshold or decision boundary to discriminate two or more than two classes. {
\emph{Anomaly-based detection} learns the normal distribution of data and detects the deviation as an anomaly or an attack. 
The common anomaly classifiers include:
K-mean \cite{RN52} (Kmean), One Class SVM \cite{RN45} (OCSVM), Gaussian Mixture Model \cite{reynolds2009gaussian} (GMM) and Hidden Markov Model \cite{beal2002infinite} (HMM).
\emph{Classification algorithms} learns data distribution of both malicious and benign to model a decision boundary that segregates two or more than two classes.
ML classifiers can be categorized into two main types: \emph{traditional ML classifiers} and \emph{deep learning classifiers}.
The common \emph{Traditional ML classifiers} include:
Support Vector Machines \cite{RN44} (SVM), Naïve Bayes \cite{RN48} (NB), K-Nearest Neighbor \cite{RN49} (KNN), Decision Trees \cite{RN50} (DT) and Logistic Regression \cite{RN51} (LR).
Most popular \emph{traditional ensemble classifiers} (combination of two or more classifiers) include Random Forest Trees  \cite{RN55} (RFT),  {Gradient Boosting (GB)} and Ada-boost \cite{RN56} classifiers.}
 {The \emph{Deep Learning (DL) classifiers} 
include
Neural Network \cite{anderson1995introduction} (NN), NN with Multi-layered Perception \cite{pal1992multilayer} (MLP), Convolutional NN \cite{krizhevsky2012imagenet} (CNN), Recurrent NN \cite{mikolov2010recurrent} (RNN), Gated RNN \cite{chung2014empirical} (GRU) and Long Short Term Memory \cite{hochreiter1997long} (LSTM).}
In the rest of this paper, we will use these acronyms instead of their full names.
\emph{Learning type} \cite{RN36} represents how a classifier learns from data. It has four main types: supervised, unsupervised, semi-supervised and re-enforcement learning \cite{RN41}.
The \emph{training} phase is responsible for building an optimum model (the model resulting in best performance over the validation set) while a \emph{ validation method} is a technique used to divide data into training and validation (unseen testing data) set \cite{RN43}; and \emph{Performance Metrics} \cite{RN36} evaluate how well a model performs. Various measures are employed to evaluate the performance of classifiers. The most popular metrics are Accuracy, 
Recall, True Positive Rate (TPR) or Sensitivity,
False Positive Rate (FPR),
True Negative Rate (TNR) or Specificity,
False Negative Rate (FNR),
Precision,
F1-Score,
Error-Rate,
Area Under the Curve (AUC), Receiver operating characteristic
Time to Train, Prediction Time and confusion matrix (please refer to \cite{RN36, RN35} for details of each metric).

\subsection{Existing Literature Reviews}
The literature on data exfiltration has been studied from various \emph{perspectives}.
In \cite{RN2}, the authors identified several challenges (e.g., lots of leakage channels and human factor) in detecting and preventing data exfiltration. They also presented data exfiltration countermeasures. 
The authors of \cite{RN3} identified and categorized 37 data exfiltration countermeasures. 
They provided a detailed discussion on the rationale behind data exfiltration attempts and vulnerabilities that lead to data exfiltration. 
In \cite{RN57}, the authors discussed the capabilities of data exfiltration countermeasures (e.g., discovering and monitoring how data is stored), techniques employed in countermeasures (e.g., content matching and learning), and various data states such as data in motion, data in rest, and data in use. 
The authors of \cite{RN58} primarily focused on the challenges in detecting and preventing data exfiltration. They highlighted encryption, lack of collaborative approach among industry, and weak access controls as the main challenges. They also proposed social network analysis and test clustering as the main approaches for efficiently detecting data exfiltration. 
In \cite{RN59}, the authors highlighted the channels used for exfiltrating data from cloud environments, the vulnerabilities leading to data exfiltration, and the types of data most vulnerable to exfiltration attacks.
The authors of \cite{RN60} reported different states of data, challenges (e.g., encryption and access control) in detecting data exfiltration, detection approaches (e.g., secure keys), and limitations/future research areas in data exfiltration. 
In \cite{RN1}, the authors provided a comprehensive review of the data exfiltration caused by external entities (e.g., hackers). They presented the attack vectors for exfiltrating data and classified around 108 data exfiltration countermeasures into detective, investigative, and preventive countermeasures.
\subsubsection{Comparison with existing literature reviews}
Compared to previous reviews, we focus on ML-based data exfiltration countermeasures. The bibliographic details of reviewed papers are given in Appendix \ref{appendix}.
Our review included papers, 
most of which have not been reviewed in previous reviews. The summary of the comparison against previous works is shown in Table~\ref{comparison}.

\begin{table*}[hbt!]
\begin{small}
\caption{Comparison of our survey with existing surveys}
\label{comparison}
\centering
\resizebox{\textwidth}{!}{\begin{tabular}{ccccccccc}
\hline
\textbf{Topic Covered} 
&\textbf{\cite{RN2}}
&\textbf{\cite{RN3}}
&\textbf{\cite{RN57}}
&\textbf{\cite{RN58}}
&\textbf{\cite{RN59}}
&\textbf{\cite{RN60}}
&\textbf{\cite{RN1}}
&\textbf{Our Survey}\\
\hline
Attack Vectors and their classification&$\times$&$\times$&$\times$&$\times$&$\times$&$\times$&\checkmark&$\times$\\

Data states with respect to data exfiltration&\checkmark&\checkmark&\checkmark&$\times$&$\times$&\checkmark&\checkmark&$\times$\\

Challenges in detecting data exfiltration&\checkmark&$\times$&$\times$&\checkmark&\checkmark&\checkmark&$\times$&$\times$\\

Vulnerabilities leading to data exfiltration&$\times$&\checkmark&$\times$&$\times$&\checkmark&$\times$&$\times$&$\times$\\
Channels used for data exfiltration&$\checkmark$&$\times$&$\times$&$\times$&\checkmark&$\times$&$\times$&$\times$\\

ML/DM based classification of countermeasures&$\times$&$\times$&$\times$&$\times$&$\times$&$\times$&\checkmark&\checkmark\\

Feature engineering for data exfiltration countermeasures&$\times$&$\times$&$\times$&$\times$&$\times$&$\times$&$\times$&\checkmark\\

Dataset used for evaluating data exfiltration countermeasures&$\times$&$\times$&$\times$&$\times$&$\times$&$\times$&$\times$&\checkmark\\

Metrics used for evaluating data exfiltration countermeasures&$\times$&$\times$&$\times$&$\times$&$\times$&$\times$&$\times$&\checkmark\\

Future Research Challenges&\checkmark&\checkmark&\checkmark&\checkmark&\checkmark&\checkmark&\checkmark&\checkmark\\
\hline
\end{tabular}}
\end{small}
\end{table*}

 \subsubsection{Comparison based on Objectives} All of the previous reviews focus on identifying the challenges (e.g., encryption and human factors) in detecting and preventing data exfiltration. Unlike previous reviews, our review focuses only on data exfiltration countermeasures. Some of the existing reviews \cite{RN1,RN2,RN3}, highlight data exfiltration countermeasures such as data identification and behavioral analysis. 
 Unlike the previous reviews, our review mainly focuses on ML-based data exfiltration countermeasures.
 \subsubsection{Comparison based on included papers} Our review included papers most of which have not been reviewed by the previous reviews; for example, there were only 3, 4, 1, and 6 papers that have been previously reviewed in \cite{RN2}, \cite{RN3},\cite{RN60} and \cite{RN1} respectively
 Unlike the previous reviews, we used the Systematic Literature Review (SLR) method guidelines. 
 There are approximately {61} papers published after the latest review (i.e., \cite{RN1}) on data exfiltration. 
 \subsubsection{Comparison based on results} Our literature review covers various aspects of ML-based data exfiltration countermeasures as compared to the previous literature reviews that included a variety of topics such as the data states, challenges, and attack vectors. 
 Whilst all of the previous reviews have identified the areas for future research, e.g., accidental data leakage and fingerprinting, the future research areas presented in our paper are different from others. Similar to our paper, one review \cite{RN1} has also identified response time as being an important quality measure for data exfiltration countermeasures. 
 However, unlike \cite{RN1}, our paper advocates the need for reporting assessment measures (e.g., training and prediction time) for data exfiltration countermeasures
\subsubsection{Our novel contributions}
In comparison to previous reviews, our paper makes the following unique contributions:
\begin{itemize}
\item Presents an in-depth and critical analysis of {92} ML-based data exfiltration countermeasures.
\item Provides rigorous analyses of the feature engineering techniques used in ML-based countermeasures.
\item Reports and discusses the datasets used for evaluating ML-based data exfiltration countermeasures.
\item Highlights the evaluation metrics used for assessing ML-based data exfiltration countermeasures.
\item Identifies several distinct research areas for future research in ML-based data exfiltration countermeasures.
\end{itemize}

\subsubsection{Research Significance}
 The findings of our review are beneficial for both researchers and practitioners. For researchers, our review has identified several research gaps for future research.
 The identification of the most widely used evaluation datasets and performance metrics along with their relative strengths and weaknesses, serve as a guide for researchers to evaluate the ML-based data exfiltration countermeasures. 
 The taxonomies of ML approaches and feature engineering techniques can be used for positioning the existing and future research in this area.
 The practitioners can use the findings to select suitable ML approaches and feature engineering techniques according to their requirements. 
 The quantitative analysis of ML approaches, feature engineering techniques, evaluation datasets, and performance metrics guide practitioners on what are the most common industry-relevant practices.  

\section{Research Methodology}
\label{section-research-methodology}
This section reports the literature review protocol that comprises of defining the research questions, search strategy, inclusion and exclusion conditions, study selection procedure and data extraction and synthesis. 

\subsection{Research Questions and Motivation}
This study aims to analyse, assess and summarise the current state-of-the-art  Machine Learning countermeasures employed to detect data exfiltration attacks. 
We achieve this by answering five Research Questions (RQs) summarised in Table~\ref{ResearchQuestions}.
\begin{table*}[hbt!]
\begin{small}
\caption{Research Questions and their motivations}
\label{ResearchQuestions}
\centering
\resizebox{\textwidth}{!}{\begin{tabular}{|c|l|l|}
\hline
 \textbf{ID} &\textbf{Research Questions} &\textbf{Motivation}\\
 \hline
    \multirow{3}{*}{RQ1} & What ML-based countermeasures are used  & The purpose is to identify the types of ML countermeasure  used for detecting  attack  \\
     &  to detect data exfiltration? & vector in DELC and  recognize their relative strength and limitations. \\ 
     \hline
     \multirow{4}{*}{RQ2} && The goal here is to identify feature-engineering methods, feature selection techniques \\ 
     & What constitutes the feature-engineering  & feature type, and number of features used to detect DE attacks.  Such an analysis is  \\ 
     & stage of  these countermeasures? &  beneficial for providing insights to practitioners and researcher about what types of \\ 
     &  &  feature engineering should be used in data exfiltration countermeasures. \\ 
     \hline
    \multirow{3}{*}{RQ3} & Which datasets are used by these  & The goal here is to identify the type of dataset used  by these solutions to train and \\ 
     & countermeasures? &  validate the ML  models. This knowledge will provide comprehension to practitioners \\
     &&  and researchers on available datasets for ML-based countermeasures.\\ 
     \hline
    \multirow{2}{*}{RQ4} & Which modelling techniques have been  & The motivation behind this is to determine what type  of modelling technique,  \\ 
     & used by these countermeasures? &  learning and classifiers are  used by ML-based data exfiltration countermeasures. \\
     \hline
    \multirow{2}{*}{RQ5} & How these countermeasures are validated  & The objective here is to identify how these countermeasures are validated and how  \\ 
     & and evaluated? & they relatively perform in relevance to an attack vectors and classifier used. \\ 
    \hline
\end{tabular}}
\end{small}
\end{table*}
\subsection{Search Strategy}
We formulated our search strategy for searching the papers based on the guidelines provided in \cite{RN63}.
The strategy was composed of the following two steps:
\paragraph{Search Method} The database-driven search method \cite{RN63} was conducted to select an initial set of papers.
It was then complemented with forward and backwards snowballing \cite{RN64}. 
\paragraph{Search String} A set of search keywords were selected based on the research questions. 
The set was composed of three main parts: data exfiltration, detection, and machine learning.
To ensure full coverage of our research objectives, all  the identified attack vectors (section~\ref{DELC}), relevant synonyms used for "data exfiltration", "detection" and "Machine learning" were incorporated.
The search terms were then combined using logical operators (AND and OR) to create a complete search string. 
The final search string was attained by conducting multiple pilot searches on two databases (IEEE and ACM) until the search string returned the papers already known to us. 
These two databases were selected because they are considered highly relevant databases for computer science literature \cite{RN65}.
\par
\begin{figure*}
  \centering
  \includegraphics[width=.99\linewidth]{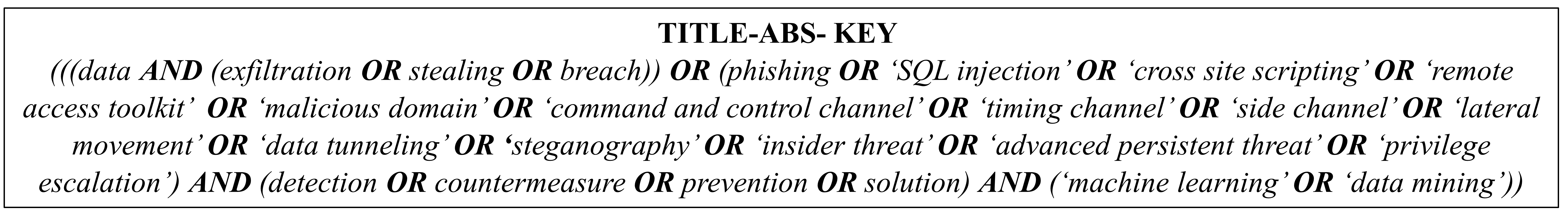}
\caption{Our Search String}
\label{fig:fig}
\end{figure*}

 During one of the pilot searches, it was noticed that exfiltration synonym like "leakage" and "theft" resulted in too many unrelated papers (mostly related to electricity or power leakage or data mining leakage and theft).
To ensure that none of the related papers was missed by the omission of these words, we searched the IEEE and ACM with a pilot search string "((data AND (theft OR leakage)) AND ("machine learning" OR ``data mining
''))”.
We inspected all the titles of the returned papers (411 IEEE and 610 ACM), we found none of these papers matched to our inclusion criteria; hence, these terms were dropped in our final search string. 
Furthermore, for the infinitive verbs in the search string, we used the stem of the word for example tunnelling to tunnel* and stealing to steal* and 
we found that it did not make any difference in our search results. 
The complete search string is shown in Fig \ref{fig:fig}.
The quotation marks used in the search string indicate phases (words that appear together). 

\subsection{Data Sources}
Six most well-known digital databases \cite{RN66} were queried based on our search string.
These databases include: IEEE, ACM, Wiley, ScienceDirect, Springer Link and Scopus.
The search string was only matched with the title, abstract and keywords.
The data sources were searched by keeping the time-frame intact, i.e., filter was applied on the year of the publication and only papers from 1st January 2009 to {31$^{st}$ December 2019} were considered. This time-frame was considered because machine learning-based methods for cyber-security and especially data exfiltration gained popularity after 2008 \cite{RN67, RN68}.
\subsection{Inclusion and Exclusion Criteria}
\label{3.4}
The assessment of the identified papers was done based on the inclusion and exclusion conditions. The selection process was steered by using these criteria to a set of ten arbitrarily selected papers. The selection conditions were refined based on the results of the pilot searches. Our inclusion and exclusion criteria are presented in Table \ref{InclusionCriteria}. The papers that provided countermeasures for detecting physical theft or dumpster diving \cite{RN1} such as hard disk theft, copying data physically or printing data were not considered.
Low-quality studies, that were published before 2017 and have a citation count of less than five or H-index of conference or journal is less than ten, were excluded. 
The three-year margin was considered because both quality metrics (citation count and H-index) are based on the citation and for papers less than three years old it is hard to compute their quality using these measures. 
\begin{table*}[hbt!]
\begin{small}
\caption{Inclusion and Exclusion Criteria}
\label{InclusionCriteria}
\centering
\resizebox{\textwidth}{!}{\begin{tabular}{|c|c|l|}
\hline
    \textbf{Criteria} & \textbf{ID} & \textbf{Description}\\
    \hline
    \multirow{4}{*}{\textbf{Inclusion}} 
    & I1 & Studies that use ML techniques (e.g., supervised learning or unsupervised learning or anomaly-based detection) 
    \\ \cline{2-3}
    &&to detect any stage or sequence of stages of data exfiltration attack. \\ \cline{2-3} 
     & I2 & The studies that are peer-reviewed. \\ \cline{2-3}
     & I3 & Studies that report contextual data (i.e., ML model, feature engineering techniques, evaluation datasets, and evaluation metrics). \\ \cline{2-3}
     & I4 & Published after 2008 and before 2020. \\ \cline{2-3}
     \hline
    \multirow{8}{*}{\textbf{Exclusion}} & E1 & Studies that use other than ML techniques (e.g., access control mechanism) to detect any stage of data exfiltration attack. \\ \cline{2-3}
     & E2 & Studies based on physical attacks (e.g., physical theft or dumpster diving) \\ \cline{2-3}
     & E3 & Short papers less than five pages. \\ \cline{2-3}
     & E4 & Non-peer reviewed studies such as editorial, keynotes, panel discussions and whitepapers. \\ \cline{2-3}
     & E5 & Papers in languages other than English. \\ \cline{2-3}
     & E6 & Studies that are published before 2016 and have citation count less than 5, journal, or conference H-index less than 10. \\ \hline
\end{tabular}}
\end{small}
\end{table*}
\subsection{Selection of Primary Studies}
\label{3.5}
\begin{figure*}[htb!]
\centerline{\includegraphics[width=\textwidth]{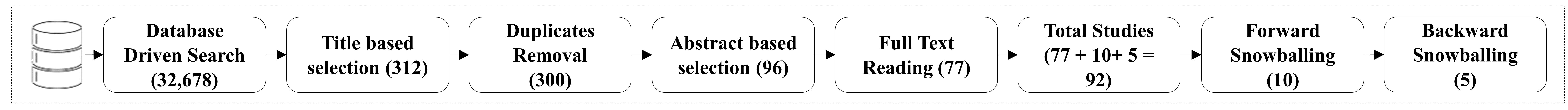}}
\caption{Paper Selection Process}
\label{figure6}
\end{figure*}

\begin{figure*}
\captionsetup[subfigure]{font=footnotesize,labelfont=scriptsize}
\begin{subfigure}{.29\textwidth}
  \centering
 \includegraphics[width=.99\linewidth]{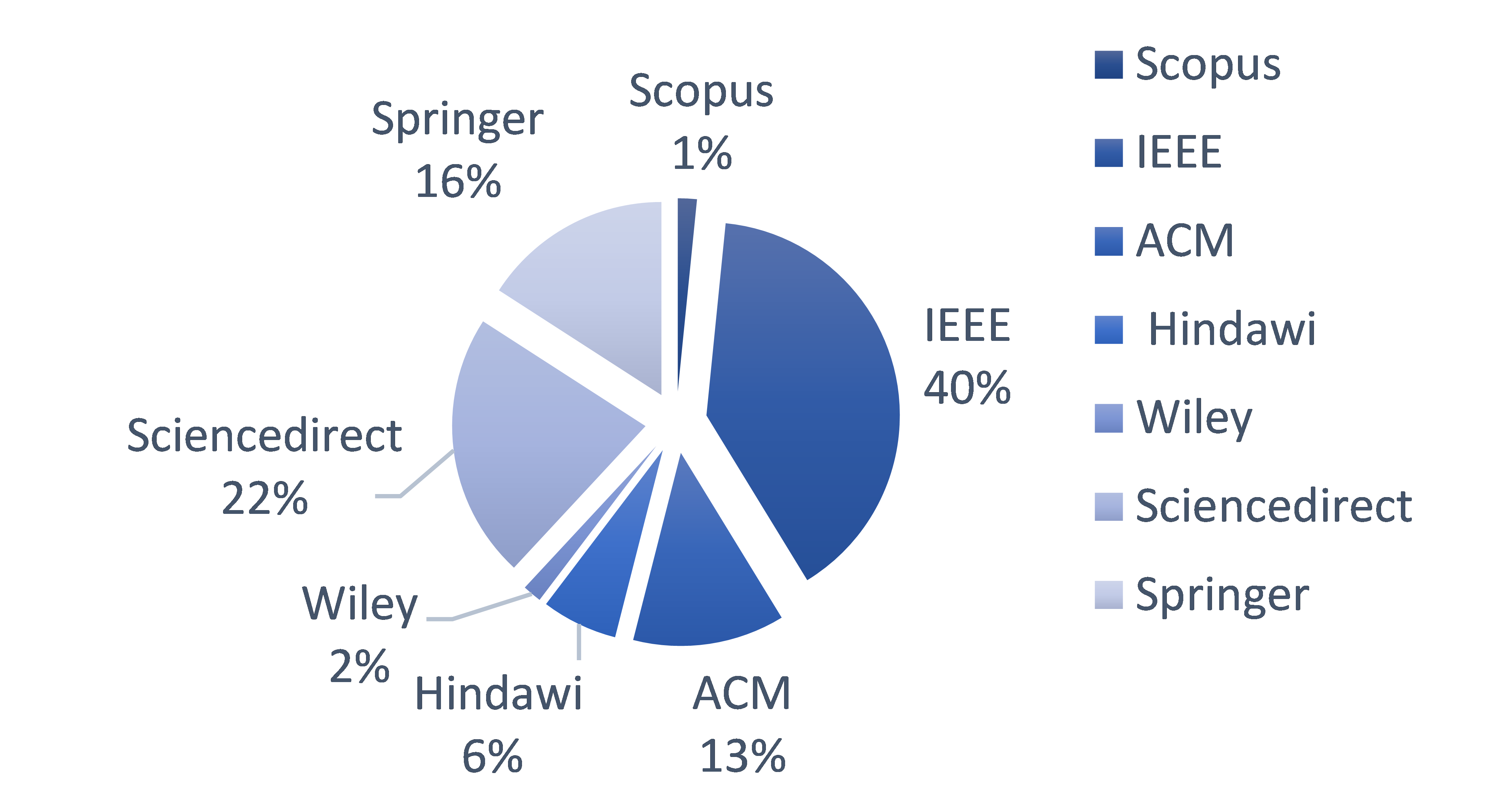}
\caption{ Distribution of primary studies over data sources}
\label{fig1:sub-first}
\end{subfigure}
\begin{subfigure}{.29\textwidth}
\begin{small}
\centering
\includegraphics[width=.99\linewidth]{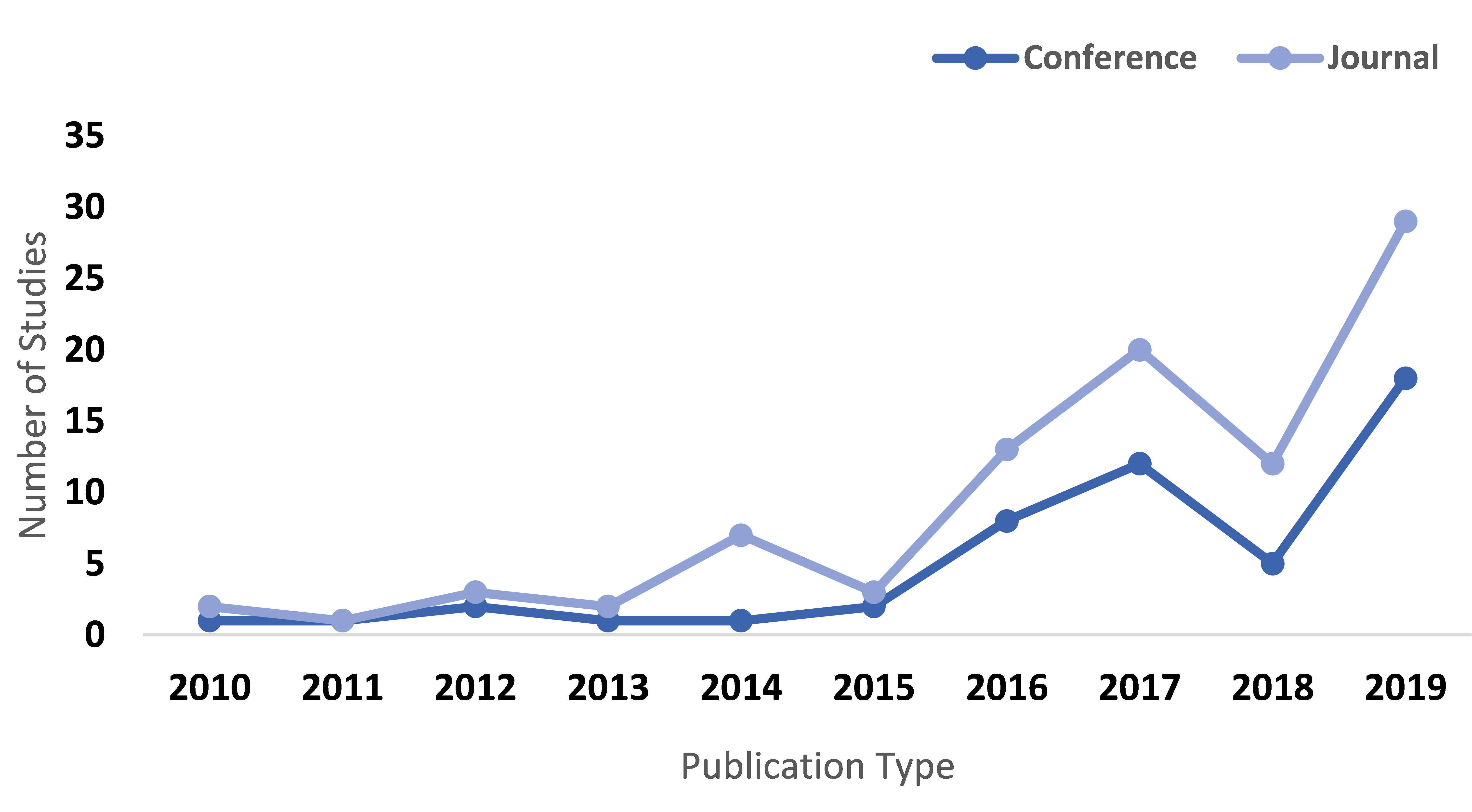}
\caption{ Distribution of primary studies over type and year of publication}
\label{fig1:sub-second}
\end{small}
\end{subfigure}
\begin{subfigure}{.32\textwidth}
\begin{small}
\centering
\includegraphics[width=.99\linewidth]{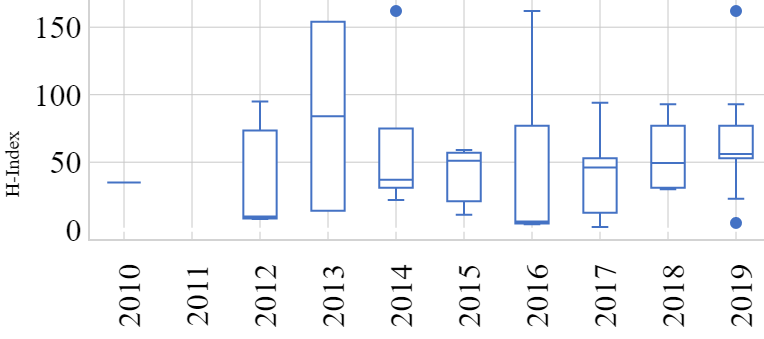}
\caption{Distribution of primary studies over H-Index and year of publication}
\label{fig1:sub-third}
\end{small}
\end{subfigure}
\caption{: Statistical Information of Primary Studies}
\label{fig1:fig1}
\end{figure*}

Fig~\ref{figure6} shows the steps we followed to select the primary studies. After initial data-driven search 32678 papers were retrieved i.e., IEEE (3112), ACM (26128), Wiley (325), Science Direct (794), Springer Link (2011) and Scopus (309).
 After following the complete process and repeatedly applying inclusion and exclusion criteria (Table~\ref{InclusionCriteria}), 92 studies were selected for the review.
Fig~\ref{fig1:fig1} shows the statistical distribution of the selected papers with respect to data sources, publication type, publication year and H-index. {It can be seen from Fig~\ref{fig1:sub-second} that ML countermeasures to detect DELC have gained popularity over the last four years, with a significant increase in the number of studies especially in the year 2017 and 2019.}
\subsection{Data Extraction and Synthesis}
\label{3.6}
\paragraph{Data Extraction}
The data extraction form was devised to collect data from the reviewed papers. Table~\ref{dataextractionform} shows the data extraction form. 
A pilot study with eight papers was done to judge the comprehensiveness and applicability of the data extraction form. The data extraction form was inspired by \cite{RN70} and consisted of three parts. 
Qualitative Data (D1 to D10): For each paper, a short critical summary was written to present the contribution and identify the strengths and weaknesses of each paper. 
Demographic information was extracted to ensure the quality of the reviewed papers. 
Context Data (D11 to D21): Context of a paper included the details in terms of the stage of data exfiltration, attack vector the study presented countermeasure for, dataset source, type of features, type of feature engineering methods, validation method, and modelling technique.
Quantitative Data (D22-D24): The quantitative data such as the performance measure, dataset size and a number of features reported in each paper were extracted.
\paragraph{Data Synthesis}
Each paper was analysed based on three types of data mentioned in Table~\ref{dataextractionform} as well as the research questions. 
The qualitative data (D1-D10) and context-based data (D11 to D21) were examined using thematic analysis \cite{RN72} while quantitative data were synthesised by using box-plots \cite{RN73} as inspired by \cite{RN70, RN74}. 
\begin{table*}[hbt!]
\begin{small}
\caption{Data type and item extracted from each study and related research questions enclosed in parenthesis}
\label{dataextractionform}
\centering
\resizebox{\textwidth}{!}{\begin{tabular}{|c|c|c|l|}
    \hline
    \textbf{Data Type} & \textbf{Id} & \textbf{Data Item} &\textbf{Description} \\ \hline
    \multirow{10}{*}{Qualitative Data} & D1 & Title & The title of the study. \\
    \cline{2-4}
     & D2 & Author & The author(s) of the study. \\ \cline{2-4}
     & D3 & Venue & Name of the conference or journal where the paper is published. \\ \cline{2-4}
     & D4 & Year Published & Publication year of the paper. \\ \cline{2-4}
     & D5 & Publisher & The publisher of the paper. \\\cline{2-4}
     & D6 & Country & The location of the conference. \\ \cline{2-4}
     & D7 & Article Type & Publication type i.e., book chapter, conference, journal. \\ \cline{2-4}
     & D8 & H-index & Metric to quantify the impact of the publication. \\ \cline{2-4}
     & D9 & Citation Count & How many citations the paper has according to google scholar. \\ \cline{2-4}
     & D10 & Summary & A brief summary of the paper along with the major strengths and weaknesses \\ \hline
    \multirow{2}{*}{Context (RQ1)} & D11 & Stage & Data exfiltration stage the study is targeting (section 2.1.2). \\ \cline{2-4}
     & D12 & Attack Vector & Data exfiltration attack vector the study is detecting (section~\ref{DELC}). \\ \hline
    \multirow{4}{*}{Context (RQ2)} & D13 & Feature Type & Type of features used by the primary studies. \\ \cline{2-4}
     & D14 & Feature Engineering Method & Type of feature engineering methods used. \\ \cline{2-4}
     & D15 & Feature Selection Method & The feature selection method used in the study. \\ \cline{2-4}
     & D16 & Dataset used & The dataset(s) used for the evaluation of the data exfiltration countermeasure. \\ \hline
    \multirow{5}{*}{Context (RQ3)} & D17 & Classifier Chosen & Which classifier is chosen after the experimentation by the study (Table 1) \\ \cline{2-4}
     & D18 & Type of Classifier & What is the type of classifier? (section~\ref{MLLC}). \\ \cline{2-4}
     & D19 & Modelling Technique & What type of modelling technique is used (Fig~\ref{figure3}). \\\cline{2-4}
     & D20 & Type of Learning & Type of learning used by the study (Fig ~\ref{figure3}). \\ \cline{2-4}
     & D21 & Validation Technique & Type of validation method used by the study. \\ \hline
    \multirow{2}{*}{Quantitative Data} & D22 & Number of Features & Total number of features/dimensions used by the study \\ \cline{2-4}
     & D23 & Dataset Size & Number of instances in the dataset \\ \cline{2-4}
     (RQ2, RQ4)&\multirow{3}{*}{D24} &     \multirow{3}{*}{Performance Measures} & Accuracy, Sensitivity /TPR /Recall,Specificity/ TNR/Selectivity,\\
     &&& FPR (\%), FNR (\%), AUC/ROC,F-measure/F-Score, Precision/Positive \\
     &&& Predictive Value, Time-to-Train the Model, Prediction-Time, Error-Rate \\ \hline
    
\end{tabular}}
\end{small}
\end{table*}
\section{Results}
\label{results}
This section presents the outcomes of our data synthesis based on RQs. 
Each section reports the result of the analysis performed followed by a summary of the findings.
\subsection{{RQ1}: ML-based Countermeasures for Detecting Data Exfiltration}
\label{RQ1}
In this section, we present the classification of the reviewed studies followed by the mapping of the proposed taxonomy based on DELC.
\subsubsection{Classification of ML-based Countermeasures}
\begin{figure*}[htb!]
\captionsetup[subfigure]{font=footnotesize,labelfont=scriptsize}
\begin{subfigure}{.23\textwidth}
  \centering
   \includegraphics[width=.95\linewidth]{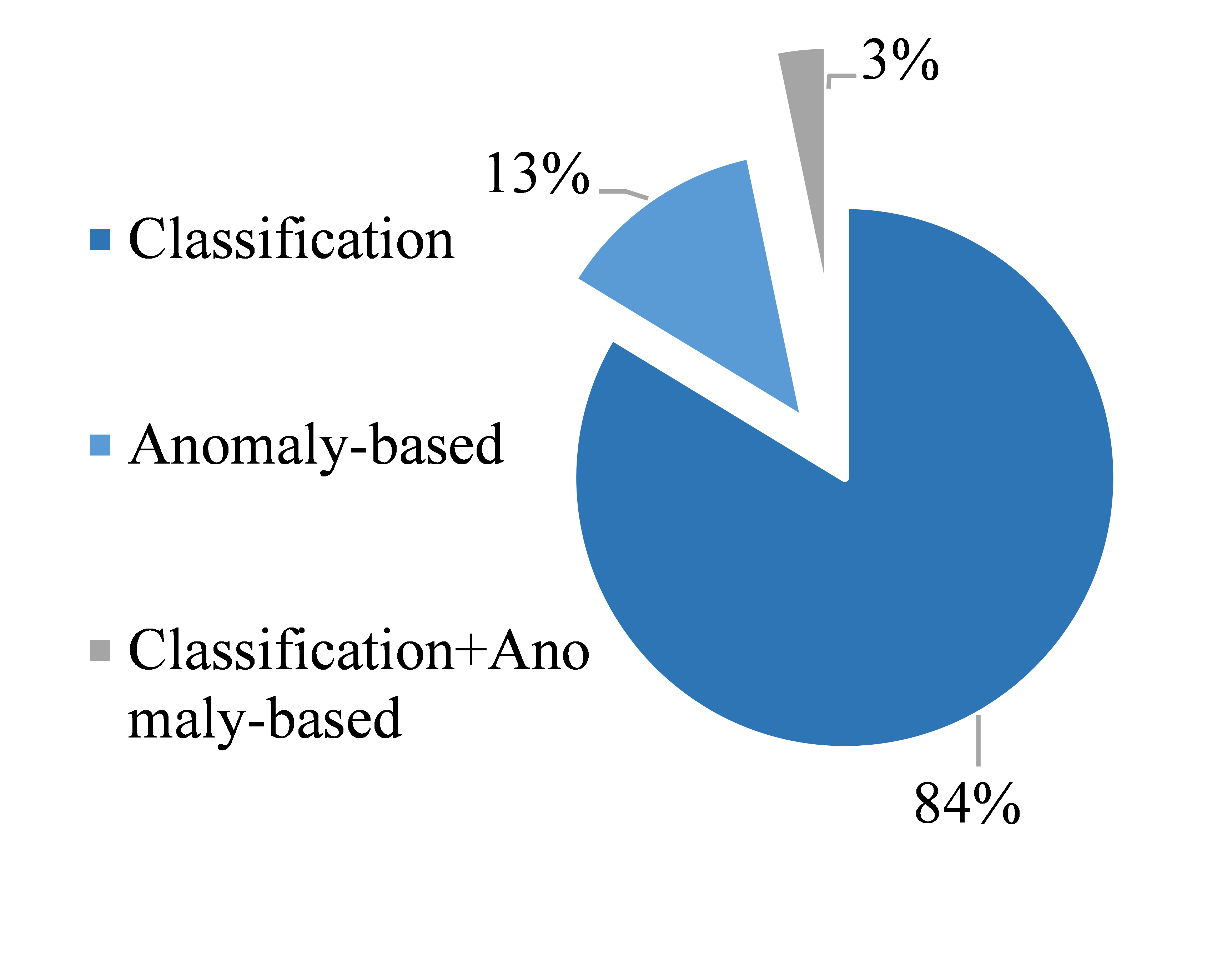}
\caption{ Distribution of Primary Studies with respect to the Modelling Technique}
\label{fig2:sub-first}
\end{subfigure}
\begin{subfigure}{.23\textwidth}
\begin{small}
\centering
\includegraphics[width=.95\linewidth]{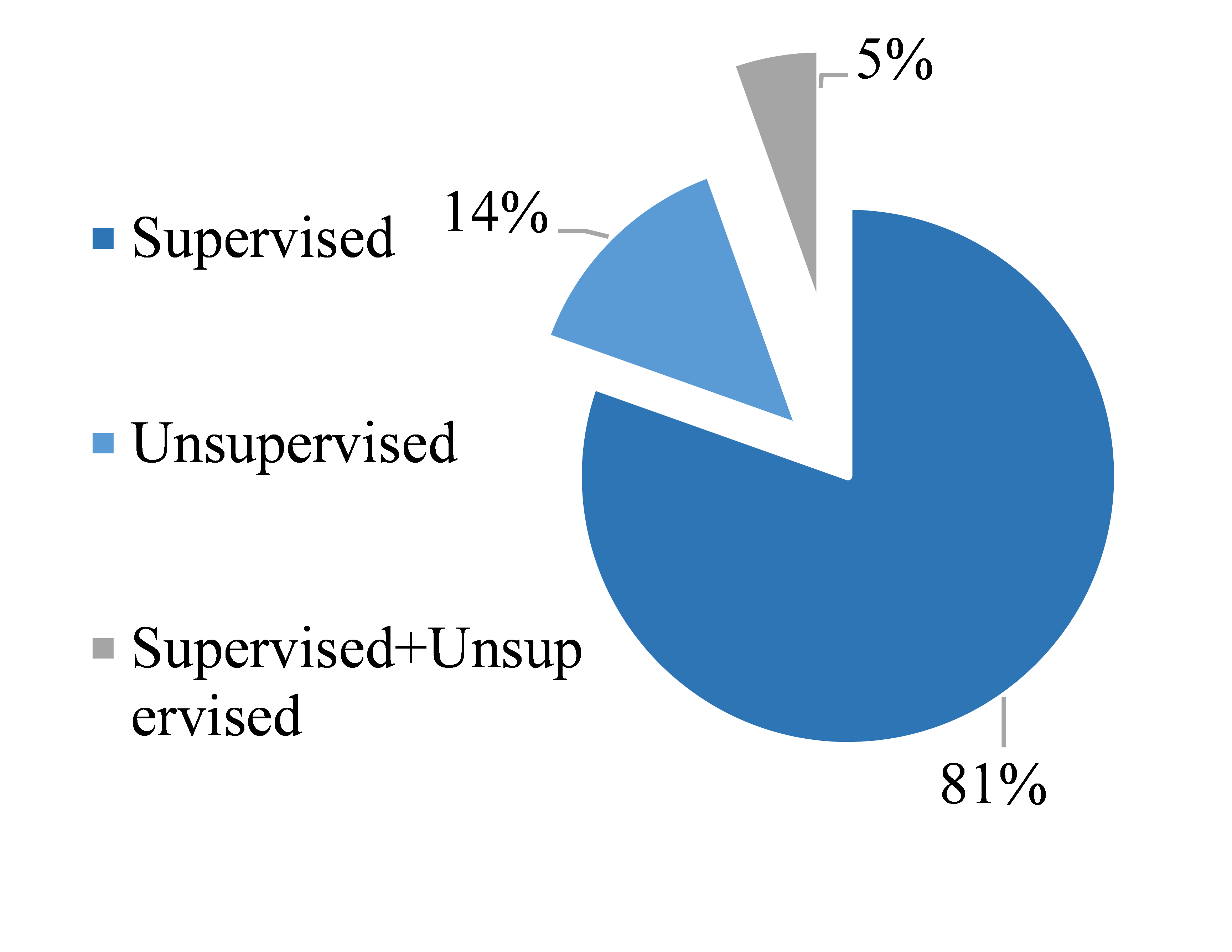}
\caption{ Distribution of Primary Studies with respect to the Type of Learning}
\label{fig2:sub-second}
\end{small}
\end{subfigure}
\begin{subfigure}{.23\textwidth}
\begin{small}
\centering
\includegraphics[width=.95\linewidth]{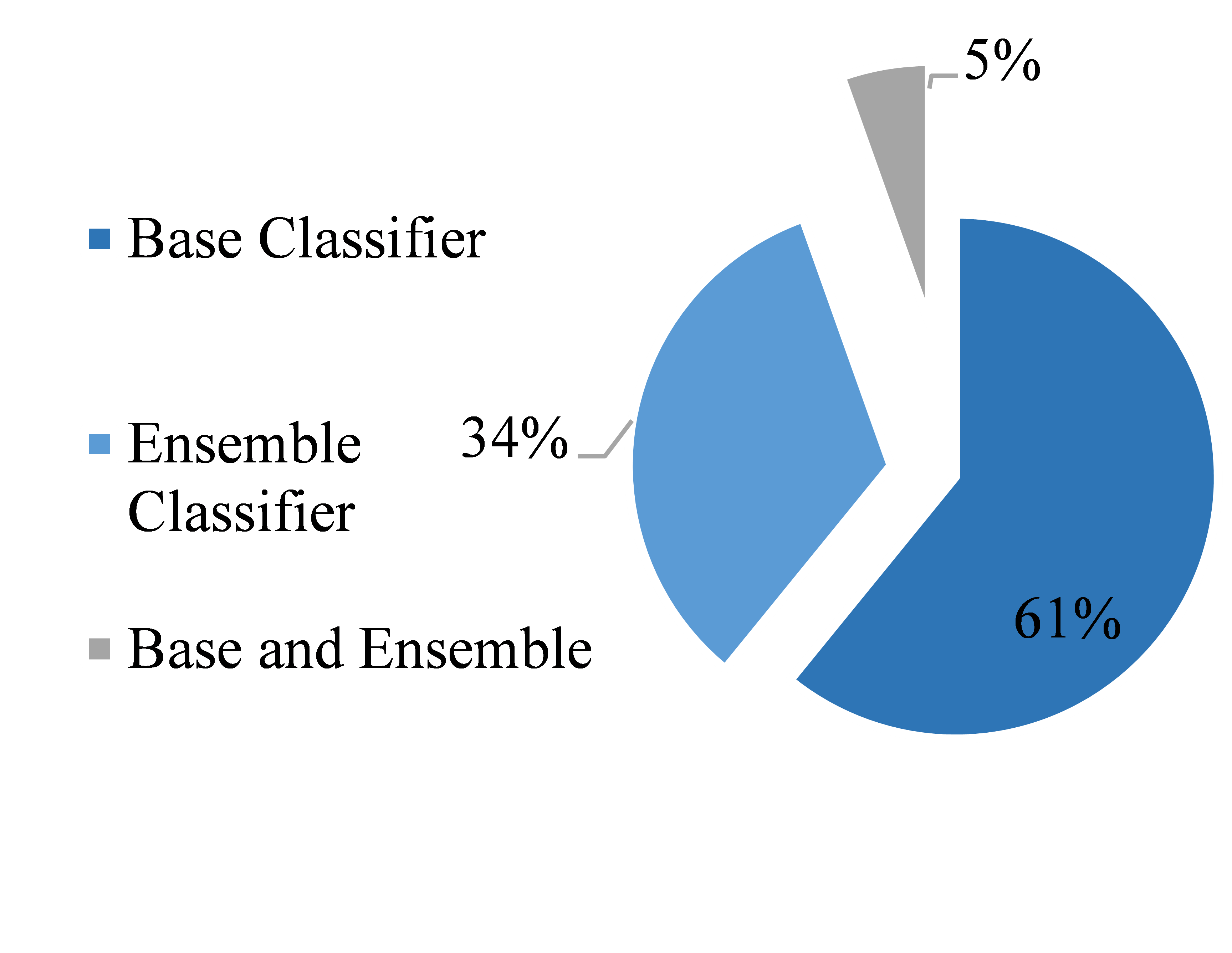}
\caption{ Distribution of Primary Studies with respect to the Type of Classifier used}
\label{fig2:sub-third}
\end{small}
\end{subfigure}
\footnotesize\begin{subfigure}{.23\textwidth}
\begin{small}
\centering
\includegraphics[width=.95\linewidth]{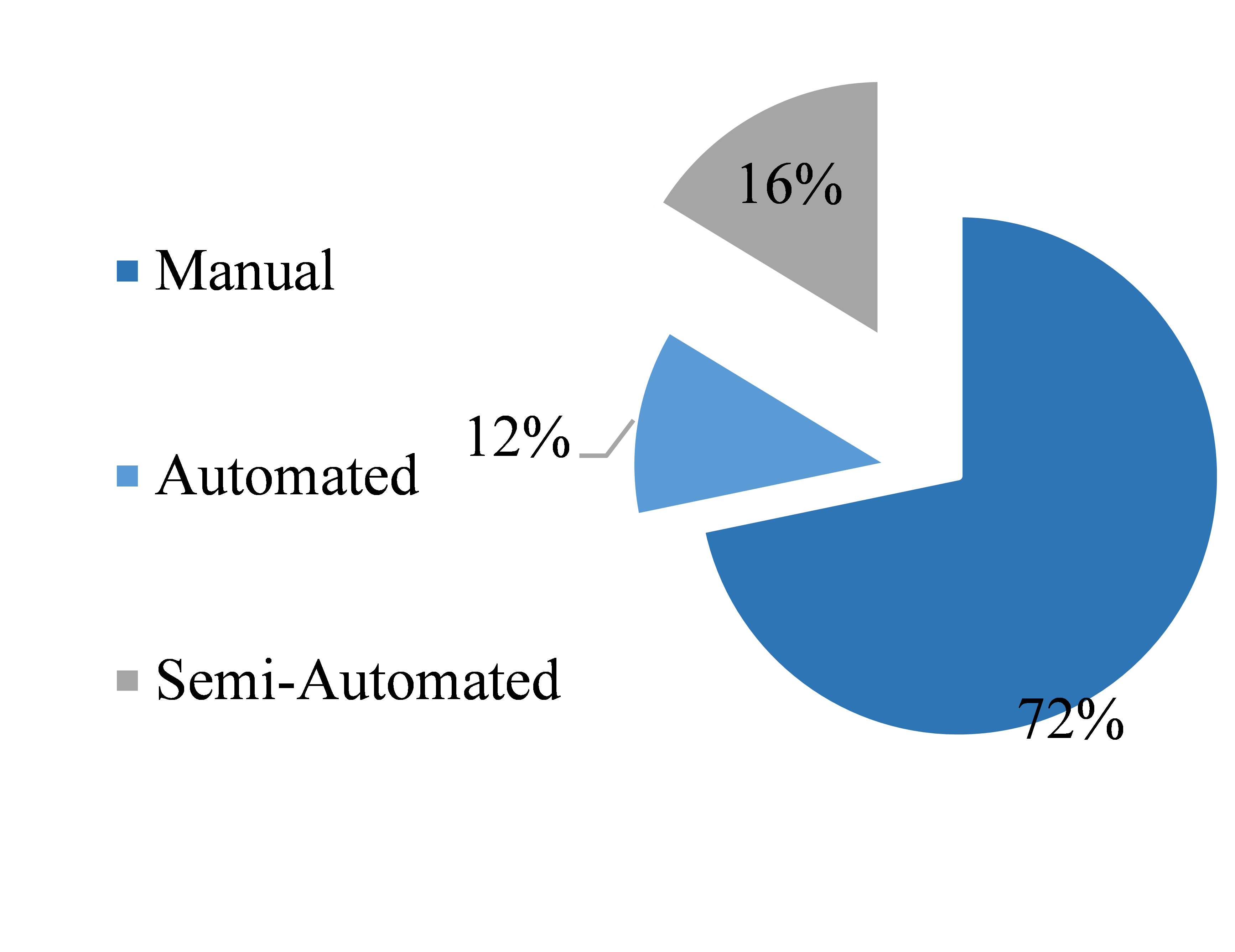}
\caption{ Distribution of Primary Studies with respect to Feature Engineering Techniques}
\label{fig2:sub-fourth}
\end{small}
\end{subfigure}
\caption{: Distribution of Primary Studies over different criterion}
\label{fig2:fig2}
\end{figure*}

Since machine learning is a data-driven approach \cite{RN75} and the success of ML-based systems is highly dependent on the type of data analysis performed to extract useful features.
We classify the countermeasures presented in the reviewed studies into two main types based on data type and data analysis technique: \emph{Data-driven } and \emph{Behavior-driven} approaches.
\paragraph{\textbf{Data-driven approaches}}
These approaches examine the content of data irrespective of data source i.e., network, system, or web application.
They are further divided into three sub-classes: Direct, Distribution and Context inspection. 
In \emph{Direct Inspection}, an analyst directly scans the content of data or consult external sources to extract useful features from data. For example, an analyst may look for lexical patterns like the presence of @ sign or not, count of hyperlinks, age of URL to extract useful features.
The \emph{Distribution Inspection} approach considers data as a distribution and performs different analyses such as statistical or temporal to extract useful features from data. These features cannot be directly obtained from data without additional computation such as average, entropy, standard deviation, and Term-frequency Inverse Document Frequency (TF-IDF).
The \emph{Context Inspection} approach analyses the structure and sequential relationship between multiple bytes, words or parts of speech to extract semantics or context of data.
This type of analysis can unveil the hidden semantics and sequential dependencies between data automatically such as identifying words or bytes that appear together (n-grams).
\paragraph{\textbf{Behavior-driven approaches}} These approaches analyze the behavior exhibited by a particular entity such as system, network, or resource to detect data exfiltration.
They are further classified into five sub-classes Event-based, Flow-based, Resource-usage based, Propagation-based, and Hybrid approaches. 
To detect the behavior (malicious or legitimate) of an entity (such as file and user), an \emph{Event-based approach} approach studies the actions of an entity by analyzing the sequence and semantics of system calls made by them e.g., login event or file deletion event.
The relationship between incoming and outgoing network traffic flow is considered by \emph{Flow-based approach} approach, e.g., ratio of incoming to outgoing payload size or the correlation of DNS request and response packets.
\emph{Resource usage-based approach}
approach aims to study the usage behavior of a particular resource, e.g., cache access pattern and CPU usage pattern.
\emph{Propagation-based approach}
approach considers multiple stages of data exfiltration across multi-host to detect data exfiltration, e.g., several hosts affected or many hosts behaving similarly. 
\emph{Hybrid approach}
approach utilizes both event-based and flow-based approaches to detect the behavior of the overall system.
Table~\ref{taxonomy} shows the classification of the reviewed studies (the numeral enclosed in a bracket depicts the number of papers classified under each category) and their strengths and weaknesses. The primary studies column shows the identification number of the study enlisted in Appendix \ref{appendix}. 
\begin{table*}[hbt!]
  \centering
  \caption{{Classification of ML-based data exfiltration countermeaures, their strengths, and weaknesses}}
  \label{taxonomy}
 \resizebox{\textwidth}{!}{\begin{tabular}{|c|c|l|l|c|}
    \hline
    \rotatebox{90}{\textbf{Type}} & \rotatebox{90}{\textbf{Sub}} \rotatebox{90}{\textbf{Type}} & \textbf{\centering {Strengths}} & \textbf{\centering {Weakness}} & \textbf{Primary Studies} \\ 
\hline
\multirow{27}{*}{\rotatebox{90}{Data-driven Approaches (44)}} &\multirow{11}{*}{\rotatebox{90}{Direct}}  \multirow{11}{*}{\rotatebox{90}{Inspection (15)}}  &\tabitem Do not require complex calculations to  &  &  \\ 
     &  & obtain features. & \tabitem Time-consuming and labour-intensive as it  &  \\ 
     
     &  & \tabitem Computationally faster as models are  & requires domain knowledge. &  \\
     
     &  & trained over more focused and smaller  & \tabitem Error-prone as it relies on human expertise. & [S5, S19, S25, S26,  \\
     \
     &  & number of features. & \tabitem Cannot extract hidden structural,  & S28, S37, S46, S47,  \\ 
     
     &  & \tabitem Can be implemented on client-side. & distributional, and sequential patterns. & S50, S51, S60, S69,  \\ 
    
     &  & \tabitem Easy to comprehend the decision boundary  & \tabitem Higher risk of performance degradation over  & S70, S71, S86] \\
     \
     &  & based on features values. & time due to concept drift and adversarial  &  \\ 
     
     &  & \tabitem Flexible to adapt to new problems as not  & manipulation. &  \\ 
     &  & dependent on type of data such as network or  &  &  \\ 
     &  & logs. &  &  \\ \cline{2-5}
     & \multirow{8}{*}{\rotatebox{90}{Distribution}}  \multirow{8}{*}{\rotatebox{90}{Inspection (12)}} & \tabitem Capable of extracting patterns and  &  &  \\ 
     &  & correlations that are not directly visible in  & \tabitem Requires complex computations. &  \\ 
     &  & data. & \tabitem Requires larger amounts of data to be  & [S3, S9, S10, S14,  \\ 
     
     &  & \tabitem More reliable when trained over large  & generalizable & S15, S17, S18, S20,  \\ 
     
     &  & amounts of data. & \tabitem Cannot extract the structural and sequential  & S22, S23, S55, S77] \\ 
     
     &  & \tabitem Easy to comprehend decision boundary  & patterns in data. &  \\ 
     
     &  & by visualizing features. &  &  \\ 
     &  & \tabitem Flexible to adapt to new problems. &  &  \\ \cline{2-5}
     & \multirow{6}{*}{\rotatebox{90}{Context}}  \multirow{6}{*}{\rotatebox{90}{Inspection (17)}} &  &  \tabitem Computationally expensive to train due to  &  \\ 
     &  & \tabitem Capable of comprehending the structural,  & complex models. & [S36, S53, S54, S58,  \\ 
      &  & sequential relationship in data. & \tabitem Resource-intensive & S59, S62, S64, S66,  \\ 
    
     &  &\tabitem  Supports automation & \tabitem Requires large amounts of data to be  & S67, S68, S75, S76,  \\ 
     
     &  & \tabitem Flexible to adapt to new problems & generalizable. & S78, S82, S87, S88,  \\ 
     &  &  &\tabitem Hard to comprehend the decision boundary. & S89] \\ 
     
     \hline
    \multirow{30}{*}{\rotatebox{90}{Behaviour-driven Approaches (48)}} & \multirow{6}{*}{\rotatebox{90}{Event}} \multirow{6}{*}{\rotatebox{90}{-based}}
    \multirow{6}{*}{\rotatebox{90}{Approach (16)}} & \tabitem Capable of capturing the malicious  & \tabitem Sensitive to the role of user and type of  &  \\ 
     &  & activities by monitoring events on a  & application e.g., the system interaction of a  & [S1, S6, S7, S12,  \\ \
     
     &  & host/server. & developer may deviate from manager or system  & S13, S31, S35, S48,  \\ 
     
     &  & \tabitem Capable of detecting insider and malware  & administrator behavior. & S49, S61, S72, S73,  \\ 
     &  & attacks. & \tabitem Require domain knowledge to identify critical  & S74, S84, S90, S91] \\ 
     &  &  & events. &  \\ \cline{2-5}
     & \multirow{6}{*}{\rotatebox{90}{Flow}}\multirow{7}{*}{\rotatebox{90}{-based }} \multirow{6}{*}{\rotatebox{90}{Approach (19)}} & \tabitem Capable of extracting correlation between  & \tabitem Dependent on the window size to detect & [S11, S21, S27, S29,  \\ 
     &  & inbound and outbound traffic.  &   exfiltration. & S32, S33, S34, S38,  \\ 
     &  & \tabitem Can detect network attacks such as data  & \tabitem Cannot detect long term data exfiltration as it   & S39, S40, S41, S42,  \\ 
     &  & tunnelling and overt channels  &requires large window sizes that is & S43, S44, S65, S80,  \\ 
     &  &  &computationally expensive. & S81, S83, S92] \\ 
     &&&&\\
     \cline{2-5}
     & \multirow{6}{*}{\rotatebox{90}{Resource}}
     \multirow{6}{*}{\rotatebox{90}{usage-based}} \multirow{6}{*}{\rotatebox{90}{Approach (4)}} &  & &  \\ 
     &  & \tabitem Can detect the misuse of a resource to  & \tabitem Highly sensitive to the type of application   & [S8, S52, S56, S63] \\ 
      &  & detect data exfiltration. &i.e., for video streaming program may  &  \\ 
     &  &  & consume more memory and CPU than other  &  \\ 
     &&&legitimate programs.&\\
     &&&&\\
     \cline{2-5}
     & \multirow{7}{*}{\rotatebox{90}{Propagation}} \multirow{7}{*}{\rotatebox{90}{based}} \multirow{7}{*}{\rotatebox{90}{Approach (3)}}&  &  & \multirow{7}{*}{[S2, S16, S24]} \\ 
     &&&&\\
      &&\tabitem Capable for detecting data exfiltration as & \tabitem Complex and time-consuming& \\
     &&large scale i.e., enterprise level.& \tabitem Hard to visualize and comprehend& \\
     &&&&\\
     &&&&\\
     
      \cline{2-5}
     & \multirow{6}{*}{\rotatebox{90}{Hybrid}} \multirow{6}{*}{\rotatebox{90}{ Approach (6) }}&   \tabitem Capable of detecting long term& \tabitem Data collection is time consuming as it  &  \\ 
     &  & data exfiltration. & requires monitoring of both host and &  \\ 
     &  & \tabitem Capable of detecting complete DELC and  & network traffic. & [S4, S30, S45, S57,  \\ 
    
     &  & complex attacks like APT. & \tabitem Suffers from the limitations of both event and. & S79, S85] \\ 
     &  &  & flow-based approaches &  \\ 
    \hline
  \end{tabular}}
\end{table*}
 Although other criteria can be applied for classification, such as supervised versus unsupervised learning, automated versus manual feature engineering, classification versus anomaly-based detection and base-classifier versus ensemble classifier. 
 These types of categorization fail to create semantically coherent, uniform and non-overlapping distribution of the selected primary studies as depicted in Fig~\ref{fig2:fig2}. 
Consequently, they limit the opportunity to inspect the state-of-the-art in a more fine-grained level.
Furthermore, our proposed classification increases the semantic coherence and uniformity between the multi-stage primary studies as depicted in Table~\ref{mapping}.
Hence, we conclude that our proposed classification is a suitable choice for analysing ML-based data exfiltration countermeasures.
\begin{table*}[hbt!]
\caption{Mapping Taxonomy with DELC stages and Attack Vectors}
\label{mapping}
\begin{small}
\resizebox{\textwidth}{!}{\begin{tabular}{|c|c|c|c|c|c|c|c|c|c|c|}
    \hline
    \textbf{Stage} & \textbf{Attack Vector}&\rotatebox{90}{\textbf{Direct}}\rotatebox{90}{\textbf{Inspection}} &\rotatebox{90}{\textbf{Context}}\rotatebox{90}{\textbf{Inspection}} & \rotatebox{90}{\textbf{Distribution}}\rotatebox{90}{\textbf{Inspection}} & \rotatebox{90}{\textbf{Event-based}}\rotatebox{90}{\textbf{Approach}}& \rotatebox{90}{\textbf{Flow-based}}\rotatebox{90}{\textbf{Approach}} & \rotatebox{90}{\textbf{Hybrid}}\rotatebox{90}{\textbf{Approach}} & \rotatebox{90}{\textbf{Propagation}}\rotatebox{90}{\textbf{based Approach}} & \rotatebox{90}{\textbf{Resource usage}}\rotatebox{90}{\textbf{-based Approach}} & \textbf{Study Ref} 
    \\ \hline
\multirow{4}{*}{Delivery(32)} & \multicolumn{1}{m{2cm}|}{\centering Phishing (24)} & \multicolumn{1}{c|}{12} 
&\multicolumn{1}{c|}{9} 
&\multicolumn{1}{c|}{2} 
&\multicolumn{1}{c|}{1}
&\multicolumn{1}{c|}{-} 
&\multicolumn{1}{c|}{-} 
&\multicolumn{1}{c|}{-} 
&\multicolumn{1}{c|}{-}
& \multicolumn{1}{m{6cm}|}{\centering[S5, S17, S19, S25,
S26, S28, S31, S36,
S37, S46, S47, S53,
S54, S58, S60, S69,
S71, S75, S76,S77, 
S82, S86, S87, S88]} 
\\
\cline{2-11}
 & \multicolumn{1}{m{2cm}|}{\centering SQL Injection (5)}
 & \multicolumn{1}{c|}{1} 
 & \multicolumn{1}{c|}{3} 
 & \multicolumn{1}{c|}{1} 
 & \multicolumn{1}{c|}{-}
 & \multicolumn{1}{c|}{-}
 & \multicolumn{1}{c|}{-}
 & \multicolumn{1}{c|}{-}
 & \multicolumn{1}{c|}{-}
 & \multicolumn{1}{m{6cm}|}{\centering[S55, S59, S62, S70, S89]}
 \\ 
 \cline{2-11}
 & \multicolumn{1}{m{2cm}|}{\centering Cross Site Scripting (3)}
 &\multicolumn{1}{c|}{2}
 & \multicolumn{1}{c|}{-}
 & \multicolumn{1}{c|}{-} 
 & \multicolumn{1}{c|}{1}
 & \multicolumn{1}{c|}{-}
 & \multicolumn{1}{c|}{-}
 &\multicolumn{1}{c|}{-}
 &\multicolumn{1}{c|}{-}
 &  \multicolumn{1}{m{6cm}|}{\centering[S50, S51, S84]}
 \\
 \hline
\multirow{1}{*}{Exploitation(7)} &
\multicolumn{1}{m{2cm}|}{\centering Malware/RAT (7)}
& \multicolumn{1}{c|}{-}
& \multicolumn{1}{c|}{-}
& \multicolumn{1}{c|}{-}
& \multicolumn{1}{c|}{4}
& \multicolumn{1}{c|}{1}
& \multicolumn{1}{c|}{1}
& \multicolumn{1}{c|}{1}
& \multicolumn{1}{c|}{-}
&  \multicolumn{1}{m{6cm}|}{\centering[S1, S13, S16, S45, S48, S61, S81]} 
\\
\hline
\multirow{2}{*}{Command \& Control(15)}
& \multicolumn{1}{m{2cm}|}{\centering Malicious Domain (5)}
& \multicolumn{1}{c|}{-}
& \multicolumn{1}{c|}{3}
& \multicolumn{1}{c|}{-}
& \multicolumn{1}{c|}{-}
& \multicolumn{1}{c|}{2}
& \multicolumn{1}{c|}{-}
& \multicolumn{1}{c|}{-}
& \multicolumn{1}{c|}{-}
& \multicolumn{1}{m{6cm}|}{\centering[S11, S40, S64, S66,S78]} 
\\ 
\cline{2-11}
& \multicolumn{1}{m{2cm}|}{\centering Overt Channels (10)} 
& \multicolumn{1}{c|}{-}
& \multicolumn{1}{c|}{-}
& \multicolumn{1}{c|}{1}
& \multicolumn{1}{c|}{1}
& \multicolumn{1}{c|}{6}
& \multicolumn{1}{c|}{2}
& \multicolumn{1}{c|}{-}
& \multicolumn{1}{c|}{-}
&  \multicolumn{1}{m{6cm}|}{\centering[S22, S39, S41, S42, S43, S44, S49, S57, S83, S85]}
\\ 
\hline
\multirow{2}{*}{Exploration(2)}
& \multicolumn{1}{m{2cm}|}{\centering Lateral Movement (1)}
& \multicolumn{1}{c|}{-}
& \multicolumn{1}{c|}{-}
& \multicolumn{1}{c|}{-}
& \multicolumn{1}{c|}{-} 
& \multicolumn{1}{c|}{-}
& \multicolumn{1}{c|}{-}
& \multicolumn{1}{c|}{1}
& \multicolumn{1}{c|}{-}
& \multicolumn{1}{m{6cm}|}{\centering[S24]}
\\ 
\cline{2-11}
& \multicolumn{1}{m{2cm}|}{\centering Privilege Escalation (0)}
& \multicolumn{1}{c|}{-}
& \multicolumn{1}{c|}{-}
& \multicolumn{1}{c|}{-}
& \multicolumn{1}{c|}{-}
& \multicolumn{1}{c|}{-}
& \multicolumn{1}{c|}{-}
& \multicolumn{1}{c|}{-}
& \multicolumn{1}{c|}{-}
& \\ 
\hline
\multirow{4}{*}{Concealment(21)}
& \multicolumn{1}{m{2cm}|}{\centering Data Tunneling (12)}
& \multicolumn{1}{c|}{-}
& \multicolumn{1}{c|}{2}
& \multicolumn{1}{c|}{3}
& \multicolumn{1}{c|}{-}
& \multicolumn{1}{c|}{7}
& \multicolumn{1}{c|}{-}
& \multicolumn{1}{c|}{-}
& \multicolumn{1}{c|}{-}
&  \multicolumn{1}{m{6cm}|}{\centering[S14, S15, S18, S21, S29, S32, S33, S34, S65, S67, S68, S80]} \\ 
\cline{2-11}
& \multicolumn{1}{m{2cm}|}{\centering Timing Channel (3)}
& \multicolumn{1}{c|}{-}
& \multicolumn{1}{c|}{-}
& \multicolumn{1}{c|}{1}
& \multicolumn{1}{c|}{-}
& \multicolumn{1}{c|}{2}
& \multicolumn{1}{c|}{-}
& \multicolumn{1}{c|}{-}
& \multicolumn{1}{c|}{-}
& \multicolumn{1}{m{6cm}|}{\centering[S23, S27, S92]} \\
\cline{2-11}
& \multicolumn{1}{m{2cm}|}{\centering Side Channel (3)}
& \multicolumn{1}{c|}{-} 
& \multicolumn{1}{c|}{-}
& \multicolumn{1}{c|}{-}
& \multicolumn{1}{c|}{-}
& \multicolumn{1}{c|}{-}
& \multicolumn{1}{c|}{-}
& \multicolumn{1}{c|}{-}
& \multicolumn{1}{c|}{3}
& \multicolumn{1}{m{6cm}|}{\centering[S8, S52, S63]} \\
\cline{2-11}
& \multicolumn{1}{m{2cm}|}{\centering Steganography (3)}
& \multicolumn{1}{c|}{-}
& \multicolumn{1}{c|}{-}
& \multicolumn{1}{c|}{3}
& \multicolumn{1}{c|}{-}
& \multicolumn{1}{c|}{-}
& \multicolumn{1}{c|}{-}
& \multicolumn{1}{c|}{-}
& \multicolumn{1}{c|}{-}
& \multicolumn{1}{m{6cm}|}{\centering[S3, S9, S20]}\\ 
\hline
\multirow{2}{*}{Multi-Stage(16)}
&\multicolumn{1}{m{2cm}|}{\centering Insider Attack (11)}
& \multicolumn{1}{c|}{-}
& \multicolumn{1}{c|}{-}
& \multicolumn{1}{c|}{1}
& \multicolumn{1}{c|}{8}
& \multicolumn{1}{c|}{-}
& \multicolumn{1}{c|}{1}
& \multicolumn{1}{c|}{-}
& \multicolumn{1}{c|}{1}
& \multicolumn{1}{m{6cm}|}{\centering[S4, S6, S7, S10, S12, S35, S56, S72, S74, S90, S91]} \\ 
\cline{2-11}
&\multicolumn{1}{m{2cm}|}{\centering APT (5)}
& \multicolumn{1}{c|}{-}
& \multicolumn{1}{c|}{-}
& \multicolumn{1}{c|}{-}
& \multicolumn{1}{c|}{1}
& \multicolumn{1}{c|}{1}
& \multicolumn{1}{c|}{2}
& \multicolumn{1}{c|}{1}
& \multicolumn{1}{c|}{-}
&[S2, S30, S38, S73, S79] \\
\hline
\end{tabular}}
\end{small}
\end{table*}
\subsubsection{Mapping of proposed taxonomy with respect to DELC}
Table~\ref{mapping} shows the mapping of the proposed taxonomy with respect to DELC.
It can be seen that 48\% of the reviewed papers report data-driven approaches. 
These approaches are most frequently (i.e., 30/48 studies) used in detecting the \emph{Delivery} stage of DELC.
In this stage, they are primarily (i.e., 12/32 and 5/5 studies respectively) employed in detecting \emph {Phishing} and \emph{SQL injection} attacks.
Furthermore, these approaches are also applied in detecting \emph{Concealment} stage with 9/21 studies classified under this category.
However, they are rarely (i.e., 5/44) used in detecting other stages of DELC. 
\par
Among data-driven approaches, \emph{Direct} inspection approach is only applied to detect the \emph{Delivery} stage in the reviewed studies, in particularly \emph{Phishing} (12/24) attacks.
One rationale behind it can be the presence of common guidelines identified by domain experts that differentiate normal website from phishing ones \cite{RN23}.
For example, in [S5, S37] examined the set of URLs and identified the presence of ``@" symbol as one of the feature with an argument that legitimate URLs don't use ``@" symbol. Subsequently, \emph{Context Inspection}
approach is applied by 17/48 data-driven studies to detect 
\emph{Phishing} (9/24), \emph{SQL Injection} (3/5), \emph{Malicious Domain} (3/5) and more recently in \emph{Data tunneling}.
For example, the authors in [S67] used bytes to represent the DNS packets and trained CNN using the sequence of bytes as input.
This representation captured the full structural and sequential information of the DNS packets to detect \emph{DNS tunnels}. 
\emph{Distribution Inspection}
on the other hand is commonly used to detect the \emph{Concealment} stage (7/21) specifically \emph{Steganography} (3/3) and \emph{Data tunnelling} (3/12) attacks.
The reason behind it is that both of these attacks use a different type of data encoding or wrapping to evade detection.
The normal and encoded data have different probability distributions that are imperceptible in \emph{direct} or \emph{context} inspection \cite{RN76}.
For instance, [S20] used inter and intra-block correlation among Discrete Cosine Transform (DCT) of the JPEG coefficient of the image to detect \emph{Steganography}.

52\% of the primary studies are based on \emph{Behavior-driven approaches}. 
In contrast to data-driven approaches, these approaches are recurrently used in detecting all the other stages of DELC except \emph{Delivery}.
This suggests that once the data exfiltration attack vector is delivered the analysis of the behavior exhibited by it is more significant than its content analysis.
An interesting observation is that these approaches are capable to detect multi-stage and sophisticated attacks like \emph{APT} (5/5) and \emph{Insider} threat (10/11).
One reason behind it is that they comprehend the behavior of a system in a particular context to create a logical inference 
that can be extended to manage complex scenarios,
e.g., how a user interacts with a system or 
how many resources are utilized by a program? 
Among behavior-driven approaches,
\emph{Event-based approach} is most evident in detecting RAT (4/7) and insider threat (8/11) detection.
This is because both of these attacks execute critical system events such as plugging a USB device or acquiring root access.
The time or sequence of the event execution is leveraged by this approach to detect data exfiltration.
For example, [S7] analyses the ``logon" event based on working days, working hours, non-working hours, non-working days to detect \emph{Insider} attack.
\par
\emph{Flow-based Approach}
is notable in detecting \emph{data tunnelling} (7/12) and \emph{Overt channels} (6/10) attacks.
The reason behind it is both of these attacks are used for exfiltrating data through the network.
The only difference is the nature of the channel used. 
\emph{Overt channels} transfer data using the conventional network protocols such as Peer to Peer or HTTP post while \emph{Data tunneling} channels use channels that are not designed for data transfer such as DNS.
Additionally, in contrast to \emph{Distribution} inspection which is also used to detect these attacks, 
this approach manually monitors the relationship between the inbound and outbound flow.
For example, [S43] used the ratio of the size of outbound packets with inbound packets to detect \emph {Overt channels} attack.
6/48 behavior-driven studies utilize 
\emph{Hybrid-based approach}.
This approach is used in detecting \emph{APT} and \emph{Overt Channels}.
It captures both the system and network behavior to detect data exfiltration.
For example, [S30] detected an \emph{APT} attack by monitoring both the system and network behavior using the frequency of system calls and the frequency of IP addresses in a session.
The authors claimed that using the hybrid approach can be used to prevent long term data exfiltration.
\emph{Propagation based approach} is an interesting countermeasure because it examines multiple stages of DELC across multi-host to detect data exfiltration. 
For instance, [S2] identifies multiple stages of DELC to detect the \emph{APT} attack. 
First, the approach detects the exploitation stage by using a threat detection module, then it uses malicious SSL certificate, malicious IP address, and domain flux detection to discover the Command and Control (C\&C) communication. After that, it monitors the inbound and outbound traffic to disclose lateral movement, and finally it uses scanning and Tor connection \cite{RN77} detection to notice the asset under threat and data exfiltration respectively.
Lastly, \emph{Resource usage-based approach}
is used by \emph{side-channel} (3/3) and \emph{Insider} (1/1) attack detection. 
A reason behind it is that for \emph{Side-channel} attacks, we only found studies that were detecting \emph{cache-based side-channel} attacks. 
In this type of attack, an attacker exploits the cache usage pattern \cite{yarom2014flush} to exfiltrate data.
Hence, this approach monitors the cache usage behavior to detect \emph{Side-channel} attacks.
For example, [S52] used a ratio of cache misses and hit as one of the features to detect \emph{Side-channel} attack.
\par
\begin{small}
\noindent\fbox{%
\parbox{\columnwidth}{%
 \textbf{\textit{Summary}}: ML countermeasures can be classified into data-driven and behavior-driven approaches. Data-driven approaches are frequently reported to detect \emph{Delivery} stage attacks while behavior-driven approaches are employed to detect sophisticated attacks such as APT and insider threat.}} \end{small}
\subsection{{RQ2}: Features Extraction Process}
\label{RQ2}
Feature engineering produces the core input for ML models. 
We have analyzed data items: D13 to D14 and D21-D22 given in Table~\ref{dataextractionform} to answer this RQ.
This question sheds light on three main techniques used in the feature extraction process, i.e., feature type, feature engineering method, and feature selection.
\subsubsection{Feature Type and Feature engineering method}
\begin{figure*}
\captionsetup[subfigure]{font=footnotesize,labelfont=scriptsize} \begin{subfigure}{.18\textwidth}
  \centering
   \includegraphics[width=\linewidth,height=7.7cm,keepaspectratio]{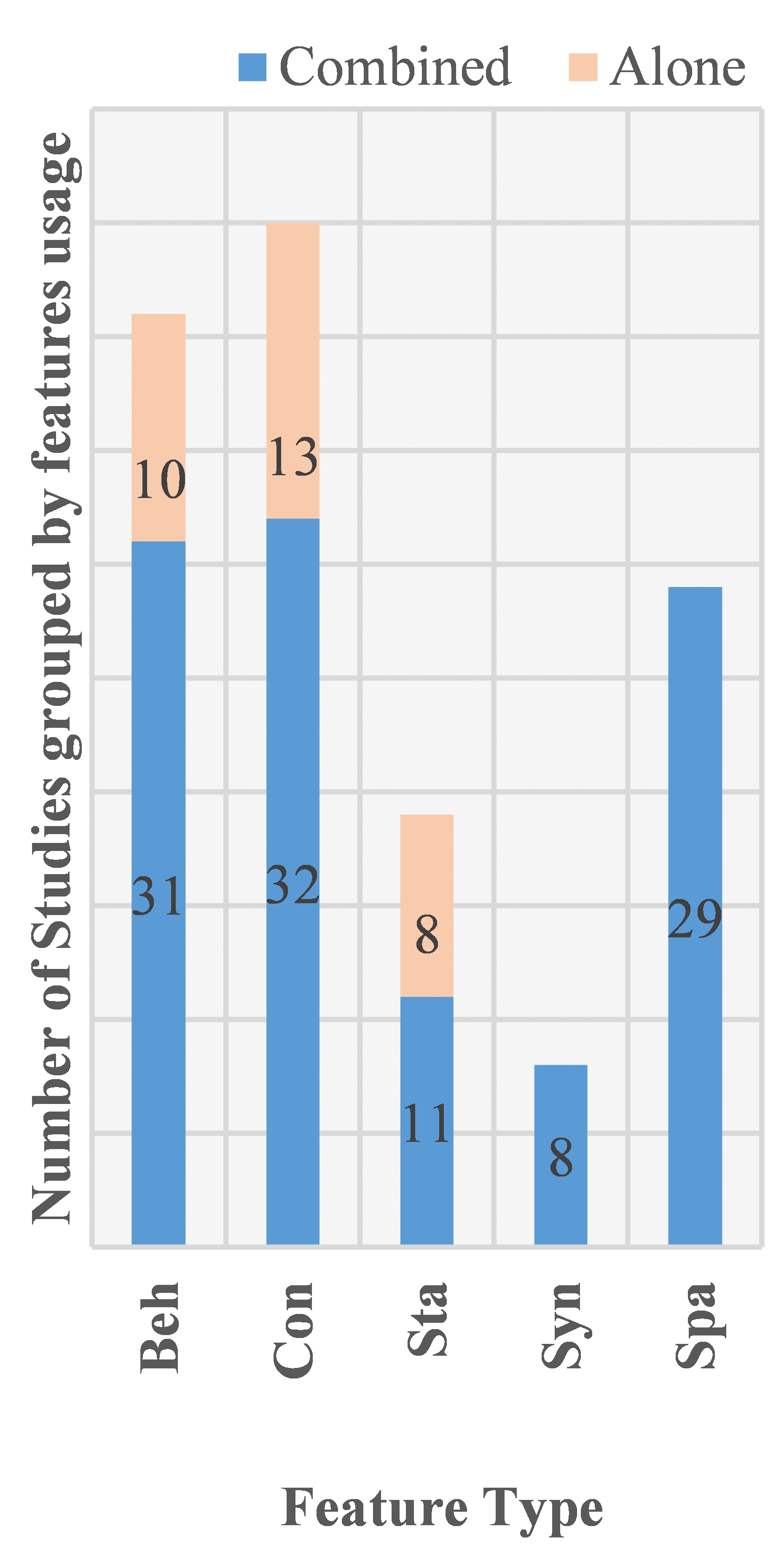}
\caption{ Feature types used in reviewed studies}
\label{fig3:sub-first}
\end{subfigure}
\begin{subfigure}{0.80\textwidth}
\begin{small}
\centering
\includegraphics[width=.99\linewidth,height=7.77cm,keepaspectratio]{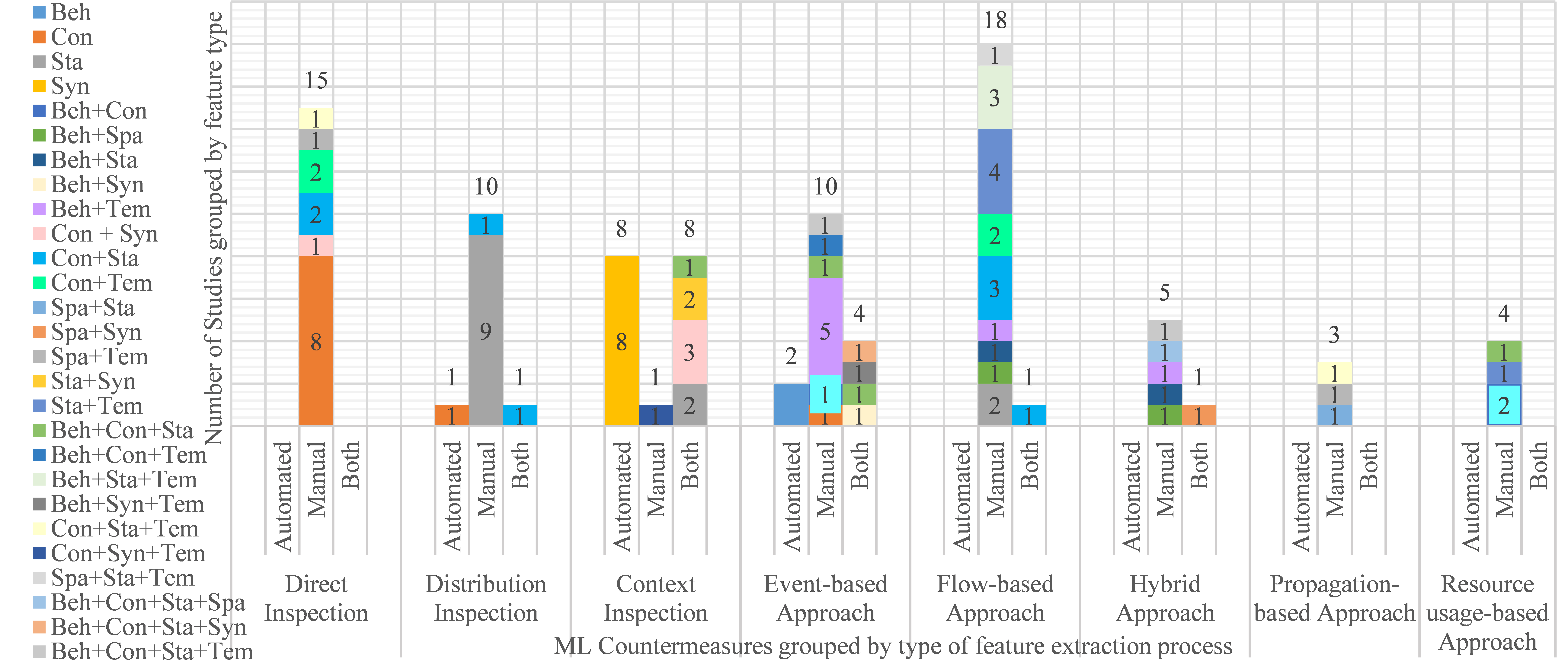}
\caption{ Mapping between Feature Types, Feature Engineering Method and ML countermeasures}
\label{fig3:sub-second}
\end{small}
\end{subfigure}
\caption{Feature Extraction Analysis here Beh, Con, Sta, Syn, Sta and Tem represent Behavioral, Content-based, Statistical, Syntactical, Statistical and Temporal features respectively}
\label{fig3:fig3}
\end{figure*}

This section discusses the type of features and feature engineering methods used by the primary studies.
The features used by the studies can be classified into six types: statistical, content-based, behavioral, syntactical, spatial and temporal. 
Table~\ref{featuredescription} briefly describes these base features and enlist their strengths and limitations.
While
Table~\ref{engineeringmethods} describes the feature engineering methods \cite{RN38} with their strength and weakness.
These features are mined single-handedly or collectively as hybrid features by different studies, as shown in Fig~\ref{fig3:sub-first}.
Furthermore, Fig~\ref{fig3:sub-second} shows the distribution of these features types with respect to ML countermeasures and feature engineering method.
\begin{table*}[hbt!]
\begin{small}
\caption{Feature type, their description, and their strengths and weaknesses}
\label{featuredescription}
\centering
\resizebox{\textwidth}{!}{\begin{tabular}{|c|c|l|l|}
\hline
\textbf{Features}& \textbf{Description} & \textbf{Strengths }& \textbf{Weaknesses} \\
\hline
\multirow{4}{*}{Statistical}& These features represent the statistical information &\tabitem Unveil hidden mathematical patterns. & \tabitem Computationally expensive to \\ 
& About data, e.g., average, standard deviation, entropy,  &\tabitem Reliable when extracted from large  &compute \cite{RN81}. \\ 
     & median, and ranking of a webpage. & population. &\tabitem Sensitive to noise \cite{RN76}\\
     &&&\\
     \hline
\multirow{4}{*}{Content} &  &\multirow{8}{*}{\tabitem Easy to extract.} & \tabitem Requires security analyst to extract features.  \\ 
     & These features denote the information present in or  &  & \tabitem Difficult to identify unless data  \\ 
     & about data and do not require any complex computation.  &  & is linear or simple in nature \cite{RN76}. \\ 
     -based& For example, presence of @ sign in a URL, IP address, &  & \tabitem Only suitable for data with regularly \\ 
     & number of hyperlinks and steal keyword. &  & repeated patterns. \\ 
     &  &  & \tabitem Short-lived and suffer from concept  \\ 
     &  &  & drift over time \cite{RN157}.   \\ \hline
\multirow{5}{*}{Behavioural }& These features signify the behaviour of an application,  &  &  \\ 
     & user or data flow. They include a sequence of API calls,  & \tabitem Offer good coverage to detect & \tabitem Unable to capture run-time attacks \\ 
     & is\_upload flag enabled or presence of POST method in  & malicious events. \cite{RN83} & \cite{RN84} and \cite{RN83}. \\ 
     & request packet, connection successful or not, record type  &  &  \\ \
     & in DNS flow and system calls made by users. &  &  \\ \hline
\multirow{5}{*}{Syntactical} & These features embody both syntax and semantics of  &\tabitem Captures the sequential, structural, & \tabitem Produce sparse, noisy and high-  \\
     & data. For example, grammatical rules like verb-object  & and semantical information from  & dimension feature vectors.  \\ 
     & pairs, n-grams, topic modelling, stylometric  & data. & \tabitem Mostly based on n-grams, which  \\ 
     & and word or character embeddings features. &  & may result in arbitrary elements  \\ 
     &  &  & \cite{RN90}. \\ \hline
    \multirow{4}{*}{Spatial}& Instead of a single host in a network, these features  &  & \tabitem Suffer from high complexity because \\ 
     & signify the information across multiple hosts. For  & \tabitem Highly useful to detect multi-stage  & they are extracted in a distributed  \\ 
     & example, the number of infected hosts or community  & attack. & manner and evidence collection is \\ 
     & maliciousness. &  & dependent on multi-hosts \\ \hline
    \multirow{5}{*}{Temporal} &  &  & \tabitem Can fail in real-time because of the  \\ 
     & These features represent the attributes of data that are  &\tabitem Effective in capturing the temporal  & volume of network stream and  \\ 
     & time dependent. For example, time interval between  & patterns in data & computational effort required to  \\ 
     & two packets, time to live and the age of a domain, time between query  &  & compute the features in a particular  \\ 
     & and response,timestamp of an action, most frequent  &  & time frame \cite{RN81}.\\
     & interval between logged in time.& &\\\hline
\end{tabular}}
\end{small}
\end{table*}
\begin{table*}[hbt!]
\caption{Feature Engineering Techniques, their description with their strengths and weaknesses}
\label{engineeringmethods}
\centering
\resizebox{\textwidth}{!}{\begin{tabular}{|c|l|l|l|}
\hline 
\textbf{Technique}
& \textbf{Description} &
\textbf{Strengths} &
\textbf{Limitations} \\ 
\hline 
\multirow{6}{*}{Manual} &   & \tabitem Return smaller number& \tabitem Time-consuming and labour-intensive. \\ 
& It requires domain  & of meaningful features. &\tabitem Error prone. \\ 
& expert to analyze data to  & \tabitem Suitable for homogeneous &\tabitem Not adaptable to new datasets as\\
& extract meaningful features. & data with linear patterns. &this approach extracts problem-dependent  \\ 
&&&features and must be re-written for new dataset. \\ 
\hline
\multirow{6}{*}{Automated} &  & \tabitem Reduce Machine Learning  &  \\
& Advance ML techniques are  & development time & Sparse and high dimensional feature vector. \\
& used to extract the features  & \tabitem Can adapt to new dataset. & \tabitem Computationally expensive. \\ 
& from data automatically. & \tabitem Suitable for complex data with  & \tabitem Require large amount of data to produce \\ 
&  & structural and sequential  & good quality features. \\ 
&  & dependencies such as network attack &\\
&&detection. &  \\
\hline
\multirow{2}{*}{Semi-} & These techniques use both  & \tabitem Suitable for complex task such&  \\
& manual and automated  &  as APT and lateral movement attack& \tabitem More time-consuming then automated. \\ 
 Automated& techniques to extract features  & detection where human intervention & \tabitem Computationally expensive than manual. \\ 
(Both)& from data. & is mandatory.   &  \\ \hline
\end{tabular}
}
\end{table*}

\paragraph{Statistical Features}
Statistical features are the most commonly used features (i.e., 45/92 studies) among all the primary studies. 
It is because the statistical analysis can yield hidden patterns in data \cite{RN78}. 
These features are used as base features by 13 studies, while 32 studies used them together with other types of features.
(34/45) reviewed studies mined these features manually to detect data exfiltration attack.
Whilst, it is intuitive to assume that \emph{Distribution} inspection countermeasures utilize these features, this hypothesis is untrue. For example, [S17] used TF-IDF \cite{RN79} to identify the most common keywords in data.
However, instead of the frequency of keywords, 
direct keywords (content-based features) are used as features to train a classifier. Moreover, statistical features are not limited to \emph{Distribution} inspection but are also used in \emph{Context} inspection as shown in Fig~\ref{fig3:fig3}.
For instance, [S53] automatically extracts features from host, path and query part of URL.
These features contain the probability of summation of each part per (for n=1 to 4) n-grams. The n-gram \cite{RN80} is a context inspection approach; however, the study used this analysis to calculate the probability of summation of n-gram frequency.
The probability was then used as a feature instead of n-grams.
\paragraph{Content-based Features}
41/92 studies employ content-based features to detect data exfiltration attacks.
The reason behind their extensive usage is the ease to extract them. 
Unlike the statistical features, these features do not require complex computation and can be extracted directly from data.
These features are used as base features by ten studies while 31 studies used them with other features.
These features are usually manually extracted (40/41) studies from data by using domain expertise.
Only one study [S17] extracted these features automatically by using TF-IDF to mine the most common keywords and then scanned these keywords directly from the content as features.
These features include IP address, gender, HTTP count, number of dots, brand name in URL and special symbols in URL.
80\% of direct inspection countermeasures used these features. 
Content-based features are mostly (18 studies) combined with statistical and temporal features to extract more fine-grained features from data as illustrated in Fig~\ref{fig3:fig3}. 
\paragraph{Behavioral Features}
These features are utilized mostly by (29/92 studies) \emph{behavior-driven} approaches because they are capable of providing a better understanding of how an attack is implemented or executed. 
Only two studies [S48, S73] used these feature solely. [S48] extracted Malicious instruction sequence pattern to detect \emph{RAT malware} whereas [S73] sequence of event codes as features to train LSTM classifier for detecting \emph{APT} attacks.
Both of these studies used automatic feature engineering techniques to extract these patterns.
In [S48], a technique named malicious sequential mining is proposed to automatically extract features from Portable Executable (PE) files using instruction set sequence in assembly language. 
{Similarly, in [S73] event logs is given as input to LSTM, which is then responsible for extracting the context of the features automatically.}
These features are usually combined with temporal features (12 studies) to represent time-based behavioral variation. 
For example, in [S12] attributes like logon during non-working hours, USB connected, or email sent in non-working hours depict these features. 
Also, behavioral features are often combined with content-based and statistical features (13 studies) to depict the activities of a system in terms of data distribution and static structure, as shown in Fig~\ref{fig3:sub-second}.
For example, in [S13], features like the address of entry point, the base of code, section name represent content-based features. In contrast, number of sections, entropy denote statistical features whereas imported DLL files, imported functions type indicate behavioral features. 
Another interesting study is [S91], which combines these features with syntactical and temporal features. For obtaining a trainable corpus, the authors first transform the security logs into the word2vec \cite{goldberg2014word2vec} representation, which are then fed to auto-encoders to detect insider threat.
\paragraph{Syntactical Features}
These are used as standalone features by 11, whereas eight studies used them combined with other features.
These features are mostly (17/19) automatically or semi-automatically extracted as can be seen in Fig~\ref{fig3:sub-second}.
These features are used by three event-based and one hybrid approach to extract the sequence of operations performed by a system, executable or user.
For example, in [S61], two types of features were extracted: n-grams from executable file unpacked to binary codes using PE tools \cite{RN88} and behavioral features, i.e., Window API Calls extracted using IDA Pro \cite{RN89}.
Whilst these features are well suited for extracting sequential structural or semantic information from data; they can result in sparse, noisy and high-dimension feature vectors. Most of these features are based on n-grams, which may result in arbitrary elements \cite{RN90}. A high-dimensional and noisy feature vector can be detrimental to detection performance.
\paragraph{Spatial Features}
Spatial features are used by flow-based (2), propagation-based (2/3) and three hybrid approaches.
These features are manually extracted by all these studies.
In [S16], the authors used a Fraction of malicious locations, number of malicious locations, files, malware, and community maliciousness as spatial features to detect RAT malware.
These features include destination diversity, the fraction of malicious locations, number of malicious locations, files, malware, community maliciousness, number of lateral movement graph passing through a node, frequency of IP, MAC, and ARP modulation. Spatial features are advanced features that can be utilized to detect complex scenarios like APT. Besides, the spatial features are not used alone by any study but are extracted with other features as depicted in Fig ~\ref{fig3:sub-first}.
\paragraph{Temporal Features}
These features act as supportive features because no study has used them alone. 
However, these features are essential to reveal the hidden temporal pattern in data that is not evident via other features. 29 studies, along with other features, use these features. 
These features extracted manually by 28/29 studies except for [S91] that transformed the security logs from four weeks in a sequence into sentence template and trained word2vec to detect \emph{Insider} attack.
\subsubsection{Feature Engineering }
Feature engineering process extracts features from data. Feature engineering methods are classified as manual, automated, and semi-automated as described in  Table 10. 51 studies used manual, and four studies used a fully automated. In comparison, eight studies used semi-automated feature engineering method to extract features. Fig 10 shows the mapping of feature type with feature engineering method. It is evident that the studies based on syntactical features either used an automated or semi-automated feature engineering method. It was predictable as syntactical features are based on text structure and NLP techniques can be utilized to extract them automatically. However, it was unanticipated that one study utilizing behavioural features used this technique. In [S48], a technique named malicious sequential mining is proposed to automatically extract features from Portable Executable (PE) files using instruction set sequence in assembly language. Other studies that used automated feature engineering method used semantic inspection countermeasure but extracted statistical or Content-based features to train the classifiers. 
\subsubsection{Feature Selection}
Feature selection is important, but an optional step in feature engineering process \cite{RN39}. 
Table~\ref{featureselection} describes, and provides strengths and weaknesses of five different methods used by our reviewed study to select the discriminant features from the initial set of features.
\begin{table*}[hbt!]
\begin{small}
\caption{Feature Selection Methods, their description with strengths and weaknesses}
\label{featureselection}
\centering
\resizebox{\textwidth}{!}{\begin{tabular}{|c|l|l|l|}
\hline
    \textbf{Methods} & \textbf{Description}&
    \textbf{Strengths} & \textbf{Weaknesses} 
    \\
    \hline
    \multirow{6}{*}{\rotatebox{90}{Filter \cite{RN39}}} & These approaches select the best features by  &  \tabitem Computational fast speed \cite{RN39},  & \tabitem Ignore classifier biases and  \\ 
     & analyzing the statistical relationship between  & \tabitem Simple to implement and comprehend & heuristics. \\ 
     & features and their label. Popular filter methods  & classifier independent & \tabitem Low performance \cite{hira2015review} as not  \\
     & include information gain, correlation analysis and  &  & classifier dependent \\ 
     & Chi-square test. &  &\tabitem  Requires manually setting the   \\ 
     &  &  & thresholds. \\ 
     \hline
    \multirow{4}{*}{\rotatebox{90}{Wrapper}}\multirow{4}{*}{\rotatebox{90}{\cite{RN91}}}& These approaches consider the classifier used  & \tabitem Good performance for specific &  \\ 
     & for classification to select the best features.  & model. & \tabitem Classifier dependent \\ 
     &  Popular wrapper method includes genetic   &\tabitem Consider features dependencies & \tabitem Computationally expensive \\ 
     & algorithm and sequential search. &  &  \\ 
     \hline
    \multirow{6}{*}{\rotatebox{90}{Embedded}}\multirow{6}{*}{\rotatebox{90}{ \cite{RN39}}}& These techniques use both advantages of filter and  & \tabitem Learn good features in parallel  & \tabitem Need for fine-grain  \\ 
     & wrapper method to select the best features. Popular  & to learning the decision  & understanding of classifier  \\ 
     & embedded methods include Random Forest Trees,  & boundary. & implementation \\ 
     & XG-Boost, Ada-boost. & \tabitem Less expensive than wrappers. & \tabitem Computationally expensive  \\ 
     &  &  & than filter. \\
     &  &  & \tabitem Classifier specific \\ 
     \hline
    \multirow{5}{*}{\rotatebox{90}{Dimension}}    \multirow{5}{*}{\rotatebox{90}{Reduction}}
    \multirow{5}{*}{\rotatebox{90}{\cite{RN92}}} & These techniques select the optimal feature set to  & \tabitem Computational fast speed \cite{RN39}  & \tabitem Not interpretable. \\ 
     & decrease the dimensions of the feature vector.  &\tabitem Transform features to low  & \tabitem Manually setting for threshold  \\ 
     & Popular methods include PCA, LDA and Iso-maps. & dimension. & is required. \\
     &  & \tabitem Classifier independent. &  \\ 
     & & & \\
     \hline
     \multirow{5}{*}{\rotatebox{90}{{Attention}}}    
    \multirow{5}{*}{\rotatebox{90}{\cite{vaswani2017attention}}} & In 2017 a mechanism for Deep Learning (DL)  & \tabitem Decrease development time  & \tabitem Not reliable \cite{jain2019attention,serrano2019attention}. \\ 
     & approaches that focus on certain factors than &\tabitem of DL methods.  &  \\ 
     & other when processing data \cite{tang2018analysis}. & \tabitem Automatically learns relevant features &  \\
     &  & \tabitem without manual thresholds. &  \\ 
     && \tabitem  Yield a significant
     performance gain \cite{galassi2019attention}. & \\
     \hline
\end{tabular}}
\end{small}
\end{table*}

\begin{figure*}
\includegraphics[width=\linewidth]{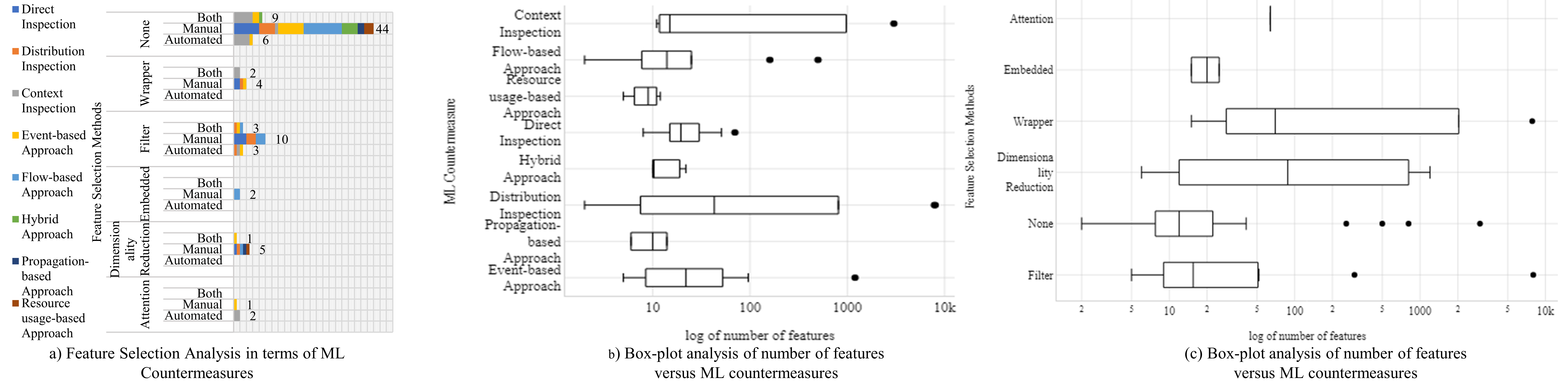}
\caption{Analysis of Feature Selection Method, Y-axis in (b,c) depicts the log of the number of features reported by the studies}
\label{figure11}
\end{figure*}
The feature selection techniques used by the reviewed studies with respect to ML-based countermeasure, feature-engineering technique and the number of the features used are shown in Fig~\ref{figure11}.
 This analysis is helpful to understand the relationship of feature selection techniques with data analysis and feature engineering process. 
It shows that 52\% studies utilized feature selection methods.
Among this \emph{filter} is the most popular (16 studies) method.
Almost all the approaches apply it except for hybrid, propagation-based and resource usage-based countermeasures.
Additionally, this approach is not only applied to features extracted manually but also applied to automatic and semi-automatically (3 studies each) extracted features. 
For example, in [S1], the authors used semi-automatically extracted features i.e., Opcodes n-grams (automated)and gray-scale images (manual), and the frequency of Dynamic Link Library (DLL) in import function (manual) to classify known versus unknown malware families. The feature set dimensionality is then reduced 
and used information gain \cite{sheen2008network} a \emph{filter} method to reduce the dimensions of the feature vector.
Similarly, [S5] extracts 16 features manually based on frequency analysis assessment criteria to detect phishing websites. 
Chi-Square \cite{sheen2008network} feature selection algorithm a \emph{filter} method is incorporated to select the features.

The second most popular technique is 
\emph{Dimensionality reduction} is used by six studies.
It is applied by all approaches except for context inspection and hybrid approaches.
Five studies employ this technique over manually extracted features while only one study [S61] used it for semi-automatically extracted features.
In [S61],  two types of features were extracted n-grams from executable file and Window API Calls.
Principle Component Analysis (PCA) is used to reduce the n-gram feature dimension, while class-wise document frequency \cite{devesa2010automatic} is utilized to select the top API call. \emph{Wrapper} method has been reported in six studies. 
It is mostly used (i.e., 5/6) in data-driven countermeasures and is employed by 4 and 2 studies that extract features manually and semi-automatically, respectively.
For instance, the study [S60] used two types of features webpage features and website embedded trojan (7) features. 
For feature selection, two \emph{wrapper} methods: RFT based Gini Coefficient \cite{singh2010feature} and stability selection \cite{meinshausen2010stability} are employed.
 {Another interesting but relatively new feature selection method is \emph{Attention} \cite{galassi2019attention}. Although, it is not primarily known as a feature selection method because this algorithm is used to select most relevant features while training DNNs, we have treated it as a feature selection method, as also suggested by \cite{galassi2019attention}. 
\emph{Attention} is used by three studies, out of which two [S87, S88] used it on automatically extracted features as a part of DNN training whereas [S74] used it on manually extracted features that are fed as input to DNN classifier.
In [S74], users action sequence are manually extracted from log files and feed into Attention-based LSTM that help a classifier to pay more or less focus to individual user actions.
} Lastly, the \emph{embedding} method is just explored by two flow-based studies [S11, S65].
In [S11], an ensemble anomaly detector of Global Abnormal trees was constructed based on weights of feature computed by information gain (filter method). 
We believe that one reason for not using the embedding method is the need for fine-grain understanding of how classifier is constructed \cite{RN93}. 
Most of the selected studies only apply ML classifiers as a black box. 
\par
In Fig~\ref{figure11}b, Fig~\ref{figure11}c, we analysed the number of features used viz-a-viz the ML countermeasure and types of features selection methods.
The y-axis shows the log of the number of features used by the reviewed studies. We have used this because the total number of features used by the studies have high variance (ranging from 2 to 8000) and hence useful visualization was not possible with direct linear dimensions.
The circular points show the outliers in the distribution while the solid line shows the quartile range and median based on the log of the number of features used by the studies. 
Fig~\ref{figure11}b analysis is based on 69/92 studies that reported the total number of features used for training the classifier.
Among the remaining 23 studies, 16 studies either extracted feature automatically or semi-automatically and 12 studies belong to context inspection ML countermeasure. 
It may be because these studies mostly rely on Natural Language Processing (NLP) techniques that produce high dimensional feature vectors \cite{RN94}. 
Furthermore, the low median of context inspection approach is just based on the four studies that reported the number of features used. 
However, for other ML countermeasures, it can be seen that the \emph{distribution} inspection studies use high feature dimensions (ranging from 258-8000).
While other countermeasure use relatively low number of features, e.g., (11/16) event-based studies utilized only $<100$, all the direct inspection approaches use $<80$ features while flow, hybrid, resource-usage and propagation-based approaches used $<25$ features.
Fig~\ref{figure11}c shows the relationship between the number of features with feature selection methods.
From the analysis it is inferred that those studies which did not use any feature selection technique and reported features dimensions (41 studies) already had low feature dimensions, i.e., having a log median of $<12$.
Among these only 4/41 of these studies used $>250$ features.
Furthermore, the studies that apply dimensionality reduction methods and wrapper still result in a relatively high dimensional space (log median is approximately equal to 100). 
\par
\begin{small}
\noindent\fbox{%
\parbox{\columnwidth}{%
 \textbf{\textit{Summary}}: In ML countermeasures to detect data exfiltration, statistical and content-based features, manual feature engineering process and filter feature selection method are most prevalent.}} 
 \end{small}
\subsection{{RQ3}: Datasets}
\label{RQ3}
\begin{figure*}[hbt!]
\includegraphics[width=\textwidth,height=6cm,keepaspectratio]{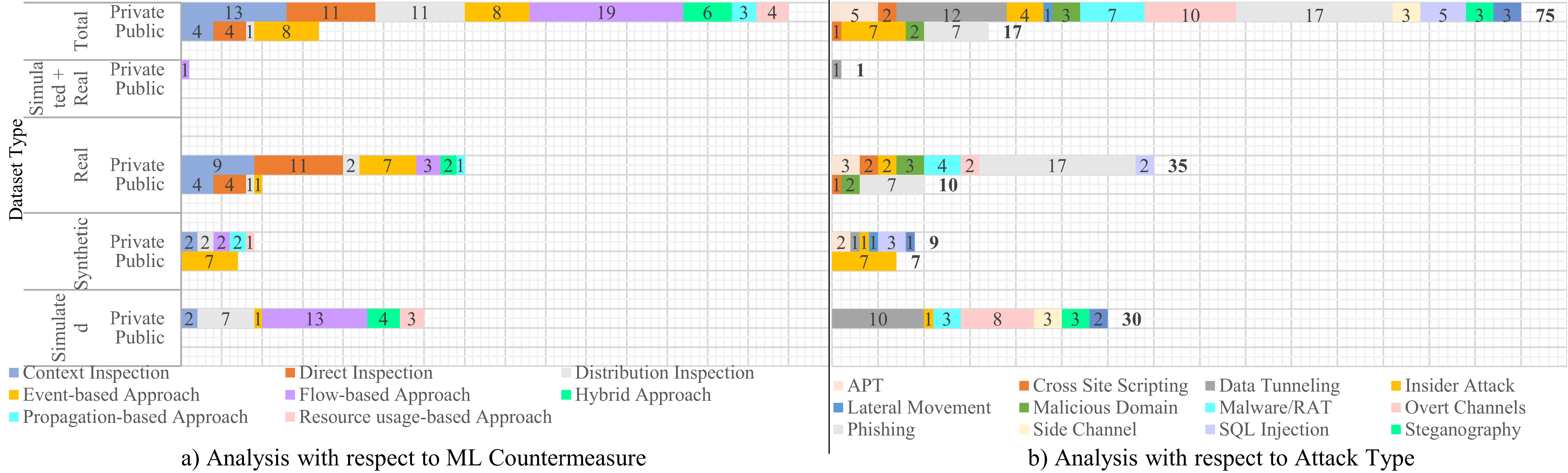}
\caption{Dataset Type and Availability with respect to ML countermeasures and Attack Type, respectively (The number shows the total studies in each category, while the bold number shows total studies in terms of y-axis)}
\label{figure12}
\end{figure*}

This question provides insight into the type of dataset used by the reviewed studies. We classify them as: 1) real, 2) simulated and 3) synthetic datasets as briefly described in Table~\ref{DatasetTypes}. 
These datasets can either be public or private. Fig~\ref{figure12} depicts the distribution of the type of dataset and their  availability (public or private) with the type of ML countermeasure and attack vectors.
\begin{table*}[hbt!]
\begin{small}
\caption{Type of Datasets, their description with strengths and weaknesses}
\label{DatasetTypes}
\centering
\resizebox{\textwidth}{!}{\begin{tabular}{|c|l|l|l| }
\hline
    \textbf{Type} & \textbf{Description} &\textbf{Strengths} & \textbf{Weaknesses} \\ \hline
    
    \multirow{6}{*}{\rotatebox{90}{Real}}
    \multirow{6}{*}{\rotatebox{90}{Datasets}}& The datasets obtained from real data,  & \tabitem Provides true distribution of  & \tabitem Can suffer privacy and confidentiality issues. \\ 
 & e.g., public repositories containing  & data. & \tabitem Heterogeneous especially for network  \\ 
     
     & actual Phishing-URLs like Phish-Tank  &  & attacks. \\ 
    
     & \cite{RN98} or original collection of spam  &  & \tabitem Cannot depict all the misuse scenarios. \\ 
    
     & emails or real malware files. Both data  &  & \tabitem Can suffer from imbalance \\
     & and environment are real. &  &  \\
     \hline
        \multirow{8}{*}{\rotatebox{90}{Simulated }}    \multirow{8}{*}{\rotatebox{90}{Datasets}} & The dataset generated by software  & \tabitem Able to reproduce balanced  &  \\ 
     & tools in a controlled environment to  & datasets. &  \\
     & simulate an attack, e.g., collected from  & \tabitem Able to generate rare misuse & \tabitem Tool specific \\ 
     & some software tools to generate an  & events. & \tabitem May not depict real distribution of data. \\ 
     & attack scenario. Data is real (e.g.,  & \tabitem Useful for attacks for which  & \tabitem Not a representation of real heterogeneous  \\ 
     & network packets) but environment is  & real data is not available. & environment. \\
     & controlled (i.e., network packets are  &  &  \\ 
     & produced by a tunnelling tool). &  &  \\ 
     \hline
        \multirow{5}{*}{\rotatebox{90}{Synthetic}}     \multirow{5}{*}{\rotatebox{90}{Datasets}}& The dataset produced by using  & \tabitem Able to generate complex  & \tabitem Not a true representation of real \\ 
     & mathematical knowledge of the  & attack scenarios. & heterogeneous environment. \\ 
     & required data. Both data and  & \tabitem Useful for complex multi- & \tabitem Requires domain expertise to generate these  \\ 
     & environment is controlled (i.e., data  & stage attacks with multiple  & datasets. \\ 
     & produced by a function f=sin(x)). & hosts. & \tabitem Hard to adapt new variations in data. \\ 
     \hline
\end{tabular}}
\end{small}
\end{table*}
\begin{table*}[hbt!]
\begin{small}
\caption{Publicly available datasets, their brief description and mapping with the reviewed studies}
\label{publicDatasets}
\resizebox{\textwidth}{!}{\begin{tabular}{|c|c|c|l|c| }
\hline
    \textbf{Attack Vector} & \rotatebox{90}{\textbf{Type}} & \textbf{Dataset}& \textbf{Description} & \textbf{Study} \\ \hline
    \multirow{20}{*}{Phishing} &\multirow{20}{*}{\rotatebox{90}{Real}} & CLAIR collection of  & This dataset contains a real collection of "Nigerian" Fraud emails from 1998  & [S17] \\
     &  & fraud email \cite{RN100} & to 2007. It contains more than 2500 fraud emails. &  \\ \cline{3-5}
     &  & URL-Dataset \cite{RN101} & The dataset is composed of approximately $2.4x10^6$ URLs and $3.2 x10^6$ features.  & [S26] \\ 
      &  &  & URLs are collected from Phish-Tank \cite{RN98}, 
     Open-Phish \cite{RN102} feeds in 2009 . &  \\ \cline{3-5}
     &  & UCI-Repository & This dataset had 11055 data samples with 30 features extracted from 4898  & [S86] \\
     &  & \cite{Dua2017, mohammad2014intelligent} & legitimate and 6157 phishing websites. &  \\  \cline{3-5}
     &  & Spam URL  & This dataset contains 7.7 million webpages from 11,400 hosts based on web  & [S31] \\ 
     &  & Corpus \cite{RN103} & crawling done on .uk domain on May 2007. Over 3,000 webpages are labelled  &  \\ 
     &  &  & by at least two domain experts as “Spam”, “Borderline” or “Not Spam”. &  \\ \cline{3-5}
     &  & Enron Email  & This dataset consists of emails from about hundred and fifty users, mostly  & [S36] \\ 
     &  & Corpus \cite{RN104,TheEnron29} & experienced management of Enron. A total of about 0.5M emails are present in  &  \\ 
     &  &  & the current (7th May 2015) dataset version. &  \\
     \cline{3-5}
     &  &  &{It is a new dataset with two version available IWSPA v1\cite{VermaRakesha2019IWSPA1} and IWSPA v2 \cite{Zeng2020IWSPA2}} & [S87] \\
     &  & {IWSPA-AP}{\cite{IWSPAAPd47, Security62IWSPA}}&  {published in 2018 and 2020 respectively. The dataset is composed of a mix of new}  &  \\ 
     &  &  & { and historical legitimate and phishing emails collected from different sources.}  &  \\ 
     \cline{3-5}
     &  & Phishing email  & \multirow{2}{*}{This dataset contains real phishing email collection from 2005 to 2018.}& [S19, S36] \\ 
     &  & corpus \cite{RN105} &  &  \\ 
    \cline{3-5}
    Malicious &  & ClueWeb09 \cite{RN96}  & The ClueWeb09 dataset contains a crawl result of 1 billion webpages containing & [S31, S51, S40]  \\ 
    Domain &  &  &  one half English webpages. The data was collected in 2009. & \\
    \cline{3-5}
    \multirow{2}{*}{XSS} &  & Leakiest \cite{RN106} & The dataset consists of 1924 instances of 72 features extracted from java-script & [S50] \\ 
     &  &  & code. The dataset was collected from leakiest tool on 30 August 2012. &  \\ \cline{3-5}
    Malware/ &  & VX-Heaven \cite{RN99} & The dataset provides a collection of window virus obtained on 18 May 2010. & [S13, S61] \\ 
    RAT&&&&\\
    \hline
    Insider  & \multirow{3}{*}{\rotatebox{90}{Synthetic}}& \multirow{2}{*}{CERT \cite{RN107}} & This dataset contains a collection of synthetic insider threat test scenarios that  & [S7, S12, S72] \\
    &&&provide both legitimate and malicious insider synthetic data. The dataset was&S74, S90, S91]\\\
     Threat &  &  & synthesized in 2011. &  \\ 
     Detection&  &  &  &  \\ 
     \cline{3-5}
     &  & DARPA 1998 & This dataset was synthesized by MIT Lincoln Laboratory in 1998 based on  & [S6] \\
     && \cite{RN108}&statistical properties of a government sites.&\\
     \hline
\end{tabular}}
\end{small}
\end{table*}

\begin{table*}[hbt!]
\begin{small}
\caption{List of threat intelligence feeds}
\label{Threatintelligence}
\centering
\resizebox{\textwidth}{!}{\begin{tabular}{|c|c|l|c| }
\hline
    \textbf{Attack Vector} & \textbf{Threat Intelligence} &\textbf{Description} & \textbf{Study} \\
    \hline
    \multirow{12}{*}{Phishing} & Millersmiles  & Maintains an active and massive collection of phishing URLs and identify  & [S5] \\ 
     &  \cite{RN116}& theft email scam data. &  \\ 
     \cline{2-4}
     & &  & [S5, S37, S46,  \\ 
     & PhishTank &  Collects and maintain phishing URL data on the internet. & S47, S53, S58,  \\ 
     & \cite{RN98}&  & S40, S69, S71],  \\ 
     &  &  & S75, S77, S88] \\ 
     \cline{2-4}
     & OpenPhish \cite{RN102}& Provide an active collection of phishing website URLs & [S28, S46, S53, \\ 
     &&& S88]\\
     \cline{2-4}
     & Anti-Phishing Alliance  & Provides a list of phishing websites in china. & [S54] \\ 
     &of China \cite{RN117}&&\\
     \cline{2-4}
     & Hacked unrestricted  & Provides a list of defaced (fake visual appearance of a website or a web  & [S60] \\ 
     & information \cite{RN118}& page) website URLs. &  \\ 
     \cline{2-4}
     & Defaced websites URLs \cite{RN119}, & Provides a list of defaced (fake visual appearance of a website or a web  & [S28, S60] \\
     &  & page) website URLs. &  \\ 
     \hline
     \multirow{2}{*}{SQL Injection} & {Exploit-DB \cite{ExploitDB}} & {A threat intelligence service for reporting vulnerabilities.} & [S70]\\
     \cline{2-4}
     & {WooYun \cite{Wooyun}}& {A threat intelligence service for reporting vulnerabilities} & [S70]\\
     \hline
    Malicious Domain & DNS-BH \cite{RN120} & Provides a blocklist of malicious domains. & [S28] \\ 
    \cline{2-4}
     & \multirow{2}{*}{{DGArchive} \cite{ DGArchiv50}} & {An online service that provide a list of malware domains generated by } & [S78] \\
     &  & {domain Generation Algorithms (DGAs).} &  \\ 
     \cline{2-4}
     & {DGA-Badar}  \cite{baderj} & {Contains a list of 45 DGA tools to generate malicious domains.} & [S65] \\ 
     \cline{2-4}
     & {OSINT-Bambenek} \cite{osint} & {Contains a list of DGA to generate malicious domains} & [S66] \\ 
     \hline
    Malware/ RAT & Contagio malware database \cite{RN122} & Contains an achieve of malware sample dumps. & [S49] \\ 
     \hline
    DNS & {Backdoor.Win32.Denis \cite{Denisand}}& {Malware that use DNS tunnelling to exfiltration data. }& [S80] \\
\cline{2-4}    
     Tunnelling & {FrameworkPos \cite{NewFrame2020}} &{ A malware that targeted the American retailer Home Depot and} & [S80] \\
     && {stole the credit card information.}&\\
     \hline
\end{tabular}}
\end{small}
\end{table*}
\begin{table*}[hbt!]
\begin{small}
\caption{Simulation Tools for Simulated Datasets}
\label{SimulationTools}
\centering
\resizebox{\textwidth}{!}{\begin{tabular}{|c|l|c|}
\hline
    \textbf{Attack Vector} & \textbf{Simulation Tools} & \textbf{Study}\\
    \hline
    SQL Injection & Amnesia testbed dataset \cite{RN123} , SQLMAP \cite{RN124} & [S59, S62, S70] \\ \hline
    Malware/RAT & ESET NOD32 \cite{RN125}, Kingsoft \cite{RN126}, Anubis \cite{RN127}, VirusTotal \cite{RN97}, & [S1, S41] \\ \hline
    APT & {Sysmon Tool \cite{Sysmonwindow}, Winlogbeat \cite{Winlogbeat}} & [S79] \\ \hline
    Overt Channels & ZeuS Tracker \cite{RN110}, Waledac \cite{RN111}, Storm \cite{RN109} & [S22, S39] \\ \hline
    Side Channel & PAPI \cite{RN129} & [S8, S63] \\ \hline
    Steganography & F5 \cite{RN113}, Model Based Steganography \cite{RN114}, Outguess \cite{RN115}, YASS \cite{RN130} & [S3, S9, S20] \\ 
    \hline
    Data  & dns2tcp \cite{RN131},BRO \cite{RN132},Iodine \cite{RN133}, dnscat \cite{RN134} and Ozymandns \cite{RN135},    & [S4, S14, S15, S21,  \\ 
    Tunnelling & CobaltStrike \cite{RN136}, {ReverseDNShell \cite{ReverseShell}} & S29, S32, S67, S68,S80] \\ 
      
    \hline
\end{tabular}}
\end{small}
\end{table*}    
\subsubsection{Real datasets}
Several studies (49\%) use real datasets, which contain real traces of actual attack data shared through various public repositories. 
A study [S51] obtained malicious XSS-based webpages dataset from XSSed \cite{RN95} and legal web pages dataset from ClueWeb09  \cite{RN96}.
Another study [S10] used a query dataset from the actual web application to detect insider threat. 
It is evident from Fig~\ref{figure12}a that all the direct inspection studies used real datasets.
However, the majority (11/15) of them were not publicly available, and only ten studies used publicly available real datasets.
Real datasets are also prominent in context inspection (i.e., 13/17) and event-based (i.e., 8/16) studies, but most of them are not publicly available.
However, none of the resource usage-based studies used real dataset. Fig~\ref{figure12}b depicts the relationship of datasets with data exfiltration attacks,
All the studies for detecting phishing, malicious domain and XSS attack use real datasets. Additionally, four studies detect RAT attacks based on real datasets. 
Table~\ref{publicDatasets} provides a list of publicly available datasets used by the studies along with a brief description.
Other studies used real but private datasets. We group them into three types:
(i) Collected privately: Some studies, e.g., [S10, S11, S16, S25, S35, S42], collected real data, e.g., by monitoring network traffic. However, these datasets are not publicly available for research reproducibility. 
For example, a study [S16] used dataset obtained from Microsoft Telemetry reports and Virus Total \cite{RN97}.
(ii) A random subset from threat intelligence feeds: 24 studies selected a random subset of data from publicly available threat feeds; however, the exact instances selected are not publicly available. 
For example, some studies [S5, S37, S46, S47, S53, S58, S40] used different and random instances from Phish-Tank \cite{RN98}.
Table~\ref{Threatintelligence} provides the list of public attack threat intelligence feeds used by the studies. 
(iii) Hybrid dataset: Five studies [S13, S19, S30, S48, S61] combined data from the public corpus or other two types of private data to formulate their datasets. 
For example, A study [S13] used 10,400 deceitful files from VX-Heaven \cite{RN99} public corpus consisting of 2600 instances of each worm, trojan, backdoor and virus and 1100 normal files samples were privately collected from Windows 7 system and Ninite.com.
\subsubsection{Simulated datasets}
32\% of the studies used simulated datasets. 
These datasets are mostly used by Flow-based approaches (11 studies) and distribution inspection studies (7) mainly to detect data tunnelling (10 studies) and overt channel (6) attacks.
All of these datasets are not publicly available.
For example, [S39] used dataset collected for almost 83 days from five legitimate peer to peer communication applications ( Skype, eMule, eDonkey, FrostWire, µTorrent, BitTorrent, and Vuze) and three malicious software tools (Storm \cite{RN109}, Zeus \cite{RN110} and Waledac \cite{RN110}. 
Table~\ref{SimulationTools} enlists the list of simulation tools used by the studies to collect malicious data.
\subsubsection{Simulated and Real Dataset}
{One recent study [S80] used both simulated tools (Iodine, ReverseDNShell) and two real-world malware: FrameworkPOS \cite{NewFrame2020} and Backdoor.Win32 Denis \cite{Denisand} to test their proposed anomaly detector (Isolation Forest Model). It is an interesting study illustrating the use of malware to exfiltrate data. However, this dataset is also not publicly available.}
\subsubsection{Synthetic Dataset}
17\% of the studies used synthetic datasets.
Data is usually synthesized by using mathematical models or injecting malicious patterns in legitimate data to synthesis an attack. 
A study [S24] obtained legitimate network flow dataset from “Comprehensive, Multi-Source Cyber-Security” \cite{RN137} and incorporates lateral movement attack traces “LANL NetFlow” logs and “Susceptible Infected Susceptible” (SIS) virus spread model \cite{RN138}. 
Seven event-based studies use this dataset type, especially to detect attacks, e.g., insider threat (8), APT (2), Lateral Movement (1) and SQL injection (3). We can infer two key points:
(i) multi-stage DE scenarios, especially, APT, is too complex to be simulated by using single or multiple software tools and require additional domain expertise.
(ii) these attack vectors are subjective, i.e., they are driven by additional sophisticated factors such as persistence in case of APT \cite{RN80}, the role of an employee in case of insider threat \cite{RN20}, weak passwords in case of lateral movement \cite{RN141}. 
Furthermore, seven studies used publicly available datasets.
Among these six studies used CERT a publicly available synthetic datasets for detecting insider threat \cite{RN107}, while one study used DARPA 1998 \cite{RN108} for detecting insider threat.
\par
\begin{small}
\noindent\fbox{%
\parbox{\columnwidth}{%
 \textbf{\textit{Summary}}: 82\% studies used privately collected, while only 17 studies used twelve publicly available datasets. Among 12 publicly available datasets, eight are one decade old, and two are synthetically created.}} 
 \end{small}

\subsection{{RQ4}: ML modelling phase}
\label{RQ4}
This section reports the results of the modelling phase analysis based on D17-D20 and D22-D23 data items in Table~\ref{dataextractionform}. 
The following subsections present the answer to this question in the light of the modelling technique and learning type and the classifier.
\subsubsection{Modelling Technique and learning type}
\begin{figure*}
 \includegraphics[width=.99\textwidth]{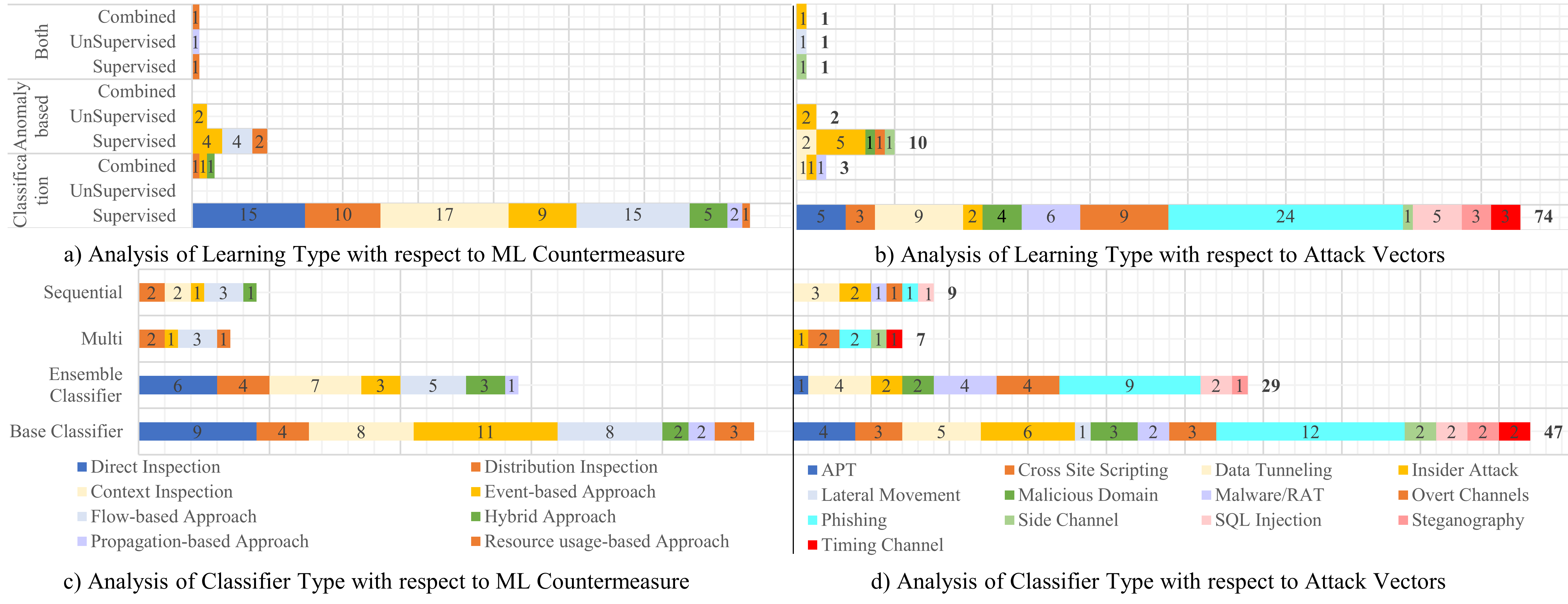}
 \caption{Analysis of ML Modelling Phase (The number shows the total studies in each category, while the bold number shows total studies in terms of y-axis)}
\label{figure13}
\end{figure*}

Fig~\ref{figure13}a and Fig~\ref{figure13}b analyse ML countermeasures and attack vectors with the type of modelling technique and their learning type used by the reviewed studies.
Whilst most of the studies (74/92) utilised supervised, only three studies used both (supervised and unsupervised) learning types and classification as a modelling technique to train the models.
Twelve studies employed anomaly-based detection; among these ten studies utilise supervised learning mechanism and trained an anomaly detectors using one class either legitimate or attack.
 {However, two studies [S72, S91] employed unsupervised anomaly detection to detect insider threat}.
{For instance, in [S72], the authors periodically trained an anomaly detector by using the psychometric score of users and their actions over time to detect insider threat.}
Anomaly-based detection is mostly applied by event-based and flow-based approaches.
One reason behind it is the malicious behavior of an insider is dependent on multi-factors, such as their job role, psychometric factors which are difficult to define to perform classification.
Three studies [S10, S24, S63] used both classification and anomaly-based methods to detect data exfiltration attacks.
This combination is applied to complex attack scenarios like lateral movement, insider threat and side-channel attacks. 
For instance, one study [S24] applied unsupervised learning to detect lateral movement by using a propagation-based approach. In this study, two classifiers were trained to detect lateral movement. First uses k-mean \cite{RN52} along with PCA to classify a host that behave like an infected host; the second uses extreme-value analysis to identify the host infected by the malicious lateral movement.
Interestingly, four of the reviewed studies [S1, S4, S10, S18] used a combination of both learning types. All these studies divided a complex problem into sub-problems and solved them sequentially by using suitable learning types. 
For instance, in [S1], the authors performed a two-tier classification to detect malware. Firstly, seven supervised classifiers (SVM, DT, RFT, NB, GB, LR and KNN) were trained to detect known vs unknown malware family using a weighted voting scheme. Subsequently, for unknown malware detection, Shared Nearest Neighbour (SNN) clustering \cite{RN143} was used to cluster unknown malware.
\begin{table*}[htb!]
\begin{small}
\caption{Mapping of ML approaches with classifiers}
\label{classifiers}
\centering
\resizebox{\textwidth}{!}{
\begin{tabular}{|c|l|c|c|c|c|c|c|c|c|l|} 
\hline
\textbf{Type} & \textbf{Classifiers}& \rotatebox{90}{\textbf{Direct}}\rotatebox{90}{\textbf{Inspection}} &
\rotatebox{90}{\textbf{Distribution}}\rotatebox{90}{\textbf{Inspection}} 
& \rotatebox{90}{\textbf{Context}}\rotatebox{90}{\textbf{Inspection}} & 
\rotatebox{90}{\textbf{Event-based}}\rotatebox{90}{\textbf{Approach}} & 
\rotatebox{90}{\textbf{Flow-based}}\rotatebox{90}{\textbf{Approach}}&
\rotatebox{90}{\textbf{Hybrid}}\rotatebox{90}{\textbf{Approach}}&
\rotatebox{90}{\textbf{Propagation}}\rotatebox{90}{\textbf{based Approach}}&
\rotatebox{90}{\textbf{Resource usage}}\rotatebox{90}{\textbf{-based Approach}}&
\multicolumn{1}{c|}{\textbf{Study}}
 \\ 
\hline 
\multirow{15}{*}{\rotatebox{90}{Traditional ML}}& 
SVM (20) & 3 &5 & 3 &2 &6 & &1 &  & \multicolumn{1}{m{8cm}|}{[S1, S2, S3, S8, S14, S17, S21, S23, S29, S31, S32, S37, S44, S55, S59, S60, S62, S65, S77, S86]} \\ 
  \cline{2-11}
     & DT (12) & 1 &  & 3 & 2 & 5 & 1 &  &  &\multicolumn{1}{m{8cm}|}{ [S1, S10, S17, S27, S30, S39, S40, S49, S51, S53, S65, S92]} \\ \cline{2-11}
     & KNN (8) & 1 &  & 0 & 2 & 5 &  &  &  & [S1, S32, S34, S39, S48, S50, S83] \\ 
     \cline{2-11}
     & LR (5) & 1 &  & 2 & 1 & 1 &  &  &  & [S1, S8, S17, S69, S82, S66] \\ 
     \cline{2-11}
     & NB (3) & 0 &  & 1 & 1 & 1 &  &  &  & [S1, S36, S41] \\ 
     \cline{2-11}
     & Fractal & 0 &  &  &  & 1 &  &  &  & [S38] \\
     & dimension (1) &  &  &  &  &  &  &  &  &  \\ 
     \cline{2-11}
     & MCAC (1) & 1 &  &  &  &  &  &  &  & [S5] \\
     \cline{2-11}
     &RFT (18) & 5&  & 2 &2 &7 &  &  & 2& \multicolumn{1}{m{8cm}|}{[S1, S13, S15, S19, S26, S28, S34, S39, S42, S43, S44, S45, S46, S47, S57, S64, S65, S92]} \\ 
     \cline{2-11}
     & Ada-boost (5) & 1 & 2 & 1 &  & 1 &  &  &  & [S22, S39, S54, S58, S70] \\ 
     \cline{2-11}
     & GB (6) &  &  & 1 & 2 & 2 &  &  & 1 & [S1, S65, S66, S70, S79, S81] \\ 
    \cline{2-11}
     & Decorate \cite{RN141} (1) &  &  &  & 1 &  &  &  &  & [S35] \\ 
     \cline{2-11}
     & \multicolumn{1}{m{4cm}|}{Adaptive Deep Forest (1)}  & 1 &  &  &  &  &  &  &  & [S70] \\
     \cline{2-11}
     & Rotation Forest (1) &  &  &  &  &  &  & 1 &  & [S16] \\
     \hline
     \multirow{6}{*}{\rotatebox{90}{Deep Learning}} & CNN (7) &  &  & 5 & 1 &  & 1 &  &  & [S67, S68, S75, S78, S85, S89, S90] \\ 
	 \cline{2-11}
     & LSTM/RNN (6) &  &  & 3 & 3 &  &  &  &  & [S66, S68, S73, S74, S75, S90] \\ 
	 \cline{2-11}
     & NN/MLP (7)& 1 &2 &  &2 &1 & 1 &  &  & [S20, S25, S61, S62, S63, S84, S92] \\ 
     \cline{2-11}
     & GRU (2) &  &  & 2 &  &  &  &  &  & [S68, S76] \\ 
	 \cline{2-11}
      & RCNN (1) &  &  & 1 &  &  &  &  &  & [S87] \\
     \cline{2-11}
     &\multicolumn{1}{m{4cm}|}{Deep Belief Network (1) } &  &  & 1 &  &  &  &  &  & [S89] \\ 
     \hline
    \multirow{13}{*}{\rotatebox{90}{Anomaly-based}} & OCSVM (5) &  &  &  & 2 & 2 &  &  & 1 & [S6, S33, S56, S72] \\ 
	\cline{2-11}
     & GMM (3) &  &  &  & 1 & 1 &  &  & 1 & [S12, S39, S52] \\
	 \cline{2-11}
     & Autoencoders (1) &  &  &  & 1 &  &  &  &  &[S91]  \\ 
	 \cline{2-11}
     & \multicolumn{1}{m{4cm}|}{Global Abnormal Forest (1)}&  &  &  &  & 1 &  &  &  & [S11] \\ 
      \cline{2-11}
     & HMM (1) &  &  &  & 1 &  &  &  &  & [S7] \\ 
	 \cline{2-11}
     & LDA (2) &  &  &  &  & 2 &  &  &  & [S8, S32] \\ 
	 \cline{2-11}
     & Kmean (4) &  &  & 1 &  & 1 &  & 1 & 1 & [S4, S18, S24, S39] \\ 
	 \cline{2-11}
     & \multicolumn{1}{m{4cm}|}{Extreme Value analysis (1)}&  &  &  &  &  &  & 1 &  & [S24] \\ 
       \cline{2-11}
     & Fisher Linear Classifier \cite{RN142} (1) &  &  & 1 &  &  &  &  &  & [S9] \\  
	 \cline{2-11}
     & Isolation Forest (3)  &  &  &  & 1 & 2 &  &  &  & [S72, S80, S83] \\
     \cline{2-11}
     & \multicolumn{1}{m{4cm}|}{Kernal Density  Estimation (2)}  &  &  &  &  & 2 &  &  &  & [S39, S83] \\ 
     \cline{2-11}
     & \multicolumn{1}{m{4cm}|}{Clustering/ Attrib Deviation (2)} &  &  & 2 &  &  &  &  &  & [S10, S17] \\ 
    \cline{2-11}
     & \multicolumn{1}{m{4cm}|}{Maximum  Entropy (1)} &  & 1 &  &  &  &  &  &  & [S77] \\ 
     \hline
\end{tabular}
}
\end{small}
\end{table*}
\subsubsection{Classifier}
This section describes the classifier selected by the reviewed primary studies.
We represent our analysis in terms of classifier structure, classifier type and classifier used by the studies.
We have used acronyms of classifiers instead of their full name (please refer to section~\ref{MLLC} for full names). 
Table~\ref{classifiers} shows the classifiers used by the reviewed studies, along with their mapping with the type of ML countermeasures. 
These classifiers are employed by multiple studies either solely or with other classifiers.
Furthermore, we have only mentioned those classifiers that were selected as best classifiers for a study because most of the studies initially tried out multiple classifiers.
Considering the reflections from learning type, we classified the classifier structure into four types: base, ensemble, sequential and multi classifiers as shown in Fig~\ref{figure13}c and Fig~\ref{figure13}d.
If a study fails to conclude the best classifier, we classify this study under multi classifier category.
Table~\ref{classifierstructures} provides a brief definition of these classifier structures and provides their strengths and weaknesses. 
\begin{table*}[hbt!]
\begin{small}
\caption{Classifier Structures, their description with strengths and weaknesses}
\label{classifierstructures}
\centering
\resizebox{\textwidth}{!}{\begin{tabular}{|c|l|l|l|}
\hline
    \textbf{Structure} & \textbf{Description} &\textbf{Strengths} & \textbf{Weaknesses} \\ \hline
    
    \multicolumn{1}{|m{3cm}|}{Base Classifiers}& 
    \multicolumn{1}{m{5cm}|}{The studies use a single learning algorithm to train their models.}
    & \multicolumn{1}{m{5cm}|}{\tabitem Easy to implement.} 
    & \multicolumn{1}{m{5cm}|}{\tabitem May suffer over-fitting.}
    \\ 
    && \multicolumn{1}{m{5cm}|}{\tabitem  Can handle large number of features and datasets}
    & \multicolumn{1}{m{5cm}|}{}
    \\
   \hline
    \multicolumn{1}{|m{3cm}|}{Ensemble Classifiers}
    &\multicolumn{1}{m{5cm}|}{The studies combine the decisions from multiple models using techniques such as bagging or boosting.}
    & \multicolumn{1}{m{5cm}|}{\tabitem Can reduce over-fitting.} 
    & \multicolumn{1}{m{5cm}|}{\tabitem Slow on large datasets and high dimensional feature vectors.}
    \\
    & \multicolumn{1}{m{5cm}|}{}
    & \multicolumn{1}{m{5cm}|}{\tabitem May result in better performance.} 
    & \multicolumn{1}{m{5cm}|}{\tabitem Computationally expensive.}
    \\
    & \multicolumn{1}{m{5cm}|}{}
    & \multicolumn{1}{m{5cm}|}{\tabitem More stable and reliable models.} 
    & \multicolumn{1}{m{5cm}|}{}
    \\ 
    \hline
    \multicolumn{1}{|m{3cm}|}{Sequential Classifiers}
    &\multicolumn{1}{m{5cm}|}{The studies using this classifier structure divides a complex problem into sub-problems solved them sequentially by using suitable learning algorithm.}
    &\multicolumn{1}{m{5cm}|}{\tabitem Can detect complex attacks such as APT and insider threat.}
    & \multicolumn{1}{m{5cm}|}{\tabitem Slower than ensemble and base classifiers.} 
    \\ 
    &\multicolumn{1}{m{5cm}|}{}
    &\multicolumn{1}{m{5cm}|}{}
    & \multicolumn{1}{m{5cm}|}{\tabitem Hard to implement and design.}\\
    \hline
    \multicolumn{1}{|m{3cm}|}{Multi Classifiers}
    &\multicolumn{1}{m{5cm}|}{The studies use multiple classifiers to perform the same task.} 
    & \multicolumn{1}{m{5cm}|}{\tabitem Can help to compare the performance of different classifiers on the same task.}
    & \multicolumn{1}{m{5cm}|}{}\\ 
 \hline
\end{tabular}}
\end{small}
\end{table*}
\paragraph{Base Classifier}
These are used by most of the studies (47). SVM is the most frequently (i.e., 20 studies) used base classifier. 
It is mostly used by \emph{distribution} inspection and \emph{flow-based} approaches.
{\emph{SVM} is a powerful \emph{traditional ML} classifier that can handle linear, non-linear, high dimensional data \cite{RN144}}.
\emph{DT} is the second recurrent base classifier used by 12 studies.
\emph{KNN}, \emph{LR} and \emph{NB} are also used by eight, five and three studies respectively. 
{\emph{Deep learning} classifiers have recently gained popularity to detect exfiltration attack vectors. 
Among these most popular are \emph{CNN} (7), \emph{NN/MLP} (7) and \emph{LSTM} (6).
These classifiers are mostly utilized by \emph{context} inspection (12) and \emph{event-based} (6) studies.}
OCSVM \cite{RN44} an \emph{anomaly-based} base classifier used by five studies. 
All these studies are based on time-dependent data, which suggests that OCSVM is a suitable choice for time series data modelling. For example, in [S56], a novel approach to detect abnormal resource usage in cloud-based infrastructures is presented.
The resource usage (CPU, memory used) behaviour was continuously monitored and was represented as a time series. 
Features were extracted from each time series, and then separate models were trained on OCSVM with several examples of legitimate time-series data for each resource (CPU, memory) for a fixed time window (e.g., a day or a week).  OCSVM, Regression using Long Short-Term Memory (LSTM) and simple statistical models were used for experimentation.
However, the author concludes the OCSVM outperformed when 5-fold cross-validation was used. Other two anomaly-based classifiers that handle time-series or sequential data well are GMM and HMM \cite{RN145}. 
OCSVM, GMM and HMM are used by four event-based approaches to detect insider threat.
\par
Inquisitively, three studies based on flow-based [S11, S38] and direct inspection [S5] based countermeasures proposed novel classifiers.
For instance, Siddiqui et al., [S38] proposed a fractal-based ML algorithm to detect APT attacks using TCP based network connection features and compared it with standard techniques. The correlation fractal is a reference dataset of the features. The algorithm first computes the correlation fractal for an APT attack and legitimate data separately to form a model for each class. For testing the input sample, the correlation fractal of that sample is compared to the previously computed attack and legitimate model. 
\paragraph{Ensemble}
29 studies constructed ensemble classifiers to detect exfiltration.
For example, A study [S32] constructed an ensemble classifier for detecting DNS tunnelling using statistical features.
Two layers of classifiers are used. The first layer is composed of three different types of supervised classifiers LDA, SVM and K-NN neighbours. The second layer of classifier takes k number of first layer classifiers and based on majority voting produces the final output. 
RFT is the most frequently (i.e., 18 studies) used ensemble classifier.
{\emph{It performs well for all types of data: real value, binary, categorical. Moreover, it reduces over-fitting when using a single DT, scales well for high dimensional data and runs DT training in parallel to speed up the training process \cite{RN55}}}. RFT is mostly used in \emph{flow-based} (7) and direct inspection (5) countermeasures.
{Gradient Boosting (GB)} and Ada-boost are also used by 6 and 5 studies, respectively.
{\emph{Deep learning} classifiers are combined in four studies [S68, S87, S88, S90] to create a custom ensemble classifier to detect exfiltration.
For instance, in [S75] CNN, Bidirectional LSTM and Independent RNN are ensembled together to detect insider threat.}
\paragraph{Sequential}
Nine studies used a sequential combination of different classifiers to accomplish the required tasks,
e.g., a study [S62] used a framework called idMAS-SQL (Intrusion Detection based on Multi-Agent System) to detect and stop SQL injection attacks using ML is presented.
For classification, a combination of NN (MLP) followed by SVM (id-CBR) was used as a classifier because it outperformed the others. Three of these studies used a combination of multiple ensembles and base classifier to detect DE.
An example of such a study is [S39]. This study divided the problem to be solved (attack or not) into small sub-problems and then applied both base and ensemble classifiers at a different level. Firstly, different base classifiers (KNN, Kmean, and GMM) were trained.
Each classifier returns a score of confidence value, 
if the score is less than the threshold for all classifiers than P2P application is unknown, if the score is greater than the threshold for different classifiers than a multi-label classifier (Ada-boost with RFT having 50 Decision Trees) was employed to detect the label of the instance.
\paragraph{Multi Classifiers}
Seven studies use multiple classifiers but failed to conclude which one is better than others. The reason behind it was that either all of the tested classifiers reported a similar performance or they performed differently based on the type of application.
For example, [S44] suggests that for large datasets RFT model is faster than SVM, but SVM is faster than RFT for small datasets.  
\subsubsection{Relation of Classifier with the Number of Features and Dataset Size}
\begin{figure*}
\centering
\includegraphics[width=\linewidth]{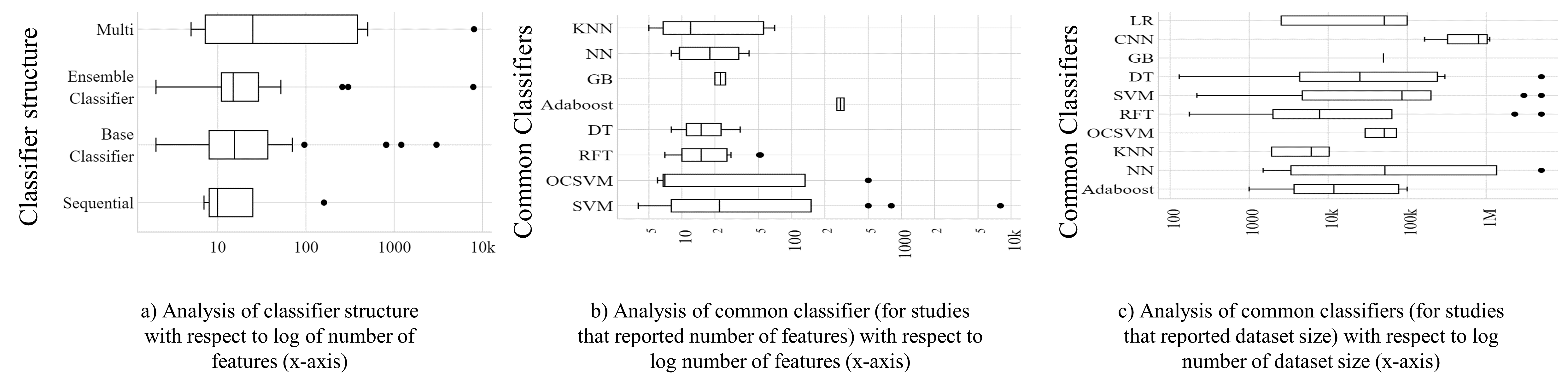}
\caption{Relationship of Classifier with Number of Features}
\label{figure14}
\end{figure*}
In Fig~\ref{figure14}, we present the results from the analysis performed to understand the relationship of the classifier with log number of features (linear range 2 to 8000) and dataset size (linear range 130 to  5000000). The log is used because of the high variance in distribution that hinders visualisation. 
From Fig~\ref{figure14}a and Fig~\ref{figure14}b, we observed that median of all the classifier structures and common classifiers except Adaboost and OCSVM lie in the log range between 10 to 100 which shows that the number of features chosen to train the model is independent of the classifier structure and classifier.
However, the studies using Adaboost classifier used a comparatively large number of features (median $log >100$) while OCSVM used less number of features (median $log<10$). Furthermore, Fig~\ref{figure14}c investigates the relationship of commonly used classifiers by the studies
and log of dataset size. 
All the classifiers except KNN, RFT are used with high dimensional datasets (median $log >10000$) to detect data exfiltration.
\par
\begin{small}
\noindent\fbox{%
\parbox{\columnwidth}{%
 \textbf{\textit{Summary}}: Base and Ensemble classifier structure is mostly (i.e., 47 and 29 studies) used by ML countermeasures to detect exfiltration. RFT and KNN are not chosen by the studies using a high dimensional dataset (median $log >10000$) due to slow training and testing process respectively.}} 
 \end{small}
\subsection{{RQ5}: Validation and evaluation of ML-based Data Exfiltration Countermeasures}
\label{RQ5}
For practical applicability of ML solutions for data exfiltration detection, the performance of the model is critical. 
This section provides results from multi-tier analysis performed on validation techniques and the performance metrics used by the primary studies. Our analysis is based on D22 to D24 data items in Table~\ref{dataextractionform}.
\subsubsection{Validation Methods}
\label{RQ5.1}
\begin{figure*}
\includegraphics[width=.99\textwidth]{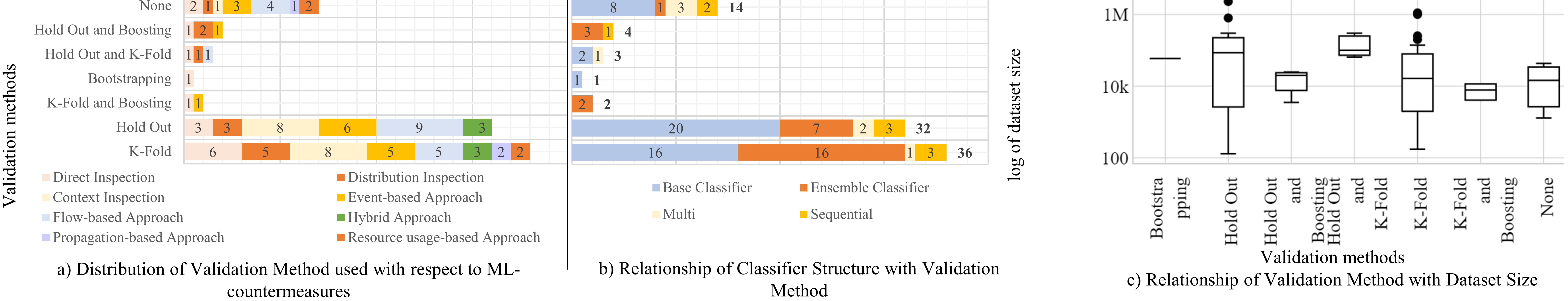}
\caption {Analysis of Validation Method}
\label{figure16}
\end{figure*}

Hold Out, K-Fold, Leave Out one and Bootstrapping are the most popular (please refer to \cite{RN35, RN36} for details) methods applied by ML approaches to validate the classifier performance. 
Fig~\ref{figure16} shows an analysis the validation methods used by the reviewed studies based on ML countermeasures, classifier structure and dataset size.
Fig~\ref{figure16}a illustrates the distribution of ML-based countermeasures with respect to validation techniques. The K-Fold method is dominant (i.e., 36/92 studies) in all the approaches.
K-Fold cross-validation is less bias estimation of accuracy \cite{RN43}, and the entire dataset is used for training and testing. 
Hold Out method is the second most common (i.e., 32/92 studies) validation methods. 
However, it is not used in the case of propagation and resource usage-based approaches. 
Three studies [S40, S51, S77] used both Hold Out and K-Fold methods; six studies [S1, S9, S13, S15, S46, S47] used Hold Out or K-Fold with boosting ensemble. 
One study [S28] used bootstrapping as a validation technique. 
Fig~\ref{figure16}b shows a relationship between the validation method with the type of classifier. It is evident that base classifiers mostly use the holdout method (20 studies);
while K-Fold and bootstrapping is a popular among studies using ensemble and more specifically for DT classifiers. 
Whereas 14 studies did not report any validation method.
We have also investigated the validation technique used in terms of training dataset size, as shown in Fig~\ref{figure16}c.
Hold Out and combination of Holdout and K-Fold method is used with large training datasets (log median approximately 100k),
whereas K-Fold is frequently used for medium datasets (log median approximately 10k). 
From the above analysis, we conclude that K-Fold and Hold Out standout among other approaches. 
\emph{However, they both have a certain \textbf{limitation}. Hold Out method can be subject to overfitting and is highly dependent on data class distribution as well as the split ratio.
Whereas the K-Fold method reduces overfitting as ‘k’ is increased, but due to its iterative nature, it requires high computational time \cite{RN43, RN68}. 
Hence, the choice of validation method is dependent on constraints such as class distribution, dataset size and split ratio.}
\subsubsection{Performance Measures}
\label{RQ5.2}
\begin{figure*}
\includegraphics[width=.99\textwidth,keepaspectratio]{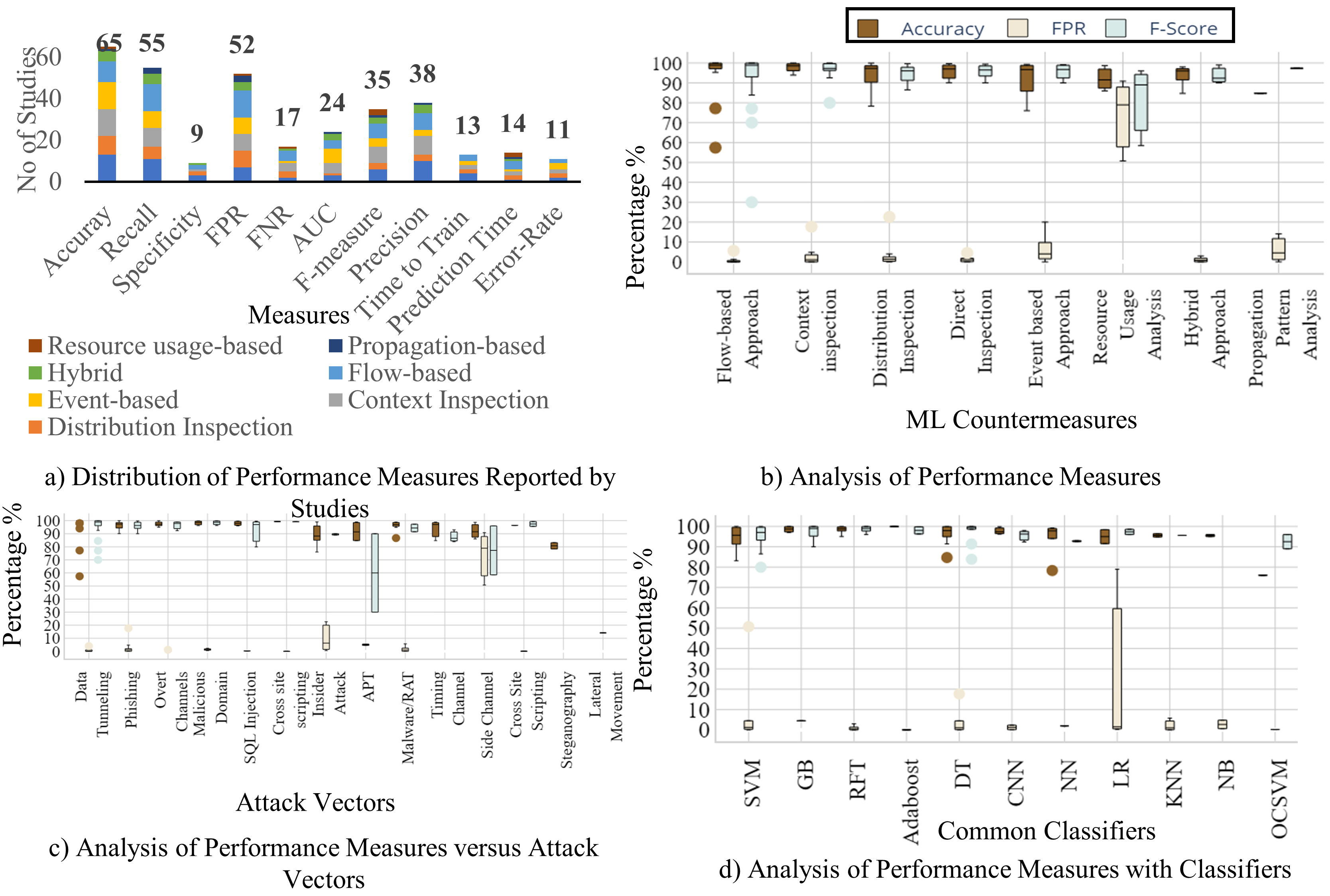}
\caption{Analysis of Common Performance Measures with reviewed studies [False Positive Rate (FPR), True Negative Rate (TNR), Area Under the Curve (AUC) and False Negative Rate (FNR)]}
\label{figure15}
\end{figure*}

In this section, we analyze the performance achieved by the models trained to detect data exfiltration attack vectors in terms of ML countermeasures, attack vectors and classifiers, as shown in Fig~\ref{figure15}. 
We have used accuracy, FPR and F-score to evaluate the performance of the reviewed studies because they collectively give a precise estimation of the performance of the models \cite{RN36}. 
Fig~\ref{figure15}b-d shows a boxplot depicting the performance reported by the reviewed studies. The x-axis with no boxplot indicates that no study reported the performance measure for that label.
\paragraph{Frequently Reported Performance Measures}
Fig~\ref{figure15}a illustrates the distribution of performance measures reported with ML countermeasures.
Accuracy (70\%), Recall (60\%) and False Positive Rate (FPR) (56\%) are the most report metric used by studies
41\% of studies used precision, while 38\% reported compound metric F-Score.
Distinctly, a few studies also reported time-to-train (14\%) and prediction-time (15\%).
While eight studies [S1, S17,
S20,
S30,
S37,
S75,
S76,
S89] 
only reported accuracy and two studies [S52, S56] used only F-Score. One study [S74] just reported AUC, and another study [S41] only evaluated based on FPR. Five studies that only reported accuracy either used holdout or no validation method. 
One study [S1] validated the results on a small dataset (2000 instances). 
Three studies [S10, S23, S41] reported both FPR and accuracy as performance measures. 
However, we argue that accuracy and FPR alone are not sufficient to evaluate a model’s performance as they are highly sensitive to distribution in training and testing datasets \cite{RN146}. 
Two interesting studies are [S80, S83], these studies fixed the FPR to 2\% and trained Isolated Forest Tree anomaly detector until the FPR is 2\%.
\paragraph{Performance in terms of ML Countermeasure}
Fig~\ref{figure15}b shows the analysis of performance with the ML countermeasures.
Such analysis aims to study what type of data analysis yield better results. 
\emph{Flow-based} approaches performed the best with a median accuracy and F-score of 99.5\% and 99.47\%.
Subsequently, \emph{Direct} inspection, \emph{Context} inspection, \emph{Distribution} inspection and \emph{Hybrid} approach achieved a median accuracy of 98.42\%, 98.61\% , 97.30\% and 96\% respectively. However, the median FPR of \emph{Direct} and \emph{Distribution} inspection based studies is 4.46\% and 1.33\% whereas \emph{Hybrid} approach and \emph{Context} inspection attained the lowest median FPR of 0.34\% and 0.72\% respectively. 
These results indicate that these approaches yield better results than other approaches.
On the other hand, \emph{Resource usage-based approach} attained the highest median FPR, i.e., 78.86\% and lowest F-score, i.e., 89\%.
\emph{Propagation} and \emph{Event} based approaches also have comparatively high median FPR i.e., $>6\%$.
\paragraph{Performance Analysis Relation to Attack Vector}
The motivation behind this analysis is to investigate how robustly ML methods can detect data exfiltration.
Fig~\ref{figure15}b shows that ML methods are quite effective in detecting \emph{Data tunnelling}, \emph{Phishing} and \emph{Overt channel} attacks achieving a median accuracy and F-score $>96\%$ and median FPR of less than 1.5\%.
Studies detecting \emph{SQL injection} and \emph{Malware/RAT} attacks also achieved high median accuracy $98\%$ and $97\%$ respectively. However, they attained a comparatively low median F-score, i.e., $88\%$ and $86\%$.
It suggests that the results of these studies can detect one class (normal or attack) better than the other.
On the other hand, studies detecting complex attacks like \emph{Insider} and \emph{APT} attacks attained a median accuracy and FPR of $<90\%$ and $6\%$ suggesting that \emph{there is a room of improving ML methods to detect these attacks}.
Albeit there is only one study [S24] on detecting lateral movement, its recall of 88.7\% and FPR of 14.1\% indicate that ML-based countermeasure may be able to detect lateral movement.
Lastly, with the median FPR 78.965\%, the side-channel attacks are difficult to detect using ML techniques. 
\paragraph{Performance in terms of Classifier}
Fig~\ref{figure15}d shows the performance of common classifiers used in data exfiltration countermeasures.
\emph{Ensemble} classifiers RFT, GB and Adaboost performed extremely well, achieving median F-score and accuracy of more than 98\% and FPR of less than 0.3\%.
After them, \emph{DT} and {\emph{Deep learning} classifier CNN also performed well with a median accuracy and F-score of 97\%. NN also attained good accuracy, but the median F-score is less than 93\%}. Among base classifiers, \emph{SVM} performed better than its one class variant and attained a median accuracy of 95.67\% and F-score of 97\%. OCSVM performed the worst with a median F-Score of 92.5\%.
From the above results, \emph{we infer that \emph{Ensemble} classifiers are more suitable for detecting data exfiltration attacks}.
\par
\begin{small}
\noindent\fbox{%
\parbox{\columnwidth}{%
 \textbf{\textit{Summary}}: All \emph{behavior-driven} except \emph{flow-based approach} yield high ($>6\%$) FPR. Ensemble and Deep learning classifiers are more suitable for detecting data exfiltration attacks with a high median accuracy of $>97\%$.}}
 \end{small}
\section{Discussion}
\label{section-discussion}
In this section, we reflect on the findings presented in the previous section to recommend best practices and highlight areas for future research.
\subsection{Need for Hybrid ML Approaches}
{We found that data-driven and behaviour-driven (Section~\ref{RQ1}) approaches are used to detect data exfiltration attacks.
However, both have some limitations (Table~\ref{taxonomy}).
Although data-driven approaches are capable of detecting single-stage attacks with high accuracy (median $>95\%$), they are unable to detect sophisticated attacks such as insider threat and APT.
On the other hand,  behaviour-based approaches can detect sophisticated attack vectors. However, they result in high false positives (median $>6\%$). 
We assert that the amalgamation of these approaches can be beneficial in reducing their scope and performance limitations. Our assertion is inspired by the success of hybrid approaches in other domains (e.g., recommended systems \cite{RN147}) where it has been demonstrated that hybrid approaches perform better than standalone data-  or behaviour-driven approaches.
The hybrid approach can uncover both global and local patterns in security event data, hence, improving data comprehension and interpretability. 
{For example, for insider threat detection, the \emph{event-based} approach can be combined with \emph{context} inspection by not only monitoring the event such as an email being sent by an insider but also examining the content of the email using \emph{context inspection}. 
Such an analysis can be utilized to detect complex attack scenarios like lateral movement, privilege escalation, and APT.}
There are a few limitations as well; unification can increase the time for data analysis and intensify data dimensionality. Such limitations can be addressed by using automated feature engineering and feature selection techniques.} 
\par
\noindent\fbox{%
 \parbox{\columnwidth}{%
 \textbf{\textit{Summary}}: Combination of both data- and behaviour-based approaches should be explored to reduce the FPR and enable detection of complex data exfiltration attacks. }} 
\subsection{Need for Propagation based Data Exfiltration Countermeasures}
{Advance attackers follow multi-stage spatial attacks (e.g., APT and insider threat) to exfiltrate data \cite{RN141}.
The detection of these attacks requires the correlation of multiple stages such as delivery and exploitation \cite{RN22, RN141, RN148}.
The correlation of security event data from multiple stages improves the detection accuracy of the countermeasure. 
It is because security events in one stage are precursors to security events in the subsequent stage. However, our SLR reveals only four studies (i.e., [S2, S16, S24]) that propose correlation/propagation-based countermeasures for detecting data exfiltration. 
It is also worth noting that these studies achieved a comparatively higher recall (above 80\%) as compared to studies that are not based on attack propagation analysis (Section~\ref{RQ5.2}).
Therefore, we recommend that more studies be conducted to develop and investigate the effectiveness of multi-step ML-based countermeasures.}
\par
\noindent\fbox{%
\parbox{\columnwidth}{%
 \textbf{\textit{Summary}}: To detect complex data exfiltration attacks, multi-step ML-based data exfiltration countermeasures should be developed.}} 
\subsection{Need for High-Quality Evaluation Datasets}
One of the key limitations of cyber-security research is the scarcity of suitable datasets \cite{RN17, RN67}; this situation is even worst for research on data exfiltration.
Our review found only 12 publicly available datasets (Table~\ref{publicDatasets}) out of which eight datasets i.e., CLAIR fraud email \cite{RN100}, URL-Dataset \cite{RN101}, Spam URL Corpus \cite{RN103}, ClueWeb09 \cite{RN96}, VX-Heaven \cite{RN99} and DARPA \cite{RN108} are approximately one-decade old. 
Such datasets are unable to represent a novel, sophisticated and evolving attack vectors for data exfiltration.
As a result, there are constant private efforts to create datasets for data exfiltration detection research which often not made public for privacy concerns, intellectually property issues, competitive advantage, and time and money spent \cite{RN150}.
That is why it is not possible to replicate the studies conducted using privately created datasets. We recommend that datasets used for the reported studies be shared to support the validation as well as future research. 
It is equally essential that the datasets are continuously updated to incorporate new attack scenarios such as APT10, lateral movement, and privilege escalation. 
Several studies used a random subset of the same dataset for evaluation. 
However, none of the reviewed studies mentioned the time and the criteria used for selecting the subsets. For example, PhishTank \cite{RN98} and OpenPhish \cite{RN102} are two well-known public repositories that keep an updated blacklist of phishing URLs.
Most of the studies detecting phishing attacks utilized these repositories, e.g., [S46, S53] used these source to collect phishing URLs but none of them mentioned when data was captured (time) or how particular samples were selected over the others (criteria). 
Such information is essential for research validity and reproducibility. For instance, phishing URLs collected in 2016 may not be valid for research in 2019 as phishing websites are short-lived, and the datasets suffer from concept and temporal drift. 
Hence, we assert that the random selection of instances of datasets should be discouraged. The available datasets and their subsets should be versioned to handle issues like the concept and temporal drift. 
The quality of datasets is a significant concern as it plays an integral role in the reliability of ML model \cite{RN70}.
This issue is raised by multiple studies \cite{RN151, RN152, RN153, RN154}, especially, for well-known datasets like KDD \cite{RN152} and DARPA \cite{RN153}.
However, there is no standard benchmark to evaluate the dataset quality in data exfiltration domain. 
In this regard, we observed that only 2/10 identified public datasets (CERT \cite{RN107} and DARPA 1998 \cite{RN108}) have documentation that describes dataset and data collection mechanism.
We believe that proper documentation is essential to assess the quality of data collection and identify any generation faults.
Hence, we assert that it is essential to document various facets of the dataset while creating the dataset. 
{Such documentation can also assist in detecting and preventing against poisoning attacks \cite{RN162} and understanding the biases in these datasets.}
There is a need to develop a common benchmark such as \cite{RN155, RN156} to evaluate the quality of datasets in data exfiltration domain. 
\noindent\fbox{%
\parbox{\columnwidth}{%
 \textbf{\textit{Summary}}: With an evolving threat landscape, new evaluation datasets for data exfiltration countermeasures should be created and constantly updated, documented and assessed.}} 
\subsection{Need for Automated Feature Extraction and Selection Techniques}
{In section~\ref{RQ2}, we found that 11/92 and 14/92 used automated and semi-automated methods to extract features from data.
In contrast, most of the studies 67/92 rely on manual process.
Since cyber-attack space evolves rapidly, these manually extracted features either become obsolete due to temporal and concept drift (referred as feature drift) \cite{RN157, RN158} or are evaded by adversaries \cite{RN59}.
Nevertheless, manual feature engineering is tedious, specific and error-prone process as it requires manual inspection of a dataset by domain experts [S21].
We assert that for detecting data exfiltration attacks ( where data is generated from several heterogeneous sources such as multiple hosts, network and file systems.) automatically extracting features is more reliable and adaptable approach.
Therefore, we recommend that future research should focus on developing new techniques to assist automatic feature extraction process such as using CNN or LSTM to extract statistical or contextual features from data, respectively.
However, the automated feature engineering process has a limitation as it results in a sparse and high dimensional feature vector.
This limitation can be addressed by using feature selection methods. 
We advocate the use of feature selection to minimize the curse of dimensionality, reduce computational time, increase ML model interpretability, reduce overfitting, improve prediction accuracy and assist machine learning models to choose the most effective and discriminant decision boundary \cite{RN59}.
Nevertheless, most of the feature selection methods suffer from the limitations and require selection of manual thresholds \cite{RN39}.
This limitation can be addressed by using attention like mechanism to select features. The attention-based mechanism has been recently introduced \cite{vaswani2017attention} in Deep learning method that is capable of automatically selecting prominent features from data. In our reviewed studies, three studies have used this mechanism to train their deep learning models.
We believe the attention can be combined with automatic feature engineering process to extract more focused and relevant features.
}
 We assert that there is a need for developing and incorporating automatic feature engineering techniques in data exfiltration and generally cyber-security domain to build more practical, accurate, and real-time detection approaches.
 \par
 \noindent\fbox{%
 \parbox{\columnwidth}{%
 \textbf{\textit{Summary}}: There is a need for developing and employing automated feature engineering and feature selection methods for ML-based data exfiltration countermeasures.}} 
\subsection{Need for Customized Validation Methods}
{The reliability of ML methods performance is dependent upon the validation methods used.
We observe that traditional validation methods like K-Fold and Hold Out have been used by most of the approaches (Section~\ref{RQ5.1}).
The blind selection of K-Fold and Hold Out in data exfiltration domain should be discouraged. 
It is because, data used for training ML-based data exfiltration countermeasures, consists of different types of features such as temporal and spatial (Section~\ref{RQ2}). 
Whilst using K-Fold and Hold Out is a suitable choice with statistical and content-based features, 
it is not suitable for time series, spatial and hierarchical data because it results in an underestimation of predictive error \cite{RN160} and overestimation of performance.
Therefore, we assert that the validation method should be applied based on the nature of data and features.
Furthermore, we emphasize the need for new validation methods for validating models based on feature types such as spatial and temporal features.}
\par
\noindent\fbox{%
\parbox{\columnwidth}{%
 \textbf{\textit{Summary}}: In order to handle different types of data (e.g., spatial and temporal), it is important to explore new validation techniques to assess ML-based data exfiltration countermeasures.}} 
\subsection{Need of Incremental Learning for Model Updates}
{Whilst model up-gradation is frequently required for data exfiltration detection to handle concept drift issue \cite{RN59, RN157, RN158}
only three studies ([S6, S16,S40]) indicated this problem. 
While, no study except [S6, S72], mentioned the procedure to retrain a model.
We can conclude that most of these studies use batch-retraining method to update their models. However, the batch method \cite{RN153} has many drawbacks. 
Firstly, a new model is trained from scratch while the previous model is discarded, which is computationally wasteful and time-consuming \cite{RN153}.
Secondly, new examples cannot be directly used to train a model until a new batch is full \cite{RN154}.
Lastly, retraining a model from scratch can open new opportunities for adversaries to poison the training process \cite{RN155}. A solution to resolve these problems is incremental learning, which has been applied in ML domain \cite{RN156, RN157}.
Incremental learning utilizes the previous knowledge as the starting point and utilizes newly available information to upgrade a model \cite{RN153, RN158}. Subsequently reducing retraining time, resources and providing quicker adaptation to changing conditions. Hence, we suggest that retraining methods be considered while designing a method to detect data exfiltration. 
Furthermore, we recommend that incremental retraining procedures be explored, customized and applied in data exfiltration domain.}
\par
\noindent\fbox{%
 \parbox{\columnwidth}{%
 \textbf{\textit{Summary}}: Model-retraining methods should be considered while designing the detection approaches and, in this regard, incremental-retraining methods should be studied to handle the limitation of batch retraining.}} 
\subsection{Need for Resilient Learning Models}
{Security of ML models is not only an open challenge in other domains \cite{RN163}, but it is also a major concern in security-based systems (e.g., intrusion detection and malware detection systems) \cite{RN167, RN168}.
Furthermore, there is significant amount of efforts done other domains such as image  \cite{RN162}, Natural Language Processing (NLP) \cite{zhang2019generating} and Intrusion detection \cite{corona2013adversarial} to create adversarial examples to test the ML models against evasion and poisoning attacks
However, we found that only seven studies [S18, S29, S35, S39, S66, S68, S81] considered this threat as an essential concern for ML countermeasures for detecting data exfiltration.
We emphasize that the risks of these attacks are incredibly high for data exfiltration countermeasures because of the monetary benefits achieved by an attackers for stealing critically sensitive information.
We recommend that future research should focus on secure ML countermeasures that are robust to adversarial attacks by employing techniques from other domain such as randomness, data sanitization and adversarial training \cite{RN155, RN162} to make their approaches resilient to these attacks.}
\par
\noindent\fbox{%
 \parbox{\columnwidth}{%
 \textbf{\textit{Summary}}: Resilience to adversarial learning should be considered while designing and implementing ML-based data exfiltration countermeasures.}} 
\subsection{Need for Benchmark Performance Measures}
{Several performance measures have been reported by the reviewed studies to evaluate the ML countermeasures (Section~\ref{RQ5}.
This diversity makes them hard to compare. Moreover, around 14\% of the reviewed studies reported a single evaluation metric to validate their approaches.
We advocate that, similar to other domains such as software defect prediction 
\cite{RN70, RN74}, multiple metrics should be used to evaluate ML-based approaches, which will enable researchers to make an apple to apple comparison with the previous data exfiltration countermeasures.
Data exfiltration inherits the problem of imbalanced class distribution from the cybersecurity domain because the number of attack scenarios is usually less than the legitimate ones. Such a class imbalance warrants careful selection of evaluation metrics. Although measures like AUC, recall, precision and F-score can help to interpret the class imbalance problem for binary classification.
However, these measures are ineffective for multi-class problems \cite{RN171}. Taking inspiration from other domains (Computer Vision \cite{RN171, RN172}), we recommend that evaluation metrics like Adjustable F-score, balance, Matthew correlation and G-means be used to effectively deal with skewed class and multi-class problem in data exfiltration domain. However, only one study [S68] reported Matthew correlation to detect data tunnelling attack. 
Furthermore, our review has revealed that 14\% and 15\% studies report training and prediction time of the classifier. 
We assert that both prediction and training time are essential as data exfiltration attacks need to detect as fast as possible \cite{RN173} and models require frequent update \cite{RN67}.
We also suggest that with the advent of adversarial ML, additional measures such as robustness of features and security of ML model (as suggested in \cite{RN170}) be considered.}
\par
\noindent\fbox{%
 \parbox{\columnwidth}{%
 \textbf{\textit{Summary}}: Due to skewed data and class distribution, we recommend that researchers and practitioners use F-score, ROC, confusion matric, G-mean, prediction time, training time, and feature and model robustness to evaluate ML-based data exfiltration countermeasures.}} 
\subsection{{Need for Real-time Detection of Data Exfiltration Attacks}}
{By real-time attack detection, we mean the data exfiltration detector collects new and targeted information from both internal (e.g., computing servers) and external sources (e.g., blacklisted IPs), and process the information almost instantaneously to generate alerts about a potential data exfiltration attack. Given the sophisticated data exfiltration attack vector, it is imperative to detect data exfiltration in real-time or at least near real-time. 
This is because the longer an attack goes undetected, the higher is the damage \cite{RN174}.
Given the fact that data exfiltration attacks are highly targeted attacks (e.g., aiming only to leak top secrets of an enterprise), the exfiltration process is quite quick especially after a hacker has located the target data. With real-time detection, the success rate of a data exfiltration attack can be reduced by around 97\% \cite{RN175}. For instance, in 2019, Wipro – a consultancy company fell victim to a data exfiltration attack aimed at exfiltrating customer data \cite{RN176}.
However, the swift response of a company to detect and mitigate an attack saved a company from a huge loss.  
In our review, only 12/92 studies report the training and prediction time.
Table~\ref{timetable}  shows the training time (for training a detection system) and the prediction time (i.e., time taken to detect an attack) for these studies.
Given the significance of time for detecting a data exfiltration attack, we believe that all researchers should consider response time (training and prediction time) as a critical quality attribute and report it in their studies. Based on Table~\ref{timetable}, the mean training time is 4.72 min, and the mean prediction time is 8.78 min for the reviewed ML-based data exfiltration countermeasures.
This time is primarily consumed in the analysis of a large size of data. 
For example, an organization as large as HP generates around one trillion security events per day \cite{RN177}. Therefore, we recommend the incorporation of big data technologies (e.g., Apache Spark and Apache Hadoop) in ML-based data exfiltration countermeasures. It is because big data technologies have proven advantages.
For example, a study \cite{RN178} by Zions incorporation revealed that the use of Apache Hadoop reduced the processing time from 20-60 min to 1 min to analyse a month of security event data.}
\par
\noindent\fbox{%
\parbox{\columnwidth}{%
 \textbf{\textit{Summary}}: Researchers should consider and report response time as an important quality attribute for ML based data exfiltration countermeasures. Moreover, researchers should explore the use of big data technologies for engineering ML based data exfiltration countermeasures.}}
 \begin{table*}[t!]
\begin{small}
\caption{{Studies that report the training time and prediction time}}
\label{timetable}
\centering
\resizebox{\textwidth}{!}{\begin{tabular}{|c|c|c|c|c| c|c|c|c|c|c|c|c|c|c|c|c|c|}
\hline
    \textbf{Study} & \textbf{S11} & \textbf{S13} & \textbf{S25} & \textbf{S43} & \textbf{S44} & \textbf{S51} & \textbf{S55} & \textbf{S59} & \textbf{S62} & \textbf{S63} & \textbf{S64} & \textbf{S65} & \textbf{S70} & \textbf{S73} & \textbf{S78} & \textbf{S92} \\ 
    \hline
    Training Time (sec)&& 0.3 & 46.13 &  & 45.5 & 3.34 & 0.77 & 4.3 &  &  & 2268 & 7.96 & 39.24 & 964 & 0.05 & 22.72 \\ 
    \hline
    Prediction Time (sec)  & 18 & 0.21 &  & 0.08 & 6180 &  & 0.0023 & 0.016 & 0.0758 & 0.64 & 1 & 0.81 &  & 123 & 0.03 & \\
    \hline
\end{tabular}}
\end{small}
\end{table*}    
\subsection{Under-addressed Quality Attributes of ML-based Data Exfiltration Countermeasures}
{We observe that most of the ML based data exfiltration countermeasures are evaluated using two quality measures i.e., accuracy  used by 70\% studies and response time  reported by 15\% studies.
However, we believe that designing and evaluating ML-based data exfiltration countermeasures only with accuracy and response time is not enough. Additional quality measures such as extensibility, customizability, interpretability, interoperability, automation of model update, security, availability, and reproducibility should be analyzed to ensure the quality and reliability of the ML-based data exfiltration countermeasures \cite{RN179}.
For instance, it is crucial to consider how the data exfiltration countermeasure interoperates in the whole security spectrum of an organization, i.e., integration of the countermeasure into the unified security orchestration platform \cite{RN180}.
Similarly, the ML model employed in the data exfiltration countermeasure should be interpretable to investigate how and why a data exfiltration attack went undetected through the countermeasure. 
Testing and verifying the quality of these ML-based software systems, including data exfiltration countermeasures is inherently challenging because traditional software testing methods are not directly applicable to these systems \cite{RN169, RN179,song2017ML} 
In this regard, there needs to be close collaboration between software and ML communities \cite{RN179}, so both work together to drive the future of ML-based cybersecurity systems such as intrusion detection system, malware detection system, and data exfiltration detection systems.}
\par
\noindent\fbox{%
 \parbox{\columnwidth}{%
 \textbf{\textit{Summary}}: Both software engineering and Machine learning communities should work together to integrate quality in ML lifecycle for ML-based data exfiltration countermeasures.}} 
\section{Threats to Validity}
\label{section-threat}
Whilst we strictly followed the guidelines provided by Kitchenham and Charters \cite{RN63} for qualitative analysis and performed quantitative analysis based on \cite{RN74}, we had similar threats like other SLRs in software engineering and machine learning domains. The findings of this SLR may have been affected by the following threats.
\begin{enumerate}
    \item \emph{Search Strategy} For any SLR, identifying all the primary studies is a challenging task. To mitigate this threat, we used six data sources which enabled us to cover a wide variety of publication sources for finding relevant papers \cite{RN49}. Furthermore, to address the issue of missing search terms in the search string, we employed three strategies: i) Search string was improved iteratively based on pilot searches. ii) We took inspiration from the search strings used in existing reviews \cite{RN1,RN20} and iii) We performed forward and backward snowballing to find other relevant papers that might have been missed by the search string. 
    \item \emph{Bias in Study Selection and Data Extraction}: This step can be prejudiced to researchers’ subjective judgement about whether the reviewed papers meet the inclusion or exclusion criteria reported in Section\ref{3.4}. To address this issue, the studies were selected through a multi-phase process (Section~\ref{3.5}). Furthermore, we performed a cross-check using a random number of selected papers to ensure that our inclusion and exclusion criteria returns the papers that we already knew to be relevant. Such a process ensures to a high degree that all relevant studies are identified and selected. In addition to study selection, the bias in extraction of data from the primary studies can also affect the findings of this SLR. To mitigate this threat, we formulated a data extraction form (see Table~\ref{dataextractionform}) to consistently extract and analyse data for answering the research questions. 
    \item \emph{Box Plots}: The data points in the boxplots reported in this SLR to synthesize the quantitative data is limited to what is reported in the reviewed studies. For example, during the investigation of feature selection in relation to number of features selected, filter method had 14 data rows while embedded had only two data row. This issue might skew the findings and so threatens the validity of our findings. To mitigate this threat, we selected median of data group for quantitative analysis. Similar to \cite{RN62,RN66}, we conclude that it is important to investigate the multiple factors involved in ML lifecycle to investigate the model performance. 
\end{enumerate}
\section{CONCLUSION}
\label{section-conclusion}
The main goal of this SLR was to analyse and synthesise the existing literature on ML-based data exfiltration countermeasures in order to identify what kind of features, datasets, modelling techniques, validation method and performance evaluation criteria are used in these studies.
A systematic literature review was conducted to fulfil the study goals and answers five research questions. After a comprehensive research protocol, 92 studies were selected as primary studies. Besides identifying the research gaps and making recommendations for future research, the main outcomes of this SLR are summarised below.
\begin{enumerate}
    \item ML-based data exfiltration countermeasures can be classified into two main groups, i.e., data-driven and behaviour-driven approaches. Behaviour-driven approaches result in more (median $>6\%$) false positives but are capable of detecting complex attack vectors like APT, lateral movement, and insider threat.
In contrast, data-driven approaches attain high performance (median accuracy and FPR $>95\%$ and $<6\%$). However, these approaches are only used frequently to detect the \emph{Delivery} stage of DELC.
    \item Six types of features are used by the identified countermeasures, i.e., statistical, structural, behavioural, spatial, syntactic, and temporal features. Among these, statistical features are most frequently used. Most of these features are extracted by using domain knowledge; 27\% of the reviewed studies used automated feature engineering process. Feature selection techniques are not used frequently by these studies.
    \item Twelve datasets used by the reviewed studies are publicly available out of which eight are one decade old.
    \item Classification is mostly used by the reviewed studies and only 16\% studies used anomaly-based detection. RFT and SVM are the most frequently used classifiers. Ensemble classifiers RFT, Gradient Boosting and Adaboost, performed better than all the other classifiers achieving F-Score of 98\%.
    \item K-Fold and Hold Out are the most frequently used validation methods. Accuracy, Recall, and FPR are most commonly reported performance measures in these studies. 
    The countermeasures are effective in detecting command and control, data tunnelling, Phishing, Overt channel and Malicious domain attacks with a high median accuracy and recall, i.e., 95\% and median FPR of less than 2\%.
\end{enumerate}
To conclude, the performance and quality of ML-based models for detecting data exfiltration is driven by the features, datasets, modelling technique and validation method. These models are still in infancy and many limitations need to be addressed, in terms of the quality and reliability of these methods. In this respect, we hope that this SLR will provide a research direction to researchers and help practitioners to adopt better practices. 
\section{Acknowledgemets}
This work is partially supported by CSIRO’s Data61, Australia and Cyber Security Research Centre Limited whose activities are partly funded by the Australian Government’s Cooperative Research Centres Programme.

\bibliographystyle{IEEEtran}
\bibliography{thesis.bib}

\begin{thebibliography}{100}
\providecommand{\url}[1]{#1}
\csname url@samestyle\endcsname
\providecommand{\newblock}{\relax}
\providecommand{\bibinfo}[2]{#2}
\providecommand{\BIBentrySTDinterwordspacing}{\spaceskip=0pt\relax}
\providecommand{\BIBentryALTinterwordstretchfactor}{4}
\providecommand{\BIBentryALTinterwordspacing}{\spaceskip=\fontdimen2\font plus
\BIBentryALTinterwordstretchfactor\fontdimen3\font minus
  \fontdimen4\font\relax}
\providecommand{\BIBforeignlanguage}[2]{{%
\expandafter\ifx\csname l@#1\endcsname\relax
\typeout{** WARNING: IEEEtran.bst: No hyphenation pattern has been}%
\typeout{** loaded for the language `#1'. Using the pattern for}%
\typeout{** the default language instead.}%
\else
\language=\csname l@#1\endcsname
\fi
#2}}
\providecommand{\BIBdecl}{\relax}
\BIBdecl

\bibitem{RN1}
F.~Ullah, M.~Edwards, R.~Ramdhany, R.~Chitchyan, M.~A. Babar, and A.~Rashid,
  ``Data exfiltration: A review of external attack vectors and
  countermeasures,'' \emph{Journal of Network and Computer Applications}, vol.
  101, pp. 18--54, 2018.

\bibitem{InfoBlox}
\BIBentryALTinterwordspacing
InfoBlox. (2020) Ddi (secure dns, dhcp, ipam). [Online]. Available:
  \url{https://bit.ly/2CbxdFH}
\BIBentrySTDinterwordspacing

\bibitem{RN2}
S.~Alneyadi, E.~Sithirasenan, and V.~Muthukkumarasamy, ``A survey on data
  leakage prevention systems,'' \emph{Journal of Network and Computer
  Applications}, vol.~62, pp. 137--152, 2016.

\bibitem{RN3}
A.~Shabtai, Y.~Elovici, and L.~Rokach, ``A survey of data leakage detection and
  prevention solutions,'' \emph{A Survey of Data Leakage Detection and
  Prevention Solutions}, 2012.

\bibitem{RN6}
\BIBentryALTinterwordspacing
ITRC, ``End of the year data breach report,'' 2018. [Online]. Available:
  \url{https:/bit.ly/2wZb9bd}
\BIBentrySTDinterwordspacing

\bibitem{RN5}
\BIBentryALTinterwordspacing
Verizon. (2018) 2018 data breach investigations report. [Online]. Available:
  \url{https://vz.to/3fqidBQ}
\BIBentrySTDinterwordspacing

\bibitem{Verizon2019}
\BIBentryALTinterwordspacing
------. (2019) Data breach investigations report. [Online]. Available:
  \url{https://vz.to/36tVx0o}
\BIBentrySTDinterwordspacing

\bibitem{RN10}
\BIBentryALTinterwordspacing
IBM, ``2018 cost of data breach study.'' 2018. [Online]. Available:
  \url{https://ibm.co/2Zu2IRp}
\BIBentrySTDinterwordspacing

\bibitem{RN9}
\BIBentryALTinterwordspacing
T.~N. Times, ``Researchers trace data theft to intruders in china.'' 2010.
  [Online]. Available: \url{https://nyti.ms/3glzxJC}
\BIBentrySTDinterwordspacing

\bibitem{RN11}
\BIBentryALTinterwordspacing
Mcafee, ``Grand theft data report: Data exfiltration study: Actors, tactics,
  and detection,'' 2020. [Online]. Available: \url{https://bit.ly/36c2XVP}
\BIBentrySTDinterwordspacing

\bibitem{RN12}
\BIBentryALTinterwordspacing
T.~Thimou. (2018) What we know about the alleged suntrust data breach.
  [Online]. Available: \url{https://bit.ly/3d6fxZE}
\BIBentrySTDinterwordspacing

\bibitem{RN13}
\BIBentryALTinterwordspacing
J.~Blankenship. (2018) Tesla sabotage: A perfect storm for insider threat.
  [Online]. Available: \url{https://bit.ly/2AWFxbr}
\BIBentrySTDinterwordspacing

\bibitem{RN19}
L.~University, ``Data exfiltration: Detecting and preventing,'' p.~40, 2014.

\bibitem{RN20}
I.~Homoliak, F.~Toffalini, J.~Guarnizo, Y.~Elovici, and M.~Ochoa, ``Insight
  into insiders and it: A survey of insider threat taxonomies, analysis,
  modeling, and countermeasures,'' vol.~99, 2019.

\bibitem{RN21}
L.~Liu, O.~De~Vel, Q.~L. Han, J.~Zhang, and Y.~Xiang, ``Detecting and
  preventing cyber insider threats: A survey,'' \emph{IEEE Communications
  Surveys and Tutorials}, pp. 1--21, 2018.

\bibitem{RN18}
\BIBentryALTinterwordspacing
F-Secure, ``Detecting and deterring data exfiltration,'' 2014. [Online].
  Available: \url{https://bit.ly/2UIj0X9}
\BIBentrySTDinterwordspacing

\bibitem{RN22}
P.~Chen, L.~Desmet, and C.~Huygens, ``A study on advanced persistent threats,''
  \emph{Communications and Multimedia Security}, vol. 8735, pp. 63--72, 2014.

\bibitem{RN23}
K.~L. Chiew, K.~S.~C. Yong, and C.~L. Tan, ``A survey of phishing attacks:
  Their types, vectors and technical approaches,'' \emph{Expert Systems with
  Applications}, 2018.

\bibitem{RN26}
H.~A.-B. Hashim, M.~M. Saudi, and N.~Basir, ``A systematic review analysis of
  root exploitation for mobile botnet detection,'' in \emph{Advanced Computer
  and Communication Engineering Technology}, 2016, pp. 113--122.

\bibitem{RN29}
L.~Bilge, S.~Sen, D.~Balzarotti, E.~Kirda, and C.~Kruegel, ``Exposure,''
  \emph{ACM Transactions on Information and System Security}, vol.~16, no.~4,
  pp. 1--28, 2014.

\bibitem{RN28}
J.~Gardiner, M.~Cova, and S.~Nagaraja, ``Command \& control: Understanding,
  denying and detecting,'' \emph{arXiv.org}, 2014.

\bibitem{RN32}
H.~P.~S. Bhasin, E.~Ramsdell, A.~Alva, R.~Sreedhar, and M.~Bhadkamkar, ``Data
  center application security: Lateral movement detection of malware using
  behavioral models,'' \emph{SMU Data Science Review}, vol.~1, no.~2, p.~10,
  2018.

\bibitem{RN33}
N.~Provos, M.~Friedl, and P.~Honeyman, ``Preventing privilege escalation.'' in
  \emph{USENIX Security Symposium}, 2003.

\bibitem{RN31}
G.~Hospodar, B.~Gierlichs, E.~De~Mulder, I.~Verbauwhede, and J.~Vandewalle,
  ``Machine learning in side-channel analysis: A first study,'' \emph{Journal
  of Cryptographic Engineering}, vol.~1, pp. 293--302, 2011.

\bibitem{RN30}
A.~K. Biswas, D.~Ghosal, and S.~Nagaraja, ``A survey of timing channels and
  countermeasures,'' \emph{ACM Computing Surveys}, vol.~50, no.~1, pp. 1--39,
  2017.

\bibitem{RN36}
E.~Alpaydin, \emph{Introduction to machine learning}.\hskip 1em plus 0.5em
  minus 0.4em\relax MIT press, 2020.

\bibitem{RN38}
U.~Khurana, F.~Nargesian, H.~Samulowitz, E.~Khalil, and D.~Turaga, ``Automating
  feature engineering,'' \emph{Transformation}, 2016.

\bibitem{hira2015review}
Z.~M. Hira and D.~F. Gillies, ``A review of feature selection and feature
  extraction methods applied on microarray data,'' \emph{Advances in
  bioinformatics}, vol. 2015, 2015.

\bibitem{RN40}
S.~Agrawal and J.~Agrawal, ``Survey on anomaly detection using data mining
  techniques,'' \emph{Procedia Computer Science}, vol.~60, pp. 708--713, 2015.

\bibitem{RN52}
V.~Ughade, N.~Mishra, and S.~Sharma, ``Improved kmean clustering with steepest
  ascent 'gradient' method for image retrieval,'' \emph{International Journal
  of Computer Applications}, 2011.

\bibitem{RN45}
K.-L. Li, H.-K. Huang, S.-F. Tian, and W.~Xu, ``Improving one-class svm for
  anomaly detection,'' in \emph{Proceedings of the 2003 International
  Conference on Machine Learning and Cybernetics (IEEE Cat. No. 03EX693)},
  vol.~5.\hskip 1em plus 0.5em minus 0.4em\relax IEEE, 2003, pp. 3077--3081.

\bibitem{reynolds2009gaussian}
D.~A. Reynolds, ``Gaussian mixture models.'' \emph{Encyclopedia of biometrics},
  vol. 741, 2009.

\bibitem{beal2002infinite}
M.~J. Beal, Z.~Ghahramani, and C.~E. Rasmussen, ``The infinite hidden markov
  model,'' in \emph{Advances in neural information processing systems}, 2002,
  pp. 577--584.

\bibitem{RN44}
S.~R. Gunn \emph{et~al.}, ``Support vector machines for classification and
  regression,'' \emph{ISIS technical report}, vol.~14, no.~1, pp. 5--16, 1998.

\bibitem{RN48}
N.~Friedman, D.~Geiger, and M.~Goldszmidt, ``Bayesian network classifiers,''
  \emph{Machine learning}, vol.~29, no. 2-3, pp. 131--163, 1997.

\bibitem{RN49}
P.~Cunningham and S.~J. Delany, ``k-nearest neighbour classifiers,''
  \emph{Multiple Classifier Systems}, 2007.

\bibitem{RN50}
S.~B. Kotsiantis, ``Decision trees: a recent overview,'' \emph{Artificial
  Intelligence Review}, vol.~39, no.~4, pp. 261--283, 2013.

\bibitem{RN51}
D.~G. Kleinbaum, K.~Dietz, M.~Gail, M.~Klein, and M.~Klein, \emph{Logistic
  regression}.\hskip 1em plus 0.5em minus 0.4em\relax Springer, 2002.

\bibitem{RN55}
J.~Ali, R.~Khan, N.~Ahmad, and I.~Maqsood, ``Random forests and decision
  trees,'' \emph{International Journal of Computer Science Issues (IJCSI)},
  vol.~9, no.~5, p. 272, 2012.

\bibitem{RN56}
T.~Hastie, S.~Rosset, J.~Zhu, and H.~Zou, ``Multi-class adaboost,''
  \emph{Statistics and Its Interface}, 2013.

\bibitem{anderson1995introduction}
J.~A. Anderson, \emph{An introduction to neural networks}.\hskip 1em plus 0.5em
  minus 0.4em\relax MIT press, 1995.

\bibitem{pal1992multilayer}
S.~K. Pal and S.~Mitra, ``Multilayer perceptron, fuzzy sets, classifiaction,''
  1992.

\bibitem{krizhevsky2012imagenet}
A.~Krizhevsky, I.~Sutskever, and G.~E. Hinton, ``Imagenet classification with
  deep convolutional neural networks,'' in \emph{Advances in neural information
  processing systems}, 2012, pp. 1097--1105.

\bibitem{mikolov2010recurrent}
T.~Mikolov, M.~Karafi{\'a}t, L.~Burget, J.~{\v{C}}ernock{\`y}, and
  S.~Khudanpur, ``Recurrent neural network based language model,'' in
  \emph{Eleventh annual conference of the international speech communication
  association}, 2010.

\bibitem{chung2014empirical}
J.~Chung, C.~Gulcehre, K.~Cho, and Y.~Bengio, ``Empirical evaluation of gated
  recurrent neural networks on sequence modeling,'' \emph{preprint
  arXiv:1412.3555}, 2014.

\bibitem{hochreiter1997long}
S.~Hochreiter and J.~Schmidhuber, ``Long short-term memory,'' \emph{Neural
  computation}, vol.~9, no.~8, pp. 1735--1780, 1997.

\bibitem{RN41}
S.~Thrun and M.~L. Littman, ``Reinforcement learning: an introduction,''
  \emph{AI Magazine}, vol.~21, no.~1, pp. 103--103, 2000.

\bibitem{RN43}
R.~Kohavi, ``A study of cross-validation and bootstrap for accuracy estimation
  and model selection,'' \emph{Proceedings of the 14th international joint
  conference on Artificial intelligence - Volume 2}, 1995.

\bibitem{RN35}
T.~M. Mitchell, \emph{The discipline of machine learning}, 2006, vol.~9.

\bibitem{RN57}
R.~Tahboub and Y.~Saleh, ``Data leakage/loss prevention systems (dlp),''
  \emph{2014 World Congress on Computer Applications and Information Systems,
  WCCAIS 2014}, 2014.

\bibitem{RN58}
P.~Raman, H.~G. Kayac{\i}k, and A.~Somayaji, ``Understanding data leak
  prevention,'' in \emph{6th Annual Symposium on Information Assurance
  (ASIA’11)}.\hskip 1em plus 0.5em minus 0.4em\relax Citeseer, 2011, p.~27.

\bibitem{RN59}
T.~Brindha and R.~S. Shaji, ``An analysis of data leakage and prevention
  techniques in cloud environment,'' \emph{2015 International Conference on
  Control Instrumentation Communication and Computational Technologies, ICCICCT
  2015}, pp. 350--355, 2016.

\bibitem{RN60}
K.~Kaur, I.~Gupta, and A.~K. Singh, ``A comparative study of the approach
  provided for preventing the data leakage,'' \emph{International Journal of
  Network Security \& Its Applications}, 2017.

\bibitem{RN63}
B.~Kitchenham and S.~Charters, ``Guidelines for performing systematic
  literature reviews in software engineering version 2.3,'' \emph{Engineering},
  2007.

\bibitem{RN64}
C.~Wohlin, ``Guidelines for snowballing in systematic literature studies and a
  replication in software engineering,'' in \emph{Proceedings of the 18th
  international conference on evaluation and assessment in software
  engineering}, 2014, pp. 1--10.

\bibitem{RN65}
A.~Hoonlor, B.~K. Szymanski, and M.~J. Zaki, ``Trends in computer science
  research,'' \emph{Commun. ACM}, vol.~56, no.~10, pp. 74--83, 2013.

\bibitem{RN66}
M.~Shahin, M.~Ali~Babar, and L.~Zhu, ``Continuous integration, delivery and
  deployment: A systematic review on approaches, tools, challenges and
  practices,'' \emph{IEEE Access}, vol.~5, pp. 3909--3943, 2017.

\bibitem{RN67}
A.~L. Buczak and E.~Guven, ``A survey of data mining and machine learning
  methods for cyber security intrusion detection,'' \emph{IEEE Communications
  Surveys \& Tutorials}, vol.~18, no.~2, pp. 1153--1176, 2016.

\bibitem{RN68}
X.~Du, ``Data mining and machine learning in cybersecurity,'' \emph{Data Mining
  and Machine Learning in Cybersecurity}, 2011.

\bibitem{RN70}
S.~Hosseini, B.~Turhan, and D.~Gunarathna, \emph{A Systematic Literature Review
  and Meta-Analysis on Cross Project Defect Prediction}, 2017.

\bibitem{RN72}
D.~S. Cruzes and T.~Dyba, ``Recommended steps for thematic synthesis in
  software engineering,'' pp. 275--284, 2011.

\bibitem{RN73}
D.~F. Williamson, R.~A. Parker, and J.~S. Kendrick, ``The box plot: a simple
  visual method to interpret data.'' \emph{Annals of internal medicine}, 1989.

\bibitem{RN74}
T.~Hall, S.~Beecham, D.~Bowes, D.~Gray, and S.~Counsell, ``A systematic
  literature review on fault prediction performance in software engineering,''
  \emph{IEEE Transactions on Software Engineering}, vol.~38, no.~6, pp.
  1276--1304, 2011.

\bibitem{RN75}
M.~Schwabacher, ``A survey of data-driven prognostics,'' in \emph{Infotech@
  Aerospace}, 2005, p. 7002.

\bibitem{RN76}
V.~Berk, A.~Giani, G.~Cybenko, and N.~Hanover, ``Detection of covert channel
  encoding in network packet delays,'' \emph{Rapport technique TR536, de
  lUniversit{\'e} de Dartmouth}, vol.~19, 2005.

\bibitem{RN77}
J.~Geddes, R.~Jansen, and N.~Hopper, ``Imux: Managing tor connections from two
  to infinity, and beyond,'' in \emph{Proceedings of the 13th Workshop on
  Privacy in the Electronic Society}, ser. WPES ’14, 2014, p. 181–190.

\bibitem{yarom2014flush}
Y.~Yarom and K.~Falkner, ``Flush+ reload: a high resolution, low noise, l3
  cache side-channel attack,'' in \emph{23rd $\{$USENIX$\}$ Security Symposium
  ($\{$USENIX$\}$ Security 14)}, 2014, pp. 719--732.

\bibitem{RN81}
W.~Lee, S.~J. Stolfo, P.~K. Chan, E.~Eskin, W.~Fan, M.~Miller, S.~Hershkop, and
  J.~Zhang, ``Real time data mining-based intrusion detection,''
  \emph{Proceedings - DARPA Information Survivability Conference and Exposition
  II, DISCEX 2001}, 2001.

\bibitem{RN157}
J.~P. Barddal, H.~M. Gomes, F.~Enembreck, and B.~Pfahringer, ``A survey on
  feature drift adaptation: Definition, benchmark, challenges and future
  directions,'' \emph{Journal of Systems and Software}, vol. 127, pp. 278--294,
  2017.

\bibitem{RN83}
A.~Moser, C.~Kruegel, and E.~Kirda, ``Limits of static analysis for malware
  detection,'' in \emph{Twenty-Third Annual Computer Security Applications
  Conference (ACSAC 2007)}.\hskip 1em plus 0.5em minus 0.4em\relax IEEE, 2007,
  pp. 421--430.

\bibitem{RN84}
Y.~Ye, T.~Li, D.~Adjeroh, and S.~S. Iyengar, ``A survey on malware detection
  using data mining techniques,'' \emph{ACM Computing Surveys}, vol.~50, no.~3,
  pp. 1--40, 2017.

\bibitem{RN90}
G.~Sidorov, F.~Velasquez, E.~Stamatatos, A.~Gelbukh, and
  L.~Chanona-Hern{\'a}ndez, ``Syntactic dependency-based n-grams as
  classification features,'' in \emph{Mexican International Conference on
  Artificial Intelligence}.\hskip 1em plus 0.5em minus 0.4em\relax Springer,
  2012, pp. 1--11.

\bibitem{RN78}
J.~Bala, ``Combining structural and statistical features in a machine learning
  technique for texture classification,'' pp. 175--183, 1990.

\bibitem{RN79}
H.~Wu and N.~Yuan, ``An improved tf-idf algorithm based on word frequency
  distribution information and category distribution information,'' in
  \emph{Proceedings of the 3rd International Conference on Intelligent
  Information Processing}, 2018, pp. 211--215.

\bibitem{RN80}
D.~Jurafsky and J.~H. Martin, ``N-grams,'' \emph{Speech and Language
  Processing}, 2014.

\bibitem{goldberg2014word2vec}
Y.~Goldberg and O.~Levy, ``word2vec explained: deriving mikolov et al.'s
  negative-sampling word-embedding method,'' \emph{preprint arXiv:1402.3722},
  2014.

\bibitem{RN88}
\BIBentryALTinterwordspacing
Aldeid. (2020) Pe tools. Aldeid. [Online]. Available:
  \url{https://bit.ly/2Ak81MF}
\BIBentrySTDinterwordspacing

\bibitem{RN89}
\BIBentryALTinterwordspacing
H.~Rays. (2020) Ida pro. [Online]. Available:
  \url{https://www.hex-rays.com/products/ida/}
\BIBentrySTDinterwordspacing

\bibitem{RN39}
G.~Chandrashekar and F.~Sahin, ``A survey on feature selection methods,''
  \emph{Computers \& Electrical Engineering}, vol.~40, no.~1, pp. 16--28, 2014.

\bibitem{RN91}
R.~Kohavi and G.~H. John, ``Wrappers for feature subset selection,''
  \emph{Artificial Intelligence}, 2002.

\bibitem{RN92}
C.~J.~C. Burges, ``Dimension reduction: A guided tour,'' \emph{Foundations and
  Trends® in Machine Learning}, 2009.

\bibitem{vaswani2017attention}
A.~Vaswani, N.~Shazeer, N.~Parmar, J.~Uszkoreit, L.~Jones, A.~N. Gomez,
  {\L}.~Kaiser, and I.~Polosukhin, ``Attention is all you need,'' in
  \emph{Advances in neural information processing systems}, 2017, pp.
  5998--6008.

\bibitem{jain2019attention}
S.~Jain and B.~C. Wallace, ``Attention is not explanation,'' \emph{preprint
  arXiv:1902.10186}, 2019.

\bibitem{serrano2019attention}
S.~Serrano and N.~A. Smith, ``Is attention interpretable?'' \emph{preprint
  arXiv:1906.03731}, 2019.

\bibitem{tang2018analysis}
G.~Tang, R.~Sennrich, and J.~Nivre, ``An analysis of attention mechanisms: The
  case of word sense disambiguation in neural machine translation,''
  \emph{preprint arXiv:1810.07595}, 2018.

\bibitem{galassi2019attention}
A.~Galassi, M.~Lippi, and P.~Torroni, ``Attention, please! a critical review of
  neural attention models in natural language processing,'' \emph{preprint
  arXiv:1902.02181}, 2019.

\bibitem{sheen2008network}
S.~Sheen and R.~Rajesh, ``Network intrusion detection using feature selection
  and decision tree classifier,'' in \emph{TENCON 2008-2008 IEEE Region 10
  Conference}.\hskip 1em plus 0.5em minus 0.4em\relax IEEE, 2008, pp. 1--4.

\bibitem{devesa2010automatic}
J.~Devesa, I.~Santos, X.~Cantero, Y.~K. Penya, and P.~G. Bringas, ``Automatic
  behaviour-based analysis and classification system for malware detection.''
  \emph{ICEIS (2)}, vol.~2, pp. 395--399, 2010.

\bibitem{singh2010feature}
S.~R. Singh, H.~A. Murthy, T.~A. Gonsalves \emph{et~al.}, ``Feature selection
  for text classification based on gini coefficient of inequality.''
  \emph{Fsdm}, vol.~10, pp. 76--85, 2010.

\bibitem{meinshausen2010stability}
N.~Meinshausen and P.~B{\"u}hlmann, ``Stability selection,'' \emph{Journal of
  the Royal Statistical Society: Series B (Statistical Methodology)}, vol.~72,
  no.~4, pp. 417--473, 2010.

\bibitem{RN93}
A.~Jovic, K.~Brkic, and N.~Bogunovic, ``A review of feature selection methods
  with applications,'' in \emph{2015 38th international convention on
  information and communication technology, electronics and microelectronics
  (MIPRO)}.\hskip 1em plus 0.5em minus 0.4em\relax Ieee, 2015, pp. 1200--1205.

\bibitem{RN94}
T.~Young, D.~Hazarika, S.~Poria, and E.~Cambria, ``Recent trends in deep
  learning based natural language processing,'' \emph{ieee Computational
  intelligenCe magazine}, vol.~13, no.~3, pp. 55--75, 2018.

\bibitem{RN98}
\BIBentryALTinterwordspacing
PhishTank. (2020) Phishtank | join the fight against phishing. [Online].
  Available: \url{https://www.phishtank.com/}
\BIBentrySTDinterwordspacing

\bibitem{RN100}
\BIBentryALTinterwordspacing
D.~Radev. (2008) Clair collection of fraud email (repository) - acl wiki.
  [Online]. Available: \url{https://bit.ly/2W5QE9f}
\BIBentrySTDinterwordspacing

\bibitem{RN101}
J.~Ma, L.~K. Saul, S.~Savage, and G.~M. Voelker, ``Identifying suspicious urls:
  an application of large-scale online learning,'' in \emph{Proceedings of the
  26th annual international conference on machine learning}, 2009, pp.
  681--688.

\bibitem{RN102}
\BIBentryALTinterwordspacing
OpenPhish. (2020) Phishing intelligence. [Online]. Available:
  \url{https://openphish.com/}
\BIBentrySTDinterwordspacing

\bibitem{Dua2017}
\BIBentryALTinterwordspacing
D.~Dua and C.~Graff. (2017) {UCI} machine learning repository. [Online].
  Available: \url{http://archive.ics.uci.edu/ml}
\BIBentrySTDinterwordspacing

\bibitem{mohammad2014intelligent}
R.~M. Mohammad, F.~Thabtah, and L.~McCluskey, ``Intelligent rule-based phishing
  websites classification,'' \emph{IET Information Security}, vol.~8, no.~3,
  pp. 153--160, 2014.

\bibitem{RN103}
\BIBentryALTinterwordspacing
C.~C.~S. Computing and W.~Mining. (2008) Web spam detection. [Online].
  Available: \url{https://chato.cl/webspam/}
\BIBentrySTDinterwordspacing

\bibitem{RN104}
B.~Klimt and Y.~Yang, ``Introducing the enron corpus,'' \emph{Machine
  Learning}, 2004.

\bibitem{TheEnron29}
\BIBentryALTinterwordspacing
A.~U. of~Economics and Business. (2020) The enron-spam datasets. [Online].
  Available: \url{https://bit.ly/30LqFrB}
\BIBentrySTDinterwordspacing

\bibitem{VermaRakesha2019IWSPA1}
R.~M. Verma, V.~Zeng, and H.~Faridi, ``Data quality for security challenges:
  Case studies of phishing, malware and intrusion detection datasets,'' in
  \emph{Proceedings of the 2019 ACM SIGSAC Conference on Computer and
  Communications Security}, 2019, p. 2605–2607.

\bibitem{Zeng2020IWSPA2}
V.~Zeng, S.~Baki, A.~E. Aassal, R.~Verma, L.~F.~T. De~Moraes, and A.~Das,
  ``Diverse datasets and a customizable benchmarking framework for phishing,''
  in \emph{Proceedings of the Sixth International Workshop on Security and
  Privacy Analytics}, ser. IWSPA ’20, 2020, p. 35–41.

\bibitem{IWSPAAPd47}
\BIBentryALTinterwordspacing
IWSPA. (2018) Iwspa-ap/data at master · barathiganesh-hb/iwspa-ap. [Online].
  Available: \url{https://bit.ly/3gz3tmg}
\BIBentrySTDinterwordspacing

\bibitem{Security62IWSPA}
\BIBentryALTinterwordspacing
------. (2018) Security and privacy analytics anti-phishing shared task.
  [Online]. Available: \url{https://dasavisha.github.io/IWSPA-sharedtask/}
\BIBentrySTDinterwordspacing

\bibitem{RN105}
\BIBentryALTinterwordspacing
Jose. (2019) Index of jose/phishing. [Online]. Available:
  \url{https://monkey.org/~jose/phishing/}
\BIBentrySTDinterwordspacing

\bibitem{RN96}
\BIBentryALTinterwordspacing
G.~V. Cormack. (2012) Waterloo spam rankings for the clueweb09 dataset.
  [Online]. Available: \url{https://plg.uwaterloo.ca/~gvcormac/clueweb09spam/}
\BIBentrySTDinterwordspacing

\bibitem{RN106}
\BIBentryALTinterwordspacing
U.~of~Birmingham. (2012) leakiest malicious javascript example dataset.
  [Online]. Available: \url{https://bit.ly/3dUn26l}
\BIBentrySTDinterwordspacing

\bibitem{RN99}
\BIBentryALTinterwordspacing
I.~Archive. (2016) Vx heaven windows virus collection. Internet Archive.
  [Online]. Available: \url{https://bit.ly/2IzmTL9}
\BIBentrySTDinterwordspacing

\bibitem{RN107}
D.~Cappelli, A.~Moore, and R.~Trzeciak, ``The cert guide to insider threats:
  How to prevent detect and respond to information technology crimes,''
  \emph{Addison-Wesley Professional}, 2012.

\bibitem{RN108}
\BIBentryALTinterwordspacing
M.~L. Laboratory. (1998) 1998 darpa intrusion detection evaluation dataset.
  [Online]. Available:
  \url{https://www.ll.mit.edu/r-d/datasets/1998-darpa-intrusion-detection-evaluation-dataset}
\BIBentrySTDinterwordspacing

\bibitem{RN116}
M.~Stevanovic and J.~M. Pedersen, ``Machine learning for identifying botnet
  network traffic,'' 2013.

\bibitem{RN117}
\BIBentryALTinterwordspacing
APAC. (2020) Apac-anti phishing alliance of china. [Online]. Available:
  \url{http://en.apac.cn/}
\BIBentrySTDinterwordspacing

\bibitem{RN118}
N.~Bagheri, A.~Eyvani, and N.~Tarabi, ``Performance evaluation of a variable
  rate application (vra) system by artificial neural network (ann) models,''
  \emph{Agricultural Engineering International: CIGR Journal}, vol.~16, pp.
  105--111, 2014.

\bibitem{RN119}
\BIBentryALTinterwordspacing
Zone-H. (2020) Unrestricted information | defacements archive. [Online].
  Available: \url{http://www.zone-h.org/archive?hz=1}
\BIBentrySTDinterwordspacing

\bibitem{ExploitDB}
\BIBentryALTinterwordspacing
E.~Database. (2020) Exploit database submission guidelines. [Online].
  Available: \url{https://www.exploit-db.com/submit}
\BIBentrySTDinterwordspacing

\bibitem{Wooyun}
\BIBentryALTinterwordspacing
Wooyun. (2020) What kind of website is wooyun? - know almost. [Online].
  Available: \url{https://www.zhihu.com/question/19993185}
\BIBentrySTDinterwordspacing

\bibitem{RN120}
\BIBentryALTinterwordspacing
RiskAnalytics. (2018) Dns-bh – malware domain blocklist. [Online]. Available:
  \url{https://www.malwaredomains.com/}
\BIBentrySTDinterwordspacing

\bibitem{DGArchiv50}
\BIBentryALTinterwordspacing
D.~Plohmann. (2020) Dgarchive - fraunhofer fkie. [Online]. Available:
  \url{https://bit.ly/2MSUiip}
\BIBentrySTDinterwordspacing

\bibitem{baderj}
\BIBentryALTinterwordspacing
J.~Bader. (2020) baderj/domain\_generation\_algorithms: Some results of my dga
  reversing efforts. [Online]. Available: \url{https://bit.ly/2XU3DNa}
\BIBentrySTDinterwordspacing

\bibitem{osint}
\BIBentryALTinterwordspacing
BambenekConsulting. (2020, June) Dga feed. [Online]. Available:
  \url{https://bit.ly/2Yqo1oi}
\BIBentrySTDinterwordspacing

\bibitem{RN122}
\BIBentryALTinterwordspacing
Contagio. (2020) Contagio malware dump. [Online]. Available:
  \url{https://urlzs.com/nYZgB}
\BIBentrySTDinterwordspacing

\bibitem{Denisand}
\BIBentryALTinterwordspacing
Kaspersky. (2018) Denis and co. | securelist. [Online]. Available:
  \url{https://securelist.com/denis-and-company/83671/}
\BIBentrySTDinterwordspacing

\bibitem{NewFrame2020}
\BIBentryALTinterwordspacing
G.~Data. (2014) New frameworkpos variant exfiltrates data via dns requests | g
  data. [Online]. Available: \url{https://bit.ly/2AXibTl}
\BIBentrySTDinterwordspacing

\bibitem{RN123}
W.~G. Halfond and A.~Orso, ``Amnesia: analysis and monitoring for neutralizing
  sql-injection attacks,'' in \emph{Proceedings of the 20th IEEE/ACM
  international Conference on Automated software engineering}, 2005, pp.
  174--183.

\bibitem{RN124}
\BIBentryALTinterwordspacing
SQLMAP. (2020) Automatic sql injection and database takeover tool. [Online].
  Available: \url{http://sqlmap.org/}
\BIBentrySTDinterwordspacing

\bibitem{RN125}
L.~Chen, M.~Ali~Babar, and H.~Zhang, ``Towards an evidence-based understanding
  of electronic data sources,'' 2010.

\bibitem{RN126}
M.~B. Mollah, M.~A.~K. Azad, and A.~Vasilakos, ``Security and privacy
  challenges in mobile cloud computing: Survey and way ahead,'' \emph{Journal
  of Network and Computer Applications}, vol.~84, pp. 38--54, 2017.

\bibitem{RN127}
N.~Baracaldo, B.~Chen, H.~Ludwig, and J.~A. Safavi, ``Mitigating poisoning
  attacks on machine learning models: A data provenance based approach,'' in
  \emph{Proceedings of the 10th ACM Workshop on Artificial Intelligence and
  Security}, 2017, pp. 103--110.

\bibitem{RN97}
V.~Braun and V.~Clarke, ``Using thematic analysis in psychology,''
  \emph{Qualitative research in psychology}, vol.~3, no.~2, pp. 77--101, 2006.

\bibitem{Sysmonwindow}
\BIBentryALTinterwordspacing
Microsoft. Sysmon - windows sysinternals | microsoft docs. [Online]. Available:
  \url{https://bit.ly/2zu3yGG}
\BIBentrySTDinterwordspacing

\bibitem{Winlogbeat}
\BIBentryALTinterwordspacing
Elastic. Download winlogbeat | ship windows event logs | elastic | elastic.
  [Online]. Available: \url{https://www.elastic.co/downloads/beats/winlogbeat}
\BIBentrySTDinterwordspacing

\bibitem{RN110}
Wikipedia, ``Operation shady rat. available at https://goo.gl/xfnjcq,'' 2017.

\bibitem{RN111}
B.~Stock, J.~G{\"o}bel, M.~Engelberth, F.~C. Freiling, and T.~Holz,
  ``Walowdac-analysis of a peer-to-peer botnet,'' in \emph{2009 European
  Conference on Computer Network Defense}.\hskip 1em plus 0.5em minus
  0.4em\relax IEEE, 2009, pp. 13--20.

\bibitem{RN109}
T.~Holz, M.~Steiner, F.~Dahl, E.~Biersack, and F.~Freiling, ``Measurements and
  mitigation of peer-to-peer-based botnets : A case study on storm worm,''
  \emph{October}, 2008.

\bibitem{RN129}
D.~Terpstra, H.~Jagode, H.~You, and J.~Dongarra, ``Collecting performance data
  with papi-c,'' in \emph{Tools for High Performance Computing 2009}, 2010, pp.
  157--173.

\bibitem{RN113}
K.~Julisch, ``Clustering intrusion detection alarms to support root cause
  analysis,'' \emph{ACM transactions on information and system security
  (TISSEC)}, vol.~6, no.~4, pp. 443--471, 2003.

\bibitem{RN114}
P.~Sallee, \emph{Model-Based Steganography}, 2011.

\bibitem{RN115}
K.~Grahn, M.~Westerlund, and G.~Pulkkis, \emph{Analytics for Network Security:
  A Survey and Taxonomy}, 2017, pp. 175--193.

\bibitem{RN130}
K.~Solanki, N.~Jacobsen, U.~Madhow, B.~S. Manjunath, and S.~Chandrasekaran,
  ``Robust image-adaptive data hiding using erasure and error correction,''
  \emph{IEEE Transactions on Image Processing}, 2004.

\bibitem{RN131}
\BIBentryALTinterwordspacing
K.~Tools. (2020) Dns2tcp, penetration testing tools. [Online]. Available:
  \url{https://bit.ly/37pv1pf}
\BIBentrySTDinterwordspacing

\bibitem{RN132}
V.~Paxson, ``Bro: A system for detecting network intruders in real-time,''
  \emph{Computer Networks}, 1999.

\bibitem{RN133}
H.-D.~J. Jeong, W.~Hyun, J.~Lim, and I.~You, ``Anomaly teletraffic intrusion
  detection systems on hadoop-based platforms: A survey of some problems and
  solutions,'' in \emph{2012 15th International Conference on Network-Based
  Information Systems}.\hskip 1em plus 0.5em minus 0.4em\relax IEEE, 2012, pp.
  766--770.

\bibitem{RN134}
R.~Alguliyev and Y.~Imamverdiyev, ``Big data: big promises for information
  security,'' in \emph{2014 IEEE 8th International Conference on Application of
  Information and Communication Technologies (AICT)}.\hskip 1em plus 0.5em
  minus 0.4em\relax IEEE, 2014, pp. 1--4.

\bibitem{RN135}
A.~Benham, H.~Read, and I.~Sutherland, ``Network attack analysis and the
  behaviour engine,'' in \emph{2013 IEEE 27th International Conference on
  Advanced Information Networking and Applications (AINA)}.\hskip 1em plus
  0.5em minus 0.4em\relax IEEE, 2013, pp. 106--113.

\bibitem{RN136}
R.~Perdisci, G.~Giacinto, and F.~Roli, ``Alarm clustering for intrusion
  detection systems in computer networks,'' \emph{Engineering Applications of
  Artificial Intelligence}, vol.~19, no.~4, pp. 429--438, 2006.

\bibitem{ReverseShell}
\BIBentryALTinterwordspacing
Infosec. (2020) Icmp reverse shell. [Online]. Available:
  \url{https://bit.ly/2XRvLQZ}
\BIBentrySTDinterwordspacing

\bibitem{RN95}
\BIBentryALTinterwordspacing
XSSed. (2012) Cross site scripting (xss) attacks information and archive.
  [Online]. Available: \url{http://www.xssed.com/}
\BIBentrySTDinterwordspacing

\bibitem{RN137}
\BIBentryALTinterwordspacing
A.~R. in~Cyber~Systems. (2015) Comprehensive, multi-source cyber-security
  events - cyber security research. [Online]. Available:
  \url{https://csr.lanl.gov/data/cyber1/}
\BIBentrySTDinterwordspacing

\bibitem{RN138}
P.~V. Mieghem, ``The <i>n</i>-intertwined sis epidemic network model,''
  \emph{Computing}, vol.~93, no. 2-4, pp. 147--169, 2011.

\bibitem{RN141}
J.~Navarro, A.~Deruyver, and P.~Parrend, ``A systematic survey on multi-step
  attack detection,'' \emph{Computers \& Security}, vol.~76, pp. 214--249,
  2018.

\bibitem{RN143}
M.~Y. Santos, J.~P. Silva, J.~Moura-Pires, and M.~Wachowicz, ``Automated
  traffic route identification through the shared nearest neighbour
  algorithm,'' \emph{Springer}, pp. 231--248, 2012.

\bibitem{RN142}
F.~Thabtah, P.~Cowling, and Y.~Peng, ``A new multi-class, multi-label
  associative classification approach,'' in \emph{Proceeding of the 4th
  International Conference on Data Mining}, 2004, pp. 217--224.

\bibitem{RN144}
D.~Meyer, ``Support vector machines,'' \emph{R News}, 2001.

\bibitem{RN145}
C.~Li and G.~Biswas, ``Applying the hidden markov model methodology for
  unsupervised learning of temporal data,'' \emph{International Journal of
  Knowledge Based Intelligent Engineering Systems}, 2002.

\bibitem{RN146}
S.~Dua and X.~Du, ``Data mining and machine learning in cybersecurity,''
  \emph{Data Mining and Machine Learning in Cybersecurity}, 2016.

\bibitem{RN147}
C.~Musto, G.~Semeraro, P.~Lops, and M.~de~Gemmis, ``Combining distributional
  semantics and entity linking for context-aware content-based
  recommendation,'' in \emph{International Conference on User Modeling,
  Adaptation, and Personalization}.\hskip 1em plus 0.5em minus 0.4em\relax
  Springer, 2014, pp. 381--392.

\bibitem{RN148}
S.~Singh, P.~K. Sharma, S.~Y. Moon, D.~Moon, and J.~H. Park, ``A comprehensive
  study on apt attacks and countermeasures for future networks and
  communications: challenges and solutions,'' \emph{Journal of Supercomputing},
  pp. 1--32, 2016.

\bibitem{RN17}
Z.~Yan, R.~Molva, W.~Mazurczyk, and R.~Kantola, \emph{Network and System
  Security: 11th International Conference, NSS Proceedings}, 2017, vol. 10394.

\bibitem{RN150}
M.~Zheng, H.~Robbins, Z.~Chai, P.~Thapa, and T.~Moore, ``Cybersecurity research
  datasets: Taxonomy and empirical analysis,'' \emph{USENIX Workshop on Cyber
  Security Experimentation and Test}, 2018.

\bibitem{RN151}
J.~McHugh, ``Testing intrusion detection systems: a critique of the 1998 and
  1999 darpa intrusion detection system evaluations as performed by lincoln
  laboratory,'' \emph{ACM Transactions on Information and System Security},
  2000.

\bibitem{RN152}
A.~M. Al~Tobi and I.~Duncan, ``Kdd 1999 generation faults: A review and
  analysis,'' \emph{Journal of Cyber Security Technology}, vol.~2, no. 3-4, pp.
  164--200, 2018.

\bibitem{RN153}
M.~V. Mahoney and P.~K. Chan, ``An analysis of the 1999 darpa/lincoln
  laboratory evaluation data for network anomaly detection,'' in
  \emph{International Workshop on Recent Advances in Intrusion
  Detection}.\hskip 1em plus 0.5em minus 0.4em\relax Springer, 2003, pp.
  220--237.

\bibitem{RN154}
N.~Moustafa and J.~Slay, ``Unsw-nb15: a comprehensive data set for network
  intrusion detection systems (unsw-nb15 network data set),'' in \emph{2015
  military communications and information systems conference (MilCIS)}.\hskip
  1em plus 0.5em minus 0.4em\relax IEEE, 2015, pp. 1--6.

\bibitem{RN162}
J.~Read, A.~Bifet, B.~Pfahringer, and G.~Holmes, ``Batch-incremental versus
  instance-incremental learning in dynamic and evolving data,'' in
  \emph{International symposium on intelligent data analysis}.\hskip 1em plus
  0.5em minus 0.4em\relax Springer, 2012, pp. 313--323.

\bibitem{RN155}
M.~Ring, S.~Wunderlich, D.~Scheuring, D.~Landes, and A.~Hotho, ``A survey of
  network-based intrusion detection data sets,'' \emph{Computers \& Security},
  2019.

\bibitem{RN156}
A.~Gharib, I.~Sharafaldin, A.~H. Lashkari, and A.~A. Ghorbani, ``An evaluation
  framework for intrusion detection dataset,'' in \emph{2016 International
  Conference on Information Science and Security (ICISS)}.\hskip 1em plus 0.5em
  minus 0.4em\relax IEEE, 2016, pp. 1--6.

\bibitem{RN158}
T.~S. Guzella and W.~M. Caminhas, ``A review of machine learning approaches to
  spam filtering,'' \emph{Expert Systems with Applications}, vol.~36, no.~7,
  pp. 10\,206--10\,222, 2009.

\bibitem{RN160}
D.~R. Roberts, V.~Bahn, S.~Ciuti, M.~S. Boyce, J.~Elith, G.~Guillera-Arroita,
  S.~Hauenstein, J.~J. Lahoz-Monfort, B.~Schr{\"o}der, W.~Thuiller
  \emph{et~al.}, ``Cross-validation strategies for data with temporal, spatial,
  hierarchical, or phylogenetic structure,'' \emph{Ecography}, vol.~40, no.~8,
  pp. 913--929, 2017.

\bibitem{RN163}
B.~Biggio and F.~Roli, ``Wild patterns: Ten years after the rise of adversarial
  machine learning,'' \emph{Pattern Recognition}, 2018.

\bibitem{RN167}
L.~Chen, Y.~Ye, and T.~Bourlai, ``Adversarial machine learning in malware
  detection: Arms race between evasion attack and defense,'' in \emph{2017
  European Intelligence and Security Informatics Conference (EISIC)}.\hskip 1em
  plus 0.5em minus 0.4em\relax IEEE, 2017, pp. 99--106.

\bibitem{RN168}
Z.~Katzir and Y.~Elovici, ``Quantifying the resilience of machine learning
  classifiers used for cyber security,'' \emph{Expert Systems with
  Applications}, 2018.

\bibitem{zhang2019generating}
W.~E. Zhang, Q.~Z. Sheng, A.~A.~F. Alhazmi, and C.~Li, ``Generating textual
  adversarial examples for deep learning models: A survey,'' \emph{CoRR,
  abs/1901.06796}, 2019.

\bibitem{corona2013adversarial}
I.~Corona, G.~Giacinto, and F.~Roli, ``Adversarial attacks against intrusion
  detection systems: Taxonomy, solutions and open issues,'' \emph{Information
  Sciences}, vol. 239, pp. 201--225, 2013.

\bibitem{RN171}
N.~V. Chawla, ``Data mining for imbalanced datasets: An overview,'' in
  \emph{Data mining and knowledge discovery handbook}.\hskip 1em plus 0.5em
  minus 0.4em\relax Springer, 2009, pp. 875--886.

\bibitem{RN172}
G.~Haixiang, L.~Yijing, J.~Shang, G.~Mingyun, H.~Yuanyue, and G.~Bing,
  ``Learning from class-imbalanced data: Review of methods and applications,''
  \emph{Expert Systems with Applications}, vol.~73, pp. 220--239, 2017.

\bibitem{RN173}
F.~Ullah and M.~Ali~Babar, ``Architectural tactics for big data cybersecurity
  analytics systems: A review,'' \emph{Journal of Systems and Software}, vol.
  151, pp. 81--118, 2019.

\bibitem{RN170}
F.~Yang, Z.~Chen, and A.~Gangopadhyay, ``Using randomness to improve robustness
  of tree-based models against evasion attacks,'' \emph{IEEE Transactions on
  Knowledge and Data Engineering}, 2020.

\bibitem{RN174}
P.~O. Obitade, ``Big data analytics: a link between knowledge management
  capabilities and superior cyber protection,'' \emph{Journal of Big Data},
  vol.~6, no.~1, p.~71, 2019.

\bibitem{RN175}
Blueliv, ``Data breach under gdpr: How threat intelligence can reduce your
  liabilities. available at shorturl.at/apcru,'' 2017.

\bibitem{RN176}
\BIBentryALTinterwordspacing
KrebsOnSecurity, ``Wipro data breach,'' 2019. [Online]. Available:
  \url{https://bit.ly/2ovvCUE}
\BIBentrySTDinterwordspacing

\bibitem{RN177}
A.~A. C{\'a}rdenas, P.~K. Manadhata, and S.~P. Rajan, ``Big data analytics for
  security,'' \emph{IEEE Security \& Privacy}, vol.~11, no.~6, pp. 74--76,
  2013.

\bibitem{RN178}
E.~Chickowski, ``A case study in security big data analysis,'' \emph{Dark
  Reading}, vol.~9, 2012.

\bibitem{RN179}
F.~Khomh, B.~Adams, J.~Cheng, M.~Fokaefs, and G.~Antoniol, ``Software
  engineering for machine-learning applications: The road ahead,'' \emph{IEEE
  Software}, vol.~35, no.~5, pp. 81--84, 2018.

\bibitem{RN180}
C.~Islam, M.~A. Babar, and S.~Nepal, ``A multi-vocal review of security
  orchestration,'' \emph{ACM Computing Surveys (CSUR)}, vol.~52, no.~2, pp.
  1--45, 2019.

\bibitem{RN169}
S.~Chen, M.~Xue, L.~Fan, S.~Hao, L.~Xu, H.~Zhu, and B.~Li, ``Automated
  poisoning attacks and defenses in malware detection systems: An adversarial
  machine learning approach,'' \emph{Computers \& Security}, vol.~73, pp.
  326--344, 2018.

\bibitem{song2017ML}
C.~Song, T.~Ristenpart, and V.~Shmatikov, ``Machine learning models that
  remember too much,'' pp. 587--601, 2017.

\bibitem{RN62}
M.~H. Bhuyan, D.~K. Bhattacharyya, and J.~K. Kalita, ``Network anomaly
  detection: Methods, systems and tools,'' \emph{IEEE Communications Surveys \&
  Tutorials}, vol.~16, no.~1, pp. 303--336, 2014.

\end{thebibliography}

\pagebreak
\clearpage
\onecolumn
\appendix
\captionsetup{labelformat=AppendixTables}
\setcounter{table}{0}
\renewcommand\thetable{\Alph{table}}
\footnotesize  
\setlength{\tabcolsep}{2pt}
\begin{longtable}{|p{0.5cm}|m{6cm}|m{5cm}|m{6cm}|p{0.05cm}|}
\caption{Selected studies in the review. Here ID denotes study identification number.} \label{appendix} \\
\hline \multicolumn{1}{|c|}{\textbf{ID}} & \multicolumn{1}{c|}{\textbf{Title}} & \multicolumn{1}{c|}{\textbf{Authors}} &
\multicolumn{1}{c|}{\textbf{Venue}} &
\multicolumn{1}{c|}{\textbf{Year}} 
 \\ \hline 
\endfirsthead
\multicolumn{5}{c}%
{{\bfseries Appendix\  A\ -- continued from previous page}} \\
\hline \multicolumn{1}{|c|}{\textbf{ID}} & \multicolumn{1}{c|}{\textbf{Title}} & \multicolumn{1}{c|}{\textbf{Authors}} &
\multicolumn{1}{c|}{\textbf{Venue}} &
\multicolumn{1}{c|}{\textbf{Year}}  \\ \hline 
\endhead
\hline \multicolumn{5}{|c|}{{Continued on next page}} \\ \hline
\endfoot
\hline
\endlastfoot
\hline
    S1 & Automatic malware classification and new malware detection using machine learning  & Liu Liu, Bao-sheng Wang, Bo Yu, Qiu-xi Zhong & Frontiers of Information Technology  \& Electronic Engineering & 2017 \\ \hline
    S2 & detection of advanced persistent threat using machine learning correlation analysis & Ibrahim Ghafir, Mohammad Hammoudeh, Vaclav Prenosil, Liangxiu Han, Robert Hegarty, Khaled Rabie, Francisco J.Aparicio-Navarro & Future Generation Computer Systems & 2018 \\ \hline
    S3 & Steganalysis of JPEG ImageBased Steganography with Support Vector Machine & Satoshi Watanabe ,Kazuki Murakami ,Tomoya Furukawa , Qiangfu Zhao & International Conference on Software Engineering, Artificial Intelligence, Networking and Parallel/Distributed Computing   (SNPD) & 2016 \\ \hline
    S4 & Use of Machine Learning in Big Data Analytics for Insider Threat Detection & Michael Mayhew , Michael Atighetchi , Aaron Adler, Rachel Greenstadt & IEEE Military Communications Conference & 2015 \\ \hline
    S5 & Phishing detection based Associative Classification data mining & Neda Abdelhamid, Aladdin Ayesh, Fadi Thabtah & Expert Systems with Applications & 2014 \\ \hline
    S6 & Supervised Learning for Insider Threat Detection Using Stream Mining & Pallabi Parveen, Zackary R Weger, Bhavani Thuraisingham, Kevin Hamlen and Latifur Khan & IEEE International Conference on Tools with Artificial Intelligence & 2011 \\ \hline
    S7 & A New Take on Detecting Insider Threats:Exploring the use of Hidden Markov Models & Tabish Rashid, Ioannis Agrafiotis, Jason R.C. Nurse & International Workshop on Managing Insider Security Threats & 2016 \\ \hline
    S8 & NIGHTs-WATCH: a cache-based side-channel intrusion detector using hardware performance counters & Maria Mushtaq, Ayaz Akram, Muhammad Khurram Bhatti, Maham Chaudhry, Vianney Lapotre, Guy Gogniat & International Workshop on Hardware and Architectural Support for Security and Privacy & 2018 \\ \hline
    S9 & Ensemble Classifiers for Steganalysis of Digital Media & Jan Kodovsky , Jessica Fridrich ,Vojtěch Holub & IEEE Transactions on Information Forensics and Security & 2012 \\ \hline
    S10 & A DataCentric Approach to Insider Attack Detection in Database Systems & Sunu Mathew, Michalis Petropoulos , Hung Q. Ngo, Shambhu Upadhyaya & International conference on Recent advances in Intrusion detection & 2010 \\ \hline
    S11 & Identifying APT Malware Domain Based on Mobile DNS Logging & Weina Niu, Xiaosong Zhang,GuoWu Yang,Jianan Zhu, Zhongwei Ren & Mathematical Problems in Engineering & 2017 \\ \hline
    S12 & Insider-threat detection using Gaussian Mixture Models and Sensitivity Profiles & Kholood Al-tabash, Jassim Happa & Computers \& Security & 2018 \\ \hline
    S13 & Investigation of Malicious Portable Executable File Detection on the Network using Supervised Learning Techniques & Rushabh Vyas , Xiao Luo , Nichole McFarland , Connie Justice & IEEE Symposium on Integrated Network and Service Management (IM) & 2017 \\ \hline
    S14 & DNS Tunneling Detection Method Based on Multilabel Support Vector Machine & Ahmed Almusawi and Haleh Amintoosi & Security and Communication Networks & 2018 \\ \hline
    S15 & Detection of Tunnels in PCAP Data by Random Forests & Anna L. Buczak ,Paul A. Hanke ,George J. Cancro, Michael K. Toma ,Lanier A. Watkins,Jeffrey S. Chavis & Cyber and Information Security Research Conference & 2016 \\ \hline
    S16 & Detection of Malicious Webmail Attachments Based on Propagation Patterns & Yehonatan Cohen. Danny Hendler, Amir Rubin & Knowledge-Based Systems & 2017 \\ \hline
    S17 & Detection of fraudulent emails by employing advanced feature abundance & Sarwat Nizamani, Nasrullah Memon,Mathies Glasdam, Dong Duong Nguye & Egyptian Informatics Journal & 2014 \\ \hline
    S18 & Detection of Exfiltration and Tunneling over DNS & Anirban Das , Min-Yi Shen , Madhu Shashanka , Jisheng Wang & IEEE International Conference on Machine Learning and Applications (ICMLA) & 2017 \\ \hline
    S19 & Classification of Phishing Email Using Random Forest Machine Learning Technique & Andronicus A. Akinyelu and Aderemi O. Adewumi & Journal of Applied Mathematics & 2014 \\ \hline
    S20 & Multiclass JPEG Steganalysis Using Extreme Learning Machine & Veenu Bhasin, Punam Bedi & International Conference on Advances in Computing, Communications and Informatics & 2013 \\ \hline
    S21 & Automated feature engineering for HTTP tunnel detection & Jonathan J.Davis, Ernest Foo & Computers \& Security & 2016 \\ \hline
    S22 & Analysis and Detection of Malicious Data Exfiltration in Web Traffic & Areej Al-Bataineh , Gregory White & International Conference on Malicious and Unwanted Software & 2012 \\ \hline
    S23 & A Support Vector Machine-based Framework for Detection of Covert Timing Channels & Pradhumna Lal Shrestha ; Michael Hempel ; Fahimeh Rezaei ; Hamid Sharif & IEEE Transactions on Dependable and Secure Computing & 2015 \\ \hline
    S24 & An Unsupervised MultiDetector Approach for Identifying Malicious Lateral Movement & Atul Bohara ; Mohammad A. Noureddine ; Ahmed Fawaz ; William H. Sanders & IEEE 36th Symposium on Reliable Distributed Systems (SRDS) & 2017 \\ \hline
    S25 & Proposed efficient algorithm to filter spam using machine learning techniques & Ali Shafigh Aski, Navid Khalilzadeh Sourati & Pacific Science Review A: Natural Science and Engineering & 2016 \\ \hline
    S26 & Detecting Malicious URLs using Machine Learning Techniques & Frank Vanhoenshoven , Gonzalo Nápoles , Rafael Falcon , Koen Vanhoof ,Mario Köppen & IEEE Symposium Series on Computational Intelligence (SSCI) & 2016 \\ \hline
    S27 & Decision Tree Rule Induction for Detecting Covert Timing Channels in TCP/IP Traffic & Félix Iglesias,Valentin Bernhardt,Robert Annessi,Tanja Zseby & International Cross-Domain Conference for Machine Learning and Knowledge Extraction & 2017 \\ \hline
    S28 & Detecting Malicious URLs Using Lexical Analysis & Mohammad Saiful Islam Mamun, Mohammad Ahmad Rathore, Arash Habibi Lashkari,Natalia Stakhanova,Ali A. Ghorbani & International Conference on Network and System Security & 2016 \\ \hline
    S29 & Detecting DNS Tunnel through Binary-Classification Based on Behaviour Features & Jingkun Liu ; Shuhao Li ; Yongzheng Zhang , Jun Xiao , Peng Chang , Chengwei Peng & IEEE Trustcom/BigDataSE/ICESS & 2017 \\ \hline
    S30 & DTB-IDS: an intrusion detection system based on decision tree using behaviour analysis for preventing APT attacks & Daesung MoonHyungjin ImIkkyun KimJong Hyuk Park & Journal of Supercomputing & 2017 \\ \hline
    S31 & A Machine Learning based Web Spam Filtering Approach & Santosh Kumar , Xiaoying Gao , Ian Welch , Masood Mansoori & International Conference on Advanced Information Networking and Applications (AINA) & 2016 \\ \hline
    S32 & DNS tunneling detection through statistical fingerprints ofprotocol messages and machine learning & M. Aiello  M. Mongelli  G. Papaleo & International Journal of Communication Systems & 2014 \\ \hline
    S33 & Detection of DNS Tunneling in Mobile Networks Using Machine Learning & Van Thuan Do,Paal Engelstad,Boning Feng, Thanh van Do & International Conference on Information Science and Applications & 2017 \\ \hline
    S34 & Detecting DNS Tunneling Using Ensemble Learning & Saeed Shafieian,Daniel Smith,Mohammad Zulkernine & International Conference on Network and System Security & 2017 \\ \hline
    S35 & Uncovering social spammers: social honeypots + machine learning & Kyumin Lee, James Caverlee, Steve Webb & Proceedings of the 33rd international ACM SIGIR conference on research and development in information retrieval & 2010 \\ \hline
    S36 & Detecting Phishing Attacks Using Natural Language Processing and Machine Learning & Tianrui Peng , Ian Harris , Yuki Sawa & International Conference on Semantic Computing (ICSC) & 2018 \\ \hline
    S37 & PHISH-SAFE: URL Features-Based Phishing Detection System Using Machine Learning & Ankit Kumar Jain,B. B. Gupta & Cyber Security. Advances in Intelligent Systems and Computing & 2018 \\ \hline
    S38 & Detecting advanced persistent threats using fractal dimension based machine learning classifcation, & Sana Siddiqui,Muhammad Salman Khan,Ken Ferens, Witold Kinsner & International Workshop on Security And Privacy Analytics & 2016 \\ \hline
    S39 & PeerRush: Mining for unwanted P2P traffic & Babak Rahbarinia, Roberto Perdisci,Andrea Lanzi,KangLi & Journal of Information Security and Applications & 2014 \\ \hline
    S40 & EXPOSURE: A Passive DNS Analysis Service to Detect and Report Malicious Domains & Leyla Bilge, Sevil Sen, Davide Balzarotti, Engin Kirda, Christopher Kruegel & ACM Transactions on Information and System Security & 2014 \\ \hline
    S41 & Detecting Remote Access Trojans through External Control at Area Network Borders & Shuang Wu , Shengli Liu , Wei Lin , Xing Zhao , Shi Chen & IEEE Symposium on Architectures for Networking and Communications Systems (ANCS) & 2017 \\ \hline
    S42 & An Approach to Detect Remote Access Trojan in the Early Stage of Communication & Dan Jiang , Kazumasa Omote & International Conference on Advanced Information Networking and Applications & 2015 \\ \hline
    S43 & A network-based framework for RAT-bots detection & Ahmed A. Awad , Samir G. Sayed , Sameh A. Salem & IEEE Annual Information Technology, Electronics and Mobile Communication Conference (IEMCON) & 2017 \\ \hline
    S44 & A Practical Experiment of the HTTP-Based RAT Detection Method in Proxy Server Logs & Mamoru Mimura , Yuhei Otsubo , Hidehiko Tanaka , Hidema Tanaka & Asia Joint Conference on Information Security (AsiaJCIS) & 2017 \\ \hline
    S45 & A host-based framework for RAT bots detection & Ahmed A. Awad , Samir G. Sayed , Sameh A. Salem & International Conference on Computer and Applications (ICCA) & 2017 \\ \hline
    S46 & Towards detection of phishing websites on client-side using machine learning based approach & Ankit Kumar JainB. B. Gupta & Telecommunication Systems & 2018 \\ \hline
    S47 & Detection of phishing websites using an efficient feature-based machine learning framework & Routhu Srinivasa Rao, Alwyn Roshan Pais & Neural Computing and Applications & 2018 \\ \hline
    S48 & Malicious sequential pattern mining for automatic malware detection & Yujie Fan,Yanfang Ye, Lifei Chen & Expert Systems with Applications: An International Journal & 2016 \\ \hline
    S49 & Detection of Mobile Applications Leaking Sensitive Data & Yavuz Canbay , Mehtap Ulker , Seref Sagiroglu & International Symposium on Digital Forensic and Security (ISDFS) & 2017 \\ \hline
    S50 & Defending Malicious Script Attacks Using Machine Learning Classifiers & Nayeem Khan,Johari Abdullah, Adnan Shahid Khan & Wireless Communications \& Mobile Computing & 2017 \\ \hline
    S51 & Prediction of Cross-Site Scripting Attack Using Machine Learning Algorithms & B. A. Vishnu,K. P. Jevitha & International Conference on Interdisciplinary Advances in Applied Computing & 2014 \\ \hline
    S52 & Cache-based side-channel attacks detection through Intel Cache Monitoring Technology and Hardware Performance Counters & Mohammad-Mahdi Bazm , Thibaut Sautereau , Marc Lacoste , Mario Sudholt , Jean-Marc Menaud & International Conference on Fog and Mobile Edge Computing (FMEC) & 2018 \\ \hline
    S53 & Wide scope and fast websites phishing detection using URLs lexical features & Ammar Yahya Daeef , R. Badlishah Ahmad , Yasmin Yacob , Ng Yen Phing & International Conference on Electronic Design (ICED) & 2016 \\ \hline
    S54 & Boosting the phishing detection performance by semantic analysis & Xi Zhang , Yu Zeng , Xiao-Bo Jin , Zhi-Wei Yan , Guang-Gang Geng & IEEE International Conference on Big Data (Big Data) & 2017 \\ \hline
    S55 & SQLiGoT: Detecting SQL injection attacks using graph of tokens and SVM & Debabrata Kar, Suvasini Panigrahi, Srikanth Sundararajan & Computers \& Security & 2016 \\ \hline
    S56 & User-profile-based analytics for detecting cloud security breaches & Trishita Tiwari , Ata Turk , Alina Oprea , Katzalin Olcoz , Ayse K. Coskun & IEEE International Conference on Big Data (Big Data) & 2017 \\ \hline
    S57 & Detecting Information Theft Based on Mobile Network Flows for Android Users & Zhenyu Cheng , Xunxun Chen , Yongzheng Zhang , Shuhao Li , Yafei Sang & International Conference on Networking, Architecture, and Storage (NAS) & 2017 \\ \hline
    S58 & Phishing website detection using Latent Dirichlet Allocation and AdaBoost & Venkatesh Ramanathan , Harry Wechsler & IEEE International Conference on Intelligence and Security Informatics & 2012 \\ \hline
    S59 & SQL injection attack classification through the feature extraction of SQL query strings using a gap-weighted string subsequence kernel & Paul R.McWhirter, Kashif Kifayat,Qi Shi, Bob Askwith & Journal of Information Security and Applications & 2018 \\ \hline
    S60 & Website Defacements Detection Based on Support Vector Machine Classification Method & Siyan Wu, Xiaojun Tong,Wei Wang, Guodong Xin, Bailing Wang, Qi Zhou & International Conference on Computing and Data Engineering & 2018 \\ \hline
    S61 & Malware Analysis and Detection Using Data Mining and Machine Learning Classification & Mozammel Chowdhury, Azizur Rahman, Rafiqul Islam & International Conference on Applications and Techniques in Cyber Security and Intelligence & 2018 \\ \hline
    S62 & idMAS-SQL: Intrusion Detection Based on MAS to Detect and Block SQL injection through data mining & Cristian I.Pinzón, Juan F.De Paz, Álvaro Herrero, Emilio Corchado, Javier Bajo, Juan M.Corchado & Information Sciences & 2013 \\ \hline
    S63 & Real time detection of cache-based side-channel attacks using hardware performance counters & Marco Chiappetta, Erkay Savas, Cemal Yilmaz & Applied Soft Computing & 2016 \\ \hline
    S64 & Detection of algorithmically generated malicious domain names using masked N-grams & Jose S, Ricardo R, Emilio S & Expert Systems With Applications & 2019 \\ \hline
    S65 & Detection of Application-Layer Tunnels with Rules and Machine Learning & Huaqing Lin, Gao Liu, Zheng Yan & International Conference on Security, Privacy and Anonymity in Computation, Communication and Storage & 2019 \\ \hline
    S66 &  Domain Generation Algorithms detection through deep neural network and ensemble & Shuaiji Li, Tao  Huang,  Zhiwei  Qin, 
Fanfang  Zhang , Yinhong  Chang
 & World Wide Web Conference & 2019 \\ \hline
    S67 & A Byte-level CNN Method to Detect DNS Tunnels & C. Liu; L. Dai; W. Cui; T. Lin &  International Performance Computing and Communications Conference (IPCCC) & 2019 \\ \hline
    S68 & A DNS Tunneling Detection Method Based on Deep Learning Models to Prevent Data Exfiltration & Jiacheng Zhang, Li Yang, Shui Yu, Jianfeng Ma & International Conference on Network and System Security & 2019 \\ \hline
    S69 & A machine learning based approach for phishing detection using hyperlinks information & Ankit Kumar Jain \& B. B. Gupta & Journal of Ambient Intelligence and Humanized Computing & 2019 \\ \hline
    S70 & A SQL Injection Detection Method Based on Adaptive Deep Forest & Q. Li; W. Li; J. Wang; M. Cheng & IEEE Access & 2019 \\ \hline
    S71 & A stacking model using URL and HTML features for phishing webpage detection & Yukun Li, Zhenguo Yang, Xu Chen, Huaping Yuan, Wenyin Liu & Future Generation Computer Systems & 2019 \\ \hline
    S72 & A trust aware unsupervised learning approach for insider threat detection & Maryam Aldairi , Leila Karimi, James Joshi &  International Conference on Information Reuse and Integration for Data Science  & 2019 \\ \hline
    S73 & Advance Persistent Threat Detection Using Long Short Term Memory (LSTM) Neural Networks & P. V. Sai Charan, T. Gireesh Kumar, P. Mohan Anand & International Conference on Emerging Technologies in Computer Engineering & 2019 \\ \hline
    S74 & Attention-Based LSTM for Insider Threat Detection & Fangfang Yuan, Yanmin Shang, Yanbing Liu, Yanan Cao, Jianlong Tan & International Conference on Applications and Techniques in Information Security & 2019 \\ \hline
    S75 & Bidirectional LSTM Malicious webpages detection algorithm based on convolutional neural network and independent recurrent neural network & Huan-huan Wang, Long Yu, Sheng-wei Tian, Yong-fang Peng \& Xin-jun Pei  & Applied Intelligence & 2019 \\ \hline
    S76 & Classifying Malicious URLs Using Gated Recurrent Neural Networks & Jingling Zhao, Nan Wang, Qian Ma, Zishuai Cheng & International Conference on Innovative Mobile and Internet Services in Ubiquitous Computing & 2018 \\ \hline
    S77 & Context-sensitive and keyword density-based supervised machine learning techniques for malicious webpage detection & Betul Altay, Tansel Dokeroglu \& Ahmet Cosar  & Soft Computing & 2019 \\ \hline
    S78 & detection method of domain names generated by DGAs based on semantic representation and deep neural network & Congyuan X, Jizhong S, Xin D & Computers \& Security & 2019 \\ \hline
    S79 & Detecting advanced persistent threat Malware using machine learning-based threat hunting & Lin, Tien-Chih; Guo, Cheng-Chung; Yang, Chu-Sing & European Conference on Cyber Warfare and Security & 2019 \\ \hline
    S80 & detection of malicious and low throughput data exfiltration over the DNS protocol & Asaf Nadler, Avi Amino, Asaf Shabtaia & Computers \& Security & 2019 \\ \hline
    S81 & Early detection of remote access Trojan by software network behavior & Masatsugu Oya, Kazumasa Omote & International Conference on Information Security and Cryptology & 2019 \\ \hline
    S82 & Everything Is in the Name A URL Based Approach for Phishing Detection & Harshal Tupsamudre, Ajeet Kumar Singh, Sachin Lodha & International Symposium on Cyber Security Cryptography and Machine Learning & 2019 \\ \hline
    S83 & Malicious Overtones: Hunting Data Theft in the Frequency Domain with One-Class Learning & Brian A. Powell

  & ACM Transactions on Privacy and Security & 2019 \\ \hline
    S84 & MLPXSS: An Integrated XSS-Based Attack Detection Scheme in Web Applications Using Multilayer Perceptron Technique & F. M. M. Mokbal; W. Dan; A. Imran; L. Jiuchuan; F. Akhtar; W. Xiaoxi & IEEE Access & 2019 \\ \hline
    S85 & MUI-defender: CNN-Driven, Network Flow-Based Information Theft Detection for Mobile Users & Zhenyu Cheng, Xunxun Chen, Yongzheng, ZhangShuhao, LiJian Xu & International Conference on Collaborative Computing: Networking, Applications and Worksharing & 2019 \\ \hline
    S86 & Phish-Hook: Detecting Phishing Certificates Using Certificate Transparency Logs & Edona Fasllija, Hasan Ferit Enişer, Bernd Prünster & International Conference on Security and Privacy in Communication Systems & 2019 \\ \hline
    S87 & Phishing Email Detection Using Improved RCNN Model With Multilevel Vectors and Attention Mechanism & Y. Fang; C. Zhang; C. Huang; L. Liu; Y. Yang & IEEE Access & 2019 \\ \hline
    S88 & Phishing URL Detection via CNN and Attention-Based Hierarchical RNN & Y. Huang; Q. Yang; J. Qin; W. Wen & IEEE International Conference On Big Data Science And Engineering (TrustCom/BigDataSE) & 2019 \\ \hline
    S89 & SQL Injection Detection Based on Deep Belief Network & Huafeng Z, Bo Z, Hui Y, Jinxiong Z, Xiaobin Y, Fangjun L & International Conference on Computer Science and Application Engineering & 2019 \\ \hline
    S90 & User Behavior Profiling using Ensemble Approach for Insider Threat Detection & M. Singh; B. M. Mehtre; S. Sangeetha & International Conference on Identity, Security, and Behavior Analysis (ISBA) & 2019 \\ \hline
    S91 & Unsupervised Insider Detection Through Neural Feature Learning and Model Optimisation & Liu Liu, Chao Chen, Jun Zhang, Olivier De Vel, and Yang Xiang & International Conference on Network and System Security & 2019 \\ \hline
    S92 & Using hierarchical statistical analysis and deep neural networks to detect covert timing channels & Omar D, Ala A, Ghassen B, Ilyes J, Athanasios V & Applied Soft Computing & 2019 \\ \hline
\end{longtable}
\end{document}